\documentclass{article}

\PassOptionsToPackage{numbers}{natbib}

\usepackage[final]{neurips_2022}

\usepackage[utf8]{inputenc} 
\usepackage[T1]{fontenc}    
\usepackage{hyperref}       
\usepackage{url}            
\usepackage{booktabs}       
\usepackage{amsfonts}       
\usepackage{nicefrac}       
\usepackage{microtype}      
\usepackage{xcolor}         

\usepackage{tikz}
\usetikzlibrary{shapes.geometric, arrows, positioning}
\usepackage{wrapfig}
\usepackage{comment}
\usepackage{bbm} 
\usepackage{amsmath}
\usepackage{amssymb} 
\usepackage{tikz}
\usepackage{algorithm}
\usepackage{algorithmic}

\usepackage{caption}
\usepackage{subcaption}
\usepackage{graphicx}

\newtheorem{definition}{{\bf Definition}}

\makeatletter
\newcommand{\printfnsymbol}[1]{%
  \textsuperscript{\@fnsymbol{#1}}%
}
\makeatother

\title{Inference and Sampling for Archimax Copulas}

\author{%
  Yuting Ng\thanks{Contributed equally. This work was supported in part by the Air Force Office of Scientific Research under award number FA9550-20-1-0397.} \\
  Duke University\\
  \texttt{yuting.ng@duke.edu} \\
  \And
  Ali Hasan\printfnsymbol{1} \\
  Duke University\\
  \texttt{ali.hasan@duke.edu} \\  
  \AND
  Vahid Tarokh \\
  Duke University\\
  \texttt{vahid.tarokh@duke.edu} \\  
}

\begin{document}
\bibliographystyle{plainnat}

\maketitle

\begin{abstract}

Understanding multivariate dependencies in both the bulk and the tails of a distribution is an important problem for many applications, such as ensuring algorithms are robust to observations that are infrequent but have devastating effects. Archimax copulas are a family of distributions endowed with a precise representation that allows simultaneous modeling of the bulk and the tails of a distribution. Rather than separating the two as is typically done in practice, incorporating additional information from the bulk may improve inference of the tails, where observations are limited. Building on the stochastic representation of Archimax copulas, we develop a non-parametric inference method and sampling algorithm. 
Our proposed methods, to the best of our knowledge, are the first that allow for highly flexible and scalable inference and sampling algorithms, enabling the increased use of Archimax copulas in practical settings. We experimentally compare to state-of-the-art density modeling techniques, and the results suggest that the proposed method effectively extrapolates to the tails while scaling to higher dimensional data. 
Our findings suggest that the proposed algorithms can be used in a variety of applications where understanding the interplay between the bulk and the tails of a distribution is necessary, such as healthcare and safety.
\end{abstract}

\section{Introduction}

Modeling dependencies between random variables is a central task in statistics, and understanding the dependence between covariates throughout the distribution is important in characterizing a distribution outside of its areas of highest density. For example, in machine learning contexts, a major topic of interest lies in enforcing dependencies between covariates, such as spatial dependence in convolutional neural networks or temporal dependence in recurrent neural networks. Copulas are functions that model the joint dependence of random variables, and they have been successfully employed in a variety of practical modeling settings due to their ease of use and intuitive properties. 

Moreover, since copulas are used to represent cumulative distribution functions (CDFs), they have been particularly useful in situations where \emph{tail events} are important -- events that have high impact but low probability. For example, in computer vision applications, local dependence modeled by convolutions may be sufficient for the bulk of the data, but for data in the tails of the distribution non-local dependencies may be present and need to be modeled. Current successful applications of copulas have largely been limited to a few basic parametric families and low dimensional settings, preventing their widespread adoption in settings where the dependencies are complicated or the dimension is large.

Archimax copulas are a class of copulas that merges a tractable form with sufficient expressiveness.
Archimax copulas effectively balance the representation of data within the bulk of the distribution while extrapolating to the tails by combining the tractability of \emph{Archimedean} copulas with the tail properties of \emph{extreme-value copulas}.

Notably, they remove the simplified symmetry assumption among covariates that is present in Archimedean copulas and the max-stable property of extreme-value copulas. 
The use of Archimax copulas has resulted in better fit to data in applications such as healthcare~\citep{mcneil2010liouville} and hydrology~\citep{mesiar2011mle2d, chatelain2020inference}.

However, existing computational methods do not allow feasible inference and sampling for Archimax copulas, preventing their widespread use.
Therein lies the motivation behind this work.
We construct efficient and flexible inference and sampling methods using deep learning techniques and discuss how they compare to traditional means of density estimation that use existing copula methods and deep generative models. 
In addition, we provide numerical studies where the proposed Archimax techniques extrapolate to the tails better than existing methods.

\subsection{Related work}


Modeling distributions is a major task in machine learning, with techniques such as generative adversarial networks (GANs)~\citep{creswell2018generative}, normalizing flows (NFs)~\citep{papamakarios2021normalizing}, and variational autoencoders (VAEs)~\citep{kingma2013auto} being the major developments for representing complex distributions. 

However, these techniques are largely used for modeling the bulk of a distribution and may not extrapolate well to out-of-distribution samples or samples within the tails, as discussed in~\citep{le2021perfect}. 
Some methods have been proposed to represent only the tails, for example in~\citep{gobet2021tail, engelke2022tail}. However, they disregard the information in the bulk and only focus on the tails. Recently, \citet{bhatia2021exgan} considered combining GANs with extreme-value theory (EVT). 
However, this and the above methods are used only for sampling, and do not provide a way of quantifying the dependence of the observation. 


Copulas are an important technique for representing distribution functions since they allow easy separation of the marginals and the joint dependence structure of a distribution. 
Copulas have been applied in machine learning wherein techniques from machine learning have been used in conjunction with traditional copula theory to model more general classes of densities with examples of such work found in~\citep{ling2020nips_ACNet, ng2021uai_genAC, tagasovska2019nips_vae}, please see Appendix~\ref{app:background_multivariate}~and~\ref{app:background_ml} for more background on copulas and application of copulas in machine learning. However, these have generally focused on simplified assumptions such as a symmetric dependence between covariates, or a hierarchy of bivariate dependencies.
Moreover, these do not extrapolate to tail distributions and do not readily appear to generalize to high dimensions.

With regards to Archimax copulas, several theoretical works have been proposed analyzing the distributions~\citep{caperaa2000archimax2d, charpentier2014ndstochastic, mesiar2013archimaxnd}. 
In~\citet{chatelain2020inference}, the authors proposed a method for inferring a stable tail dependence function (stdf) when given an Archimedean generator. However the method assumes knowledge of the Archimedean generator or infers a one-parameter Archimedean generator from pairwise Kendall's taus. Past applications of Archimax copulas were of low dimensions, such as dimensions 2, 3 and 3 in the studies of river flow rates~\citep{mesiar2011mle2d}, rainfall~\citep{chatelain2020inference} and nutrient intake~\citep{mcneil2010liouville}.

\paragraph{Our contributions}
We propose methods for filling the gaps in the existing literature by: 
\begin{enumerate}
    \item Developing methods for inferring both the Archimedean generator and stdf;
    \item Developing methods for sampling from Archimax copulas;
    \item Providing flexible representations for both the radial and spectral components. 
\end{enumerate}

Specifically, deep generative models are used to represent the distributions of the radial and spectral components. By taking an expectation, these characterize the Archimedean generator and stdf.

\section{Background}\label{sec:background}
Copulas are given by separating the marginal distributions from the joint dependence of a random variable. 
Specifically, consider
\begin{equation}
\label{eq:def_sklar}
    F(\mathbf{x}) =C(F_1(x_1),\cdots,F_d(x_d)),
\end{equation}
where $F$ is a $d$-variate cumulative distribution function (CDF) of the random variable $\mathbf{X} = (X_1, \cdots, X_d) \in \mathbb{R}^d$, $F_j$ is the $j$th univariate margin, and $C$ is the copula describing the dependence between the uniform random variables $\mathbf{U}=(U_1,\cdots,U_d)=(F_1(X_1),\cdots,F_d(X_d))\in[0,1]^d$. Moreover, if the marginals $F_j$ are continuous, then the copula $C$ is unique~\citep{sklar1959}.\footnote{notation: random variables in uppercase, observations in lowercase, scalars are not bold, vectors are bold.}

As discussed in the introduction, Archimax copulas describe a generalization of Archimedean and extreme-value copulas.
They are defined as:
\begin{definition}[Archimax copula]
\label{thm:archimax}
An Archimax copula is given by
\begin{equation}
\label{eq:def_archimax}
    C(\mathbf{u}) = \varphi(\ell(\varphi^{-1}(u_1),\cdots,\varphi^{-1}(u_d))),
\end{equation}
where $\varphi : [0, \infty) \to [0,1]$ is an \emph{Archimedean generator} and $\ell : [0, \infty)^d \to [0,\infty)$ is a \emph{stable tail dependence function (stdf)}~\citep{caperaa1997cfg, charpentier2014ndstochastic, mesiar2013archimaxnd}.  
\end{definition}
From the definition, the Archimax copula is completely characterized by these two functions and the objective during inference lies in estimating both the \emph{stdf} and the \emph{Archimedean generator}.
Both functions have specific properties that must be fulfilled in order to obtain a valid Archimax copula. 

\begin{figure}[t]
    \centering
    \includegraphics[width=0.14\textwidth]{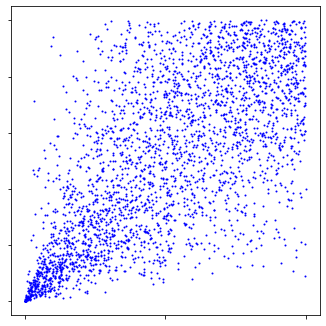}
    \includegraphics[width=0.14\textwidth]{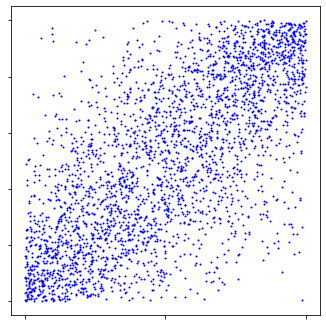}
    \includegraphics[width=0.14\textwidth]{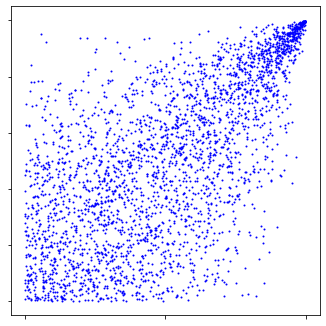}
    \includegraphics[width=0.14\textwidth]{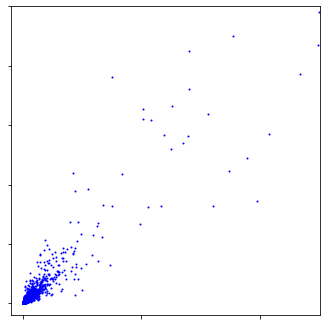}
    \includegraphics[width=0.14\textwidth]{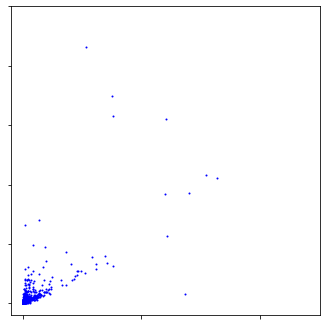}
    \includegraphics[width=0.14\textwidth]{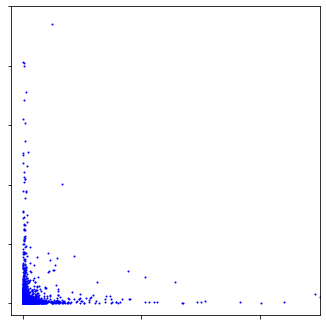}
    \caption{Examples of different radial envelopes and asymptotic dependencies.}
    \label{fig:background}
\end{figure}

The \emph{stdf} is defined as: 
\begin{definition}[stable tail dependence function (stdf)]
\label{def:stdf}
A $d$-variate stdf $\ell:[0,\infty)^d\to[0,\infty)$ is given by: 
\begin{equation}
\label{eq:def_stdf_spectral}
    \ell(\mathbf{x}) = d \int_{\Delta_{d-1}} \max_{j\in\{1,\cdots,d\}}\{x_jw_j\}\,\mathrm{d}F_\mathbf{w}(\mathbf{w}) = d \, \mathbb{E}_\mathbf{W}\left[\max_{j\in\{1,\cdots,d\}}\{x_jW_j\}\right],
\end{equation}
with a spectral random variable $\mathbf{W}\in\Delta_{d-1}$ satisfying moment constraints $\mathbb{E}_\mathbf{W}[W_j]=1/d$ for $j\in\{1,\cdots,d\}$~\citep{dehaan1984spectral, huang1992stdfspectral, ressel2013stdfspectral}.
\end{definition}
Intuitively, the \emph{stdf} dictates the asymptotic dependence between covariates, with examples given in~\figurename~\ref{fig:background}.
Notably, the definition completely relies on the distribution of $\mathbf{W}$, and the \emph{stdf} is homogeneous of order one, i.e. $\ell(cx_1,\cdots,cx_d)=c\,\ell(x_1,\cdots,x_d)$ for $c>0$.

On the other hand, the \emph{Archimedean generator} is defined by:
\begin{definition}[Archimedean generator]
\label{def:arch_gen}
An Archimedean generator $\varphi:[0,\infty)\to[0,1]$ is the Williamson $d$-transform of a random variable $R>0$:
\begin{equation}
\label{eq:def_varphi_Williamson}
    \varphi(x)=\mathcal{W}_R(x) = \int_x^\infty \left(1-\frac{x}{r}\right)^{d-1} dF_R(r) = \mathbb{E}_R\left[\left(1-\frac{x}{R}\right)_+^{d-1}\right],
\end{equation}
where $(y)_+ := \max(0,y)$ for $y \in \mathbb{R}$~\citep{mcneil2009dmonotone, williamson1956dmonotone}.
\end{definition}
The Archimedean generator has a one-to-one correspondence with the distribution of $R$~\citep{mcneil2009dmonotone} and dictates the shape of the radial envelope applied across all covariates, with examples given in~\figurename~\ref{fig:background}.

With these definitions in mind, we leverage~\citet[Theorem 3.3]{charpentier2014ndstochastic} for inference and sampling, which established that any random vector 
\begin{equation}
\label{eq:def_archimax_stochastic}
    \mathbf{X}\stackrel{d}{=}R\times(S_1,\cdots,S_d),\;\mathbf{X}\in[0,\infty)^d,
\end{equation}
has dependence that follows an Archimax copula, where $R$ and $\mathbf{S}$ are independent and known as the \emph{radial} and \emph{simplex} components, with supports $R>0$, $\mathbf{S}\in[0,1]^d$, $\ell(\mathbf{S})\leq 1$. The marginals of $\mathbf{X}$ is given by $\varphi$ such that the random vector
\begin{equation}
\label{eq:def_archimax_stochastic_U}
    \mathbf{U}\stackrel{d}{=}(\varphi(X_1),\cdots,\varphi(X_d)),\;\mathbf{U}\in[0,1]^d,
\end{equation}
follows an Archimax copula. Moreover, every Archimax copula given in~\eqref{eq:def_archimax} has a decomposition given by~\eqref{eq:def_archimax_stochastic} and~\eqref{eq:def_archimax_stochastic_U}. The \emph{radial} component $R$ is the same $R$ in Definition~\ref{def:arch_gen}, and the \emph{simplex} component $\mathbf{S}$ has a one-to-one correspondence with the \emph{spectral} component $\mathbf{W}$ from Definition~\ref{def:stdf}. 
Therefore, representing the $R$ and $\mathbf{W}$ in Definitions~\ref{def:arch_gen} and~\ref{def:stdf} respectively provides complete understanding of the distribution. 

\section{Method}

As established in Section~\ref{sec:background}, the \emph{stdf} and \emph{Archimedean generator} are expectations of functions of the \emph{spectral} component $\mathbf{W}$ and \emph{radial} component $R$ respectively. 
We first model these expectations where $\mathbf{W}$ and $R$ are discrete random variables with finite support~\citep{fougeres2013dsm, genest2011inferencend}. We then let them be outputs of generative networks in the limit of infinite support. 
We begin by describing the inference algorithm for the stdf followed by the inference algorithm for the Archimedean generator. 
Through sampling the stdf we define the relationship between $\mathbf{S}$ and $\mathbf{W}$ which we make use of in the inference for the Archimedean generator.
We finally show how the representations we use can be easily adapted for sampling from Archimax copulas. 
A flow chart describing the relationship between stdf and Archimedean generator parameter estimation is in~\figurename~\ref{fig:flow}. 

\tikzstyle{block-stdf} = [rectangle, draw, fill=blue!20, 
    text width=19em, text centered, node distance=2cm and 1cm, rounded corners, minimum height=4em]
\tikzstyle{block-gen} = [rectangle, draw, fill=red!20, 
    text width=16em, text centered, node distance=2cm and 1cm, rounded corners, minimum height=4em]

\tikzstyle{line} = [draw, -latex']
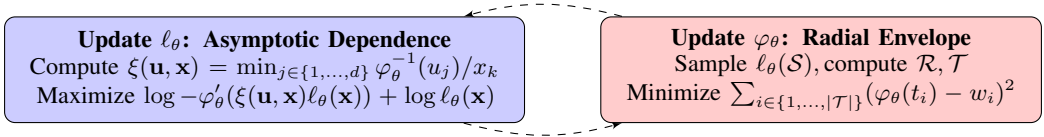
\begin{figure}[h]
    \centering
    \begin{tikzpicture}
    \node [block-stdf] (update-w) {\small\textbf{Update $\mathbf{\ell_\theta}$: \textbf{Asymptotic Dependence}} \\
    Compute $\xi(\mathbf{u},\mathbf{x}) = \min_{j\in\{1,\ldots,d\}} \varphi_\theta^{-1}(u_j) /x_k$ \\
    Maximize $\log-\varphi_\theta'(\xi(\mathbf{u},\mathbf{x}) \ell_\theta(\mathbf{x})) + \log \ell_\theta(\mathbf{x})$};
    \node [block-gen, right = of update-w] (update-r) {\small\textbf{Update $\mathbf{\varphi_\theta}$: \textbf{Radial Envelope}} \\
    Sample $\ell_\theta(\mathcal{S}),\, $compute $\mathcal{R}, \mathcal{T}$ \\ Minimize $\sum_{i \in\{1,\ldots, |\mathcal{T}|\}}(\varphi_\theta(t_i) - w_i)^2$ };

    \path [line,dashed] (update-r) edge[bend right=15] node [left] {} (update-w);
    \path [line,dashed] (update-w) edge[bend right=15] node [left] {} (update-r);

\end{tikzpicture}
    \caption{Flow chart describing relationship between stdf and Archimedean generator estimation.}
    \label{fig:flow}
\end{figure}

\subsection{Stable tail dependence function inference and sampling}\label{sec:inferW}

For this section, we suppose that the Archimedean generator $\varphi( \cdot ) $ is known and the goal is to infer or sample from the stdf. 

\subsubsection{Inference for stable tail dependence function}

We consider a stdf following Definition~\ref{def:stdf}, where $\ell$ is specified by the spectral decomposition with spectral component $\mathbf{W}$~\citep{dehaan1984spectral, huang1992stdfspectral, ressel2013stdfspectral}.
To summarize the inference for the stdf: we first establish a representation of the stdf through the empirical expectation of samples $\mathbf{w}$ from a generative network $G_\mathbf{W}$. 
We then define the likelihood of transformed data observations and optimize the parameters of $G_\mathbf{W}$ to maximize this likelihood. 

Following~\citet{chatelain2020inference}, when given the Archimedean generator $\varphi$ we can define the transformation $\xi(\mathbf{u},\mathbf{x})$ for an observation $\mathbf{u}$ and a pseudo-observation $\mathbf{x}=(x_1,\cdots,x_d)\in\Delta_{d-1}$:
\begin{equation}
\label{eq:def_inferW_xi}
    \xi(\mathbf{u},\mathbf{x})=\min\{\varphi^{-1}(u_1)/x_1,\cdots,\varphi^{-1}(u_d)/x_d\},
\end{equation}
a transformation used for estimating extreme-value copulas~\citep{caperaa1997cfg, pickands1981pdf}.

In the case of Archimax copulas, and by using the homogeneity property of the stdf~\eqref{eq:def_stdf_spectral}, the CDF of the random variable $\xi(\mathbf{U},\mathbf{x})$ is expressed as:
\begin{equation} 
\label{eq:def_inferW_xi_sdf}
    P(\xi(\mathbf{U},\mathbf{x})\leq x)= 1-P(\xi(\mathbf{U},\mathbf{x})>x)=1-C(\varphi(x\mathbf{x}))=1-\varphi(x\ell(\mathbf{x})).
\end{equation}
The full derivation is given in Appendix~\ref{app:inferW}.

Following~\citet{hasan2021uai_deepevc}, we differentiate the CDF of $\xi(\mathbf{U},\mathbf{x})$ with respect to $x$ and obtain the log-likelihood of the transformed observation as:
\begin{equation}
\label{eq:def_inferW_xi_nll}
    \log\mathcal{L}(x;\varphi,\ell)=\log\left(-\varphi'(x\,\ell(\mathbf{x}))\right) + \log(\ell(\mathbf{x})).
\end{equation}

Recalling Definition~\ref{def:stdf}, let $\ell_\theta$ be parameterized by a generative network $G_\mathbf{W}$ such that:
\begin{equation}
\label{eq:def_stdf_gnn}
    \ell_\theta(\mathbf{x}) = \frac{d}{l}\sum_{k=1}^l \left[ \max_{j\in\{1,\cdots,d\}} x_jw_{kj} \right],
\end{equation}
where $\mathbf{w}_k = (w_{k1},\cdots,w_{kd})$ for $k=\{1,\cdots,l\}$ are $l$ samples from $G_\mathbf{W}$ with dimension $d$ and output activation given by the $\text{Softmax}(\cdot)$ function to respect the support $\mathbf{W}\in\Delta_{d-1}$. The moment constraints $\mathbb{E}_\mathbf{W}[W_j]=1/d$ for $j=\{1,\cdots,d\}$ are scale conveniences and not necessities~\citep{fougeres2013dsm}. To approximate the moment constraints, we penalize the residual $\sum_{j=1}^d(\sum_{k=1}^l w_{kj}/l - 1/d)^2$. 

We may then train the generative network $G_\mathbf{W}$ such that $\ell_\theta$ approximates $\ell$ by minimizing the negative log-likelihood of transformed observations $\xi_(\mathbf{u},\mathbf{x})$, with $\mathbf{x}$ uniformly sampled on the unit simplex $\Delta_{d-1}$.  
The full technique is presented in Algorithm~\ref{alg:estimating_W_GNN} in Appendix~\ref{app:algorithms}.

The inverse $\varphi^{-1}$ in the computation of $\xi(\mathbf{u},\mathbf{x})$ in~\eqref{eq:def_inferW_xi} can be computed numerically by Newton-Raphson. The derivative $\varphi'$ in the computation of $\log\mathcal{L}(x;\varphi,\ell)$ in~\eqref{eq:def_inferW_xi_nll} can be calculated explicitly from Definition~\ref{def:arch_gen} as:
\begin{equation}
    \varphi'(x) = \mathbb{E}_R\left[\left(d-1\right)\left(-\frac{1}{R}\right)\left(1-\frac{x}{R}\right)^{d-2}_+\right].
\end{equation}
Putting these computations together results in the likelihood of the stdf given the generator.

\subsubsection{Sampling the simplex component}\label{sec:sampleS}

We now consider sampling the simplex component $\mathbf{S}$. 
We noted in the introduction that there is a one-to-one relationship between $\mathbf{S}$ and $\mathbf{W}$. Using this relationship and the result of~\citet{charpentier2014ndstochastic}, the stdf may also be written in terms of the simplex component through the following definition:
\begin{definition}[survival distribution of simplex component]\label{def:sdfS}
The survival distribution function (SDF) of the simplex component $\mathbf{S}$ is:
\begin{equation}\label{eq:def_sdfS}
    P(\mathbf{S}>\mathbf{s}) = \max(0,1-\ell(s_1,\cdots,s_d))^{d-1},
\end{equation}
\end{definition}
which relates to the so-called generalized Pareto copulas defined as:
\begin{definition}[generalized Pareto copula]
A generalized Pareto copula is defined as the copula of a generalized Pareto distribution with support on $(-\infty,0]^d$ and CDF specified by $\ell$ as:
\begin{equation}
    P(X_1< x_1,\cdots, X_d< x_d) = \max(0,1-\ell(-x_1,\cdots,-x_d)). \label{eq:def_gpc}
\end{equation}
Moreover, it has stochastic representation as:
\begin{equation}
    \mathbf{X}\stackrel{d}{=}-\left(\frac{U}{W_1},\cdots,\frac{U}{W_d}\right),\label{eq:def_gpc_stochastic}
\end{equation}
where $U$ is uniformly distributed on $[0,1]$, independent of spectral component $\mathbf{W}=(W_1,\cdots,W_d)$ from the spectral decomposition of $\ell$ in Definition~\ref{def:stdf}~\citep{buishand2008gpc2d, falk2019gpc}.
\end{definition}

Therefore, we can frame the sampling of the simplex component through the sampling from a generalized Pareto copula, such that, given samples $(x_{11}, \cdots, x_{1d}), \cdots, (x_{(d-1)1}, \cdots, x_{(d-1)d})$ of size $d-1$ from a generalized Pareto distribution with CDF in~\eqref{eq:def_gpc}, we may obtain a sample of $\mathbf{s}$, where coordinates $s_j$ for $j=\{1,\cdots,d\}$ are computed as~\citep{charpentier2014ndstochastic}:
\begin{equation}\label{eq:calc_s_from_x}
    s_j=-\max(x_{1j},\cdots,x_{(d-1)j}).
\end{equation}
The coordinate-wise maxima is taken over $d-1$ samples to get the $d-1$ exponent in the SDF of $\mathbf{S}$, corresponding to the $d-1$ exponent in the expression of the Archimedean generator $\varphi$.

Due to our convenient representation of $\ell$ as an expectation of a function of $\mathbf{W}$ from generative network $G_\mathbf{W}$, we are able to sample a generalized Pareto distribution from its stochastic representation. 
The full technique is presented in Algorithm~\ref{alg:sampling_S_GNN} in Appendix~\ref{app:algorithms}. 

\subsection{Archimedean generator inference and sampling}\label{sec:inferR}

We now assume that the stdf is known and the goal is to infer the Archimedean generator. 
We consider a general representation of the Archimedean generator following Definition~\ref{def:arch_gen}, where $\varphi$ is a $d-$monotone function specified by the Williamson $d-$transform of the radial component $R$~\citep{charpentier2014ndstochastic, mcneil2009dmonotone}. 

To provide an outline for the overall approach: we first consider the so-called Kendall distribution, which describes an integral transform of the copula. 
We then use the fact that the empirical Kendall distribution converges to the true Kendall distribution as the empirical copula converges to the true copula, providing a means of estimation. 
The Kendall distribution is formally defined as:
\begin{definition}[Kendall distribution]
\label{def:kendall_dist}
Let $C$ be a copula, and let the CDF of the random variable $\mathbf{U}\in[0,1]^d$ be the copula $C$. Define the random variable $W:=C(\mathbf{U})$. The Kendall distribution of $C$ is the multivariate probability integral transform given by the CDF of $W$:
\begin{equation}
\label{eq:def_kendall}
    K(w)=P(C(\mathbf{U})\leq w), \quad w \in [0, 1].
\end{equation}
\end{definition}

In the case of Archimax copulas, using the stochastic representation~\eqref{eq:def_archimax_stochastic} and the homogeneity property of the stdf~\eqref{eq:def_stdf_spectral}, the Kendall distribution is expressed as: 
\begin{equation}
\label{eq:def_kendall_archimax}
    K(w)=P(\varphi(R\,\ell(S_1,\cdots,S_d))\leq w).
\end{equation}
The full derivation is provided in Appendix~\ref{app:inferR_general}.
The empirical Kendall distribution $K_n$ for $n$ given observations $(u_{11},\cdots,u_{1d}),\cdots,$ $ (u_{n1},\cdots,$ $u_{nd})$ is defined as:
\begin{equation}
\label{eq:def_kendall_empiricalk}
    K_n(w)=\frac{1}{n}\sum_{i=1}^n\mathbbm{1}\{w_i\leq w\},\\
\end{equation}
where
\begin{equation}
\label{eq:def_kendall_empiricalw}
    w_i=\frac{1}{n+1}\sum_{k=1}^n\mathbbm{1}\{u_{k1}<u_{i1},\cdots,u_{kd}<u_{id}\}.
\end{equation}

In the case of symmetric dependence in the non-parametric inference of Archimedean copulas~\citep{genest2011inferencend}, $\mathbf{S}$ is uniformly distributed on the simplex $\Delta_{d-1}$ and $\ell(S_1,\cdots,S_d)=S_1+\cdots+S_d\equiv 1$.
Then, $R$ is discrete with the cardinality of the support the same as $W$ and $K_n$ is the Kendall distribution of a unique Archimedean copula~\citep{genest2011inferencend}.

In the non-parametric inference of Archimax copulas, we define the random variables:
\begin{equation}
\label{eq:def_inferR_ZT}
    Z:=\ell(S_1,\cdots,S_d)\quad \text{and}\quad T:= RZ \quad
\end{equation}
where $T$ is a discrete random variable with support the same size as $W$ and $K_n$ is the Kendall distribution of an Archimax copula. 
Note that $R$ and $Z$ are independent as $R$ and $\mathbf{S}$ are independent. 

Using $K_n$, we reconstruct the support of $R$, given $\ell$ and the support of $\mathbf{S}$, which in turn provides the support of $Z$.
With the final objective of letting $R$ and $\mathbf{S}$ be independent and identically distributed (iid) outputs of generative networks, we first (linearly) interpolate $K_n$ to be equispaced, such that each $w_i$ has the same probability $k_i=1/(n_rn_z)$, where $n_r$ and $n_z$ are the chosen sizes of supports for $R$ and $Z$. 
We describe the reconstruction procedure for the general case with non-iid random variables including cases where $r_jz_l=r_{j'}z_{l'}$ in Appendix~\ref{app:inferR_general}. 
The sizes of supports $n_r$ and $n_z$ are chosen empirically with examples given in Appendix~\ref{app:inferR_nrnz}. 

Suppose the supports of the distributions of $R, Z, W, T$ are finite and respectively denoted by: 
$\mathcal{W}=\{w_1,w_2,\cdots,w_{n_rn_z}\}$, 
$\mathcal{R}=\{r_1,r_2,\cdots,r_{n_r}\}$,
$\mathcal{Z}=\{z_1,z_2,\cdots,z_{n_z}\}$,
$\mathcal{T}=\{r_j z_l:r_j\in\mathcal{R},z_l\in\mathcal{Z}\}=\{t_1,t_2,\cdots,t_{n_rn_z}\}$,
where the $n_rn_z$ elements of $\mathcal{W}$ are sorted in decreasing (non-increasing) order, and the $n_rn_z$ elements of $\mathcal{T}$ are sorted in increasing (non-decreasing) order. This reverse ordering is due to $\varphi$ being a decreasing function.

We minimize the mean sum of square residuals motivated by the uniform convergence of the empirical process $\sqrt{n}(K_n-K)$ as $n\to\infty$ established by~\citet{barbe1996kendall}:
\begin{equation}
\label{eq:inferR_residuals}
    \frac{1}{n_rn_z}\sum_{i=1}^{n_rn_z}(w_i - \varphi_\theta(t_i))^2
\end{equation}
where, following Definition~\ref{def:arch_gen},
\begin{equation}
\label{eq:inferR_varphiti}
    \varphi_\theta(t_i) = \mathcal{W}_{\mathcal{R}}(t_i)=\frac{1}{n_r} \sum_{j=1}^{n_r}  \left(1-\frac{t_i}{r_j}\right)_+^{d-1}.
\end{equation}

The finite support assumption of $R$ and $Z$ is not necessary, and we consider a modification where the supports $\mathcal{R}, \mathcal{Z}$ are specified by samples from generative networks. The main objective is to learn the parameters of generative network $G_R$ given the empirical Kendall distribution $K_n$ and samples of $Z$.
Since scaling the support $\mathcal{R}$ by a constant $c>0$ does not change the copula, we add a regularization term for $\mathbb{E}_R[R] = 1$. The algorithm can be understood as an alternating minimization algorithm, where the map between $\mathcal{R}$ and $\mathcal{Z}$ to $\mathcal{W}$ via $\mathcal{T}$, and the support $\mathcal{R}$ are updated in an alternating fashion. The full technique is presented in Algorithm~\ref{alg:estimating_R_GNN} in Appendix~\ref{app:algorithms}. 

The learned $G_R$ also provides a source of samples for the radial component $R$, which we use to generate full samples from the Archimax copula.

\subsection{Inference and sampling for Archimax copulas}
We finally summarize the combination of inference and sampling for both the Archimedean and simplex component to obtain the full algorithm for inference and sampling for Archimax copulas. 
This culminates into an iterative technique that successively updates each component.

\subsubsection{Inference for Archimax copulas}

We initialize with $\varphi(x)=\exp(-x)$ and infer $\ell$ to learn $G_\mathbf{W}$, following Section~\ref{sec:inferW}. 
This special combination of $\varphi$ and $\ell$ corresponds to extreme-value dependence with the max-stable property. To aid inference for this initialization step, we pre-process our data to have extreme-value dependence. 
We do so by computing the \emph{block maxima}, a technique from extreme-value copulas where we group observations and take the coordinate-wise maximas within each group. 
Specifically, given observations $(x_{11},\cdots,x_{1d}),\cdots,$ $ (x_{n1},\cdots,$ $x_{nd})$, the block-maximas
$(m_{11},\cdots,m_{1d}),\cdots,$ $ (m_{k1},\cdots,$ $m_{kd})$ for $k$ blocks of size $n/k$ are computed as: 
\begin{equation}\label{eq:def_blockmaxima}
    m_{ij} = \max (x_{(n/k)(i-1)+1\,j},\cdots, x_{(n/k)(i)\,j}),
\end{equation}
for $i=\{1,\cdots,k$\} and $j=\{1,\cdots,d\}$. 
To determine the block size, with larger block sizes $n/k$ better approximating extreme-value dependence and more blocks $k$ for more observations, we test the block maximas for extreme-value dependence via the max-stable property using the test by~\citet{kojadinovic2011_evctest}, and select the first $k$ where the null hypothesis of extreme-value dependence is not rejected, starting from $k=n$, with details given in Appendix~\ref{app:initialization}.\footnote{An alternative initialization scheme based on testing different one-parameter families of Archimedean generators with the log-likelihood of transformed observations $\xi$ is also given in Appendix~\ref{app:initialization}.} 

Given an estimate of $\ell$ and a learned $G_\mathbf{W}$, we are able to generate many samples of $\mathbf{S}$, thereby providing a way to compute $Z=\ell(\mathbf{S})$. 
We then infer $\varphi$, learning $G_R$, following Section~\ref{sec:inferR}. 
At this point, we repeat the estimation for $G_\mathbf{W}$ with the updated $\varphi$ to improve our estimate for $\ell$. 
The algorithms may be iterated as needed with suggested convergence criteria such as Cramér–von Mises (CvM) distance between successively estimated copulas. 

\subsubsection{Sampling for Archimax copulas}

Given samples of the radial and simplex components, the stochastic representation of Archimax copulas~\eqref{eq:def_archimax_stochastic} gives a straightforward method for sampling Archimax copulas~\citep{charpentier2014ndstochastic}.
Specifically, we sample $\mathbf{S}$ and $R$, then multiply and normalize by $\varphi$; see Algorithm~\ref{alg:sampling_archimax_GNN} in Appendix~\ref{app:algorithms} for details. 

\section{Experiments}
To understand the modeling and sampling capabilities of the proposed algorithms, we conduct a number of empirical studies. 
We compare the proposed method to a number of existing copula based and deep learning based methods for density estimation. 
Our experiments relate to the main focus of the proposed method where we are interested in understanding the dependencies between the variables in both the bulk and the tail. 
In that sense, we conduct a number of experiments where we wish to extrapolate to the tail.
Further experimental details may be found in Appendix~\ref{app:experiments}. 

The metric we use to compare the methods is based on the Cram\'er-von Mises (CvM) statistic, which computes the $L^2$ distance between the empirical copula and the estimated copula. 
This statistic is commonly used to determine differences between distributions in goodness-of-fit tests~\citep{remillard2007_gof}. We use a version where we compare the empirical copula of true samples to the empirical copula of generated samples. Specifically, it can be written as: $CvM = \int (C_{*,n}(\mathbf{u}) - C_{\theta,n}(\mathbf{u}))^2\mathrm{d} \mathbf{u}$, where $C_{*,n}$ is the empirical copula of true samples and $C_{\theta,n}$ is the empirical copula of generated samples~\citep{remillard2007_gof}.

\paragraph{Inference for Archimedean generator}

Our first set of experiments involve inference for the Archimedean generator following the proposed method in Section~\ref{sec:inferR}.
We consider data with dimension $d=10$ and sample size of $n=1000$ given the true stdf $\ell$. To the best of our knowledge, our proposed method is the first method for non-parametric inference of flexible Archimedean generators in Archimax copulas. As such, there are no baselines for comparison. Instead, we compute the results in terms of the map $\lambda(w) = \varphi^{-1}(w)/(\varphi^{-1}(w))', w\in(0,1)$ due to its scale invariant property and known asymptotic variance, which was described as a useful metric for how well $\varphi$ is fit in~\citep{genest1993inference2d, genest2011inferencend}. We evaluated our proposed method on the Clayton (C), Frank (F), Joe (J) and Gumbel (G) generators, representing different radial envelopes, for Kendall's tau of $\tau=\{0.2,0.5\}$, representing different associations. The stdf comes from the family of negative scaled extremal Dirichlet (NSD), a flexible class of stdf~\citep{belzile2017nsd}. All estimates were within the asymptotic variance of $\lambda(w), w\in(0,1)$. The next best method is comparing to a Clayton generator estimated by pairwise Kendall's tau~\citep{chatelain2020inference}. We give the results in terms of MSE to $\lambda$ in Table~\ref{tab:givenGT}, and plot estimates of $\lambda(w)$ in Figure~\ref{fig:estimating_varphi_stdfGT} in Appendix~\ref{app:results_estimating_varphi_stdfGT}. Additional experiment results, including small sample performance $n=200$, and choices of support sizes $n_r$ and $n_z$, are in Appendix~\ref{app:results_estimating_varphi_stdfGT}.

\paragraph{Inference for stable tail dependence function and sampling for simplex component}

We now consider the reverse scenario where we wish to estimate the stdf $\ell$ given the true Archimedean generator $\varphi$. We compute the integrated relative absolute error (IRAE) between the estimated $\ell_\theta$ and the true $\ell$. The IRAE is given by $    IRAE(\ell,\ell_\theta) = \frac{1}{|\Delta_{d-1}|}\int_{\Delta_{d-1}}|\ell(\mathbf{x})-\ell_\theta(\mathbf{x})|/\ell(\mathbf{x}) \, \mathrm{d}\mathbf{x}$~\citep{chatelain2020inference}. For the same experiment settings as above, results are in Table~\ref{tab:givenGT}. Additional experiment results, including time taken, are in Appendix~\ref{app:results_estimating_S_varphiGT}. 

\begin{table}[h]
\caption{Inference of $\varphi$ given true $\ell$ and inference of $\ell$ given true $\varphi$\vspace{5pt}} \label{tab:givenGT}
\centering \footnotesize
\begin{tabular}{lllllllll}
\toprule
 & \textsc{C 0.2}      & \textsc{C 0.5}     & \textsc{F 0.2}      & \textsc{F 0.5}      & \textsc{J 0.2}      & \textsc{J 0.5}    & \textsc{G 0.2}      & \textsc{G 0.5} \\\midrule
 \textsc{MSE \tiny $ \times 10^{-3}$}~\citep{chatelain2020inference} & \textbf{0.01} & \textbf{0.04} & 0.9 & 9 & 1 & 9 & 0.7 & 8\\
 \textsc{MSE {\tiny $ \times 10^{-3}$} (Ours)} & 0.2 & 0.2 & \textbf{0.1} & \textbf{0.1} & \textbf{0.3} & \textbf{0.1} & \textbf{0.2} & \textbf{0.1}\\\midrule
\textsc{IRAE $\pm0.01$}~\citep{chatelain2020inference}      
  & 0.05       & 0.11      & 0.04       & 0.04       & 0.05       & 1.00     & 0.06       & 0.15   \\
\textsc{IRAE $\pm0.01$ (Ours)}     
  & 0.06 & 0.12 & 0.04 & 0.05 & 0.06 & \textbf{0.07} & 0.08 & 0.15 \\\bottomrule
\end{tabular}
\end{table}

\paragraph{Modeling nutrient intake}


The USDA studied the nutrient intake of women~\citep{usda1985nutrient}. 
One particular task is understanding the dependencies between the intake of different nutrients. We can model this using an Archimax copula to understand the dependencies between the nutrients in the bulk and the tail. 
To assess this, we fit and compare models from the literature on representing distributions and compute the CvM goodness-of-fit statistic for each model. 
Specifically, we break our comparison into two different types of models: copula based models and deep network based models.
For the copula models, for methods marked with *, we use the $\varphi$ described in Section~\ref{sec:inferR} and for methods marked with $\dagger$ we use the $\ell$ described in Section~\ref{sec:sampleS}.
For the deep generative models, we use standard methods based on the Wasserstein GAN~\citep{arjovsky2017wasserstein}, masked autoregressive flow (MAF)~\citep{papamakarios2017masked}, and variational autoencoders (VAE)~\citep{kingma2013auto}. 
The results are presented in Table~\ref{tab:nutrient}.
The proposed method (Gen-AX) with the Archimax has the lowest CvM distance among all the competing methods, suggesting that the proposed method is recovering the true dependency structure. 
The state of the art Clayton Archimax (C-AX) did not perform well possibly due to the difficulty in scaling the single parameter Clayton generator to higher dimension. The Archimedean copula (AC *) possibly benefited from the use of our proposed $\varphi$.
We additionally provide examples of samples versus the ground truth in Appendix~\ref{app:results_nutrient} for the different methods as well as an explanation of the different abbreviations. 

\begin{table}[h]
\caption{Goodness-of-fit to nutrient intake data and 100d NSD copula \vspace{5pt}}
\label{tab:nutrient}
\resizebox{\textwidth}{!}{%
\begin{tabular}{@{}lllllllllllll@{}}
\toprule
     & \multicolumn{7}{c}{\textsc{Copulas}}                  &        & \multicolumn{3}{c}{\textsc{Deep Nets}} & \multicolumn{1}{c}{\textsc{Ours}} \\ \midrule
    & GC    & RV  & CV  & DV  & AC *  & HAC & EV $\dagger$  & C-AX $\dagger$  & GAN   & MAF    & VAE    & Gen-AX *$\dagger$                 \\
Nutrient \textsc{CvM} {\footnotesize $ \times 10^{-3}$} & 0.081 & 0.3 & 0.2 & 0.6 & 0.030 & 0.2 & 0.4  & 0.6    & 0.033  & 0.036  & 0.053  & \textbf{0.026}           \\ \midrule 
C-NSD \textsc{CvM} {\footnotesize $ \times 10^{-5}$} & - & - & - & - & - & - & - & - & 16 & \textbf{3} & 5 & \textbf{3} \\
\bottomrule
\end{tabular}%
} \vskip -10pt
\end{table}

\paragraph{Extrapolating to extreme rainfall}

Archimax copulas were initially developed as a tool to study the behaviour of methods that estimate the joint distribution of extreme events~\citep{caperaa2000archimax2d}. The extreme-value copula that arises in the limit can be understood from the stdf $\ell$ and the index of regular variation of the Archimedean generator $\varphi$. 
Unlike extreme-value copulas which emerge from the limiting distribution of extreme events, the motivation for the use of Archimax copulas is to model extreme data, where observations are rare, from a mix of moderately less extreme data, where observations are relatively more abundant.

In this experiment, we consider another realistic dataset which models the monthly rainfall in French Britanny as studied by~\citet{chatelain2020inference}. We are interested in testing how well the proposed method can extrapolate to the extremes from non-extreme data. Specifically, we analyze the monthly rainfall data, which did not pass the test of extreme-value dependence~\citep{kojadinovic2011_evctest, chatelain2020inference}.

We first train the models on the full dataset. We then generate many samples from the trained model, compute the block maxima, then estimate the extremal dependence from the block maxima using the CFG estimator~\citep{caperaa1997cfg}. The results are presented in Table~\ref{tab:extreme} where we compare the proposed method to deep generative models. Additional experiment details and results, including plots of samples from the bulk and the extremes, are in Appendix~\ref{app:results_CNSD}. 
We did not compare to copula based models since classical copulas are generally Gumbel or independent in the extremes and thus not suitable for this application. However, for the purpose of modeling monthly rainfall without extrapolating to extremes, we compared to a variety of copula based models, including skew-t copulas which is a class of flexible asymmetric copulas~\citep{demarta2005tcopula, kollo2010skewtcopula, yoshiba2018skewtmle, smith2012skewtmcmc}.
The full details and results are given in Appendix~\ref{app:results_CNSDmontly}.

\paragraph{Out-of-distribution detection}

Using the same realistic dataset as above, we added outliers generated uniformly at random on the unit cube. The AUC and F1 scores for outlier detection based on likelihoods are given in Table~\ref{tab:ood}. Additional experiment details and figures are in Appendix~\ref{app:results_ood}.

\vspace{5pt}
\begin{minipage}[t]{0.6\textwidth}
\centering \footnotesize
\captionof{table}{Goodness-of-fit to dependence in the extremes \vspace{5pt} }\label{tab:extreme}
\begin{tabular}{@{}llllll@{}}
\toprule
 IRAE $\pm$ 0.01    & \textsc{GAN}  & \textsc{MAF}  & \textsc{VAE}  & \textsc{C-AX} & \textsc{Ours} \\ \midrule
\textsc{C-NSD} & 0.52 & 0.42 & 0.16 & 0.12 & \textbf{0.03} \\
\textsc{F-NSD} & 0.10 & 0.15 & \textbf{0.04} & \textbf{0.04} & \textbf{0.04} \\
\textsc{J-NSD} & \textbf{0.03} & 0.48 & \textbf{0.03} & 0.08 & \textbf{0.03} \\
\textsc{G-NSD} & 0.07 & 0.38 & 0.08 & 0.16 & \textbf{0.04} \\\bottomrule
\end{tabular}
\end{minipage}
\begin{minipage}[t]{0.4\textwidth}
\centering \footnotesize
\captionof{table}{Out-of-distribution detection \vspace{5pt} }\label{tab:ood}
\begin{tabular}{@{}llll@{}}
\toprule
               & \textsc{MAF}  & \textsc{VAE}  & \textsc{Ours} \\ \midrule
\textsc{AUC}   & 0.82 & 0.37 & \textbf{0.92} \\
\textsc{F1}    & 0.48 & 0.04 & \textbf{0.72} \\\bottomrule
\end{tabular}
\end{minipage}
\vspace{5pt}

\paragraph{High dimensional modeling}
We finally consider an experiment where we infer and sample data from a 100-dimensional Clayton-NSD Archimax copula. 
Scaling to high dimensions is an important property of the proposed method since many existing copula methods fail to scale beyond lower dimensions. 
As such, we only compare with deep generative models since the existing copula models resulted in numerical errors during optimization.
We report the CvM statistic in the second line of Table~\ref{tab:nutrient}. 
We additionally provide examples of the samples in Appendix~\ref{app:results_100d}. 

\section{Conclusion}

We developed highly flexible and scalable inference and sampling algorithms, facilitating the use of Archimax copulas in practical settings.
We experimentally compare to state-of-the-art density modeling techniques, and the results suggest that the proposed method effectively extrapolates to tails while scaling to higher dimensional data. 
The methods are especially useful in scenarios requiring extrapolations to the tails while also incorporating data from the bulk. 
\paragraph{Limitations and future work}

A single Archimedean generator $\varphi$ to describe the radial envelope across all coordinates may not be sufficiently expressive for certain datasets. For these cases, hierarchical Archimax copulas may be more appropriate and a direction of future work~\citep{hofert2018harchimax}. Other directions for future work include modifying the generator architectures to allow modeling of temporal dependence and also application of Archimax copulas to describe dependencies of non-tabular data, such as via graph neural networks~\citep{ma2020iclr_copulagnn}.

\paragraph{Potential negative societal impacts}

Model misspecification may lead to misspecification of risks, leading to potentially catastrophic outcomes in areas such as healthcare, safety and finance. Risks may be mitigated by confirming a reasonable fit between the observations generated by the model and the data.

\appendix

\section{Algorithms and Background}

\subsection{Algorithms}\label{app:algorithms}

We provide algorithms to describe the algorithmic contributions in the main paper. 
Algorithm~\ref{alg:estimating_W_GNN} describes the method for learning the stable tail dependence function (stdf) given the Archimedean generator. 
Algorithm~\ref{alg:sampling_S_GNN} describes sampling the simplex component with a learned spectral measure. 
Algorithm~\ref{alg:estimating_R_GNN} describes learning the Archimedean generator with a given stdf. 
Finally, a sampling algorithm for the full Archimax copula is provided in Algorithm~\ref{alg:sampling_archimax_GNN} assuming a learned generator and stdf.
The code is attached in the supplementary material.

\begin{algorithm}[H]
\caption{Learn stable tail dependence function (stdf)}\label{alg:estimating_W_GNN}
\begin{algorithmic}
\STATE \textbf{input} observations $\{\mathbf{u}_i:i=1,...,m\}$.
\STATE \textbf{input} Archimedean generator $\varphi_\theta$.
\STATE \textbf{initialize} $G_\mathbf{W}$. \\
\algorithmicdo \hskip 2pt \algorithmicwhile $\,\text{loss}=\text{NLL}+\text{reg}$ \text{\textbf{not converged}}:
\item \hskip 15pt \textbf{sample} $\{\mathbf{x}_j:j=1,...,n\}$ from $\text{Unif}(\Delta_{d-1})$.  \\
\item \hskip 15pt \textbf{compute} $\{\xi(\mathbf{u}_i,\mathbf{x}_j):i=1,...,m,\;j=1,...,n\}$ from~\eqref{eq:def_inferW_xi}.
\item \hskip 15pt \textbf{sample} $\{\mathbf{w}_k:k=1,...,l\}$ from $G_\mathbf{W}$.  \\
\item \hskip 15pt \textbf{compute} $\{\ell_\theta(\mathbf{x}_j): j=1,...,n\}$ from~\eqref{eq:def_stdf_gnn}.
\item \hskip 15pt \textbf{compute} $\text{NLL}=\frac{1}{mn} \sum_{i=1,j=1}^{i=m,j=n} \log \mathcal{L}(\xi(\mathbf{u}_i,\mathbf{x}_j);\varphi_\theta,\ell_\theta)$ from~\eqref{eq:def_inferW_xi_nll}. 
\item \hskip 15pt \textbf{compute} $\text{reg}=\sum_{j=1}^d(\sum_{k=1}^l w_{kj}/l - 1/d)^2$.
\item \hskip 15pt \textbf{descent} $\text{argmin}_{G_\mathbf{W}} \, \text{NLL}+\text{reg}$.\\
\algorithmicend \hskip 2pt \algorithmicwhile
\RETURN learned $G_\mathbf{W}$.
\end{algorithmic}
\end{algorithm}

\begin{algorithm}[H]
\caption{Sample simplex component}\label{alg:sampling_S_GNN}
\begin{algorithmic}
\STATE \textbf{input} learned $G_\mathbf{W}$.\\
\algorithmicdo \hskip 2pt \algorithmicwhile $\,$ $\ell_\theta({\mathbf{s}})>1$:
\item \hskip 15pt \textbf{sample} $\{u_i:i=1,...,d-1\}$ from $\text{Unif}(0,1)$.\\
\item \hskip 15pt \textbf{sample} $\{(w_{i1},...,w_{id}):i=1,...,d-1\}$ from $G_\mathbf{W}$.\\
\item \hskip 15pt \textbf{compute} $\{(x_{i1},...,x_{id}):i=1,...,d-1\}$ from~\eqref{eq:def_gpc_stochastic}.
\item \hskip 15pt \textbf{compute} $\mathbf{s}=(s_1,...,s_d)$ from~\eqref{eq:calc_s_from_x}.
\item \hskip 15pt \textbf{compute} $\ell_\theta(\mathbf{s})$ from~\eqref{eq:def_stdf_gnn}.\\
\algorithmicend \hskip 2pt \algorithmicwhile
\RETURN a sample $\mathbf{s}=(s_1,...,s_d)$.
\end{algorithmic}
\end{algorithm}

\begin{algorithm}[H]
\caption{Learn Archimedean generator}\label{alg:estimating_R_GNN}
\begin{algorithmic}
\STATE \textbf{input} Kendall observations $\mathcal{W} = \{w_i: i=1,...,n_rn_z\}$.
\STATE \textbf{input} learned $G_Z$ such that $Z:=\ell_\theta(\mathbf{S})$, $\mathbf{S}$ sampled from learned $G_\mathbf{W}$ with Algorithm~\ref{alg:sampling_S_GNN}.
\STATE \textbf{initialize} $G_R$.
\STATE \textbf{sort} $\mathcal{W}$ in decreasing (non-increasing) order.\\
\algorithmicdo \hskip 2pt \algorithmicwhile $\;\text{MSE}=\sum_{i=1}^{n_rn_z} (w_i-\varphi_\theta(t_i))^2/(n_rn_z) > \epsilon$ :
\item \hskip 15pt \textbf{sample} $\mathcal{R}=\{r_1,...,r_{n_r}\}$ from $G_R$. 
\item \hskip 15pt \textbf{sample} $\mathcal{Z}=\{z_1,...,z_{n_z}\}$ from $G_Z$. 
\item \hskip 15pt \textbf{compute} $\mathcal{T}=\{r_jz_l, \, r_j\in\mathcal{R},z_l\in\mathcal{Z}\}$.
\item \hskip 15pt \textbf{sort} $\mathcal{T}$ in increasing (non-decreasing) order.
\item \hskip 15pt \textbf{compute} $(\varphi_\theta(t_1),...,\varphi_\theta(t_{n_rn_z}))$ from~\eqref{eq:inferR_varphiti}.
\item \hskip 15pt \textbf{descent} $\text{argmin}_{G_R} \sum_{i=1}^{n_rn_z} (w_i-\varphi_\theta(t_i))^2/(n_rn_z)+(\sum_{j=1}^{n_r}r_j/n_r-1)^2$.\\
\algorithmicend \hskip 2pt \algorithmicwhile
\RETURN learned $G_R$.
\end{algorithmic}
\end{algorithm}

\begin{algorithm}[H]
\caption{Sample Archimax copulas}\label{alg:sampling_archimax_GNN}
\begin{algorithmic}
\STATE \textbf{input} learned $G_R$, $G_\mathbf{W}$
\STATE \textbf{sample} $r, r_1, \cdots, r_{n_r}$ from $G_R$.
\STATE \textbf{sample} $\mathbf{s}=(s_1,\cdots,s_d)$ from $G_\mathbf{W}$ with Algorithm~\ref{alg:sampling_S_GNN}.
\STATE \textbf{compute} $\mathbf{u} = (\varphi_\theta(rs_1),\cdots,\varphi_\theta(rs_d))$ from~\eqref{eq:inferR_varphiti}.
\RETURN a sample $\mathbf{u}=(u_1,\cdots,u_d)$.
\end{algorithmic}
\end{algorithm}

\subsection{Stable tail dependence function inference and sampling}\label{app:inferW_sampleS}

\subsubsection{Inference for stable tail dependence function}\label{app:inferW}

\paragraph{Pickands transformation}

The Pickands transformation in~\eqref{eq:def_inferW_xi} for a particular test point $\mathbf{x}\in\Delta_{d-1}$ results in the transformed observation $\xi(\mathbf{U},\mathbf{x})$ with survival distribution function (SDF): 
\begin{align*}
            P(\xi(\mathbf{U},\mathbf{x})>x) 
        &= P(\min\{\varphi^{-1}(U_1)/x_1,\cdots,\varphi^{-1}(U_d)/x_d\}>x), \\
        &= P(\varphi^{-1}(U_1)>xx_1,\cdots,\varphi^{-1}(U_d)>xx_d), \\
        &= P(U_1< \varphi(xx_1),\cdots,U_d< \varphi(xx_1)), && \varphi \text{ is decreasing}, \\
        &= \varphi(\ell(xx_1,\cdots,xx_d)), && C(\mathbf{u}):=\varphi(\ell(\varphi^{-1}(\mathbf{u}))),\\
        &= \varphi(x\ell(x_1,\cdots,x_d)), && \ell(c\mathbf{x})=c\ell(\mathbf{x}),c>0,\\
        &= \varphi(x\ell(\mathbf{x})).
\end{align*}

In the case of extreme-value copulas, $\varphi(x):=\exp(-x)$, $\varphi^{-1}(u):=-\log(u)$ and $C(\mathbf{u}):=\exp(-\ell(\allowbreak-\log(u_1),\cdots,-\log(u_d)))$, such that the $d-$dimensionsal observations distributed according to the extreme-value copula are transformed into $1-$dimensional observations distributed according to an exponential with rate $\ell(\mathbf{x})$. The stdf at a particular test point $\ell(\mathbf{x})$ is then estimated from the mean of the transformed observations, with endpoint corrections, as~\citep{pickands1981pdf}:
\begin{equation}\label{eq:stdfhat_pickands}
    \hat{\ell}(x) = \left(\sum_{i=1}^n-\log\left(\frac{i}{n+1}\right)\right)/\left(\sum_{i=1}^n\xi(\mathbf{u}_i,\mathbf{x})\right).
\end{equation}

A modification to the Pickands estimator, using the transformation $\log(\xi(\mathbf{U},\mathbf{x}))$, leads to the CFG estimator~\citep{caperaa1997cfg}:
\begin{equation}\label{eq:stdfhat_CFG}
    \log(\hat{l}(\mathbf{x}))=\frac{1}{n}\sum_{i=1}^n\log\left(-\log\left(\frac{i}{n+1}\right)\right)-\frac{1}{n}\sum_{i=1}^n\log(\xi(\mathbf{u}_i,\mathbf{x})).
\end{equation}

A modification to the CFG estimator was made by \citet{chatelain2020inference} for the case of Archimax copulas:
\begin{equation}\label{eq:stdfhat_CFGmodified}
    \log(\hat{l}(\mathbf{x}))=\frac{1}{n}\sum_{i=1}^n\log\left(\varphi^{-1}\left(\frac{i}{n+1}\right)\right)-\frac{1}{n}\sum_{i=1}^n\log(\xi(\mathbf{u}_i,\mathbf{x})).
\end{equation}

In the main paper, we provide the likelihood of a transformed observation $\xi(\mathbf{u},\mathbf{x})$ in~\eqref{eq:def_inferW_xi_nll} and directly train the generative network to maximize this likelihood. We do not have an additional endpoint correction step since the stochastic form of $\ell_\theta$ in~\eqref{eq:def_stdf_gnn} is a valid stdf. 
On the other hand, the Pickands and CFG estimators above have corrected endpoints but may not be valid stdfs~\citep{hasan2021uai_deepevc}.
In the additional experimental results in Appendix~\ref{app:results_estimating_S_varphiGT} our estimator, although based on the Pickands estimator, performs better than the Pickands estimator. This may be due to our estimator representing the class of valid stdfs, a phenomenon noted in~\citep{segers2008projectdsm} where projecting to the class of valid stdfs reduced estimation error. 
This direct method of training may perform better than the alternative method that first estimates $\hat{\ell}$ then trains the generative network to match $\hat{\ell}$. A future direction may be to improve the robustness of our MLE estimator with the CFG modification.

\paragraph{Architecture of generative network for spectral component}

$G_\mathbf{W}$ is a generic multilayer perceptron with layers: $(\text{Linear}(d,d_h)$, $\text{BatchNorm}(d_h)$, $\text{ReLU}()$, $\text{Linear}(d_h,d_h)$, $\text{BatchNorm}(d_h)$, $\text{ReLU}()$, $\text{Linear}(d_h, d)$, $\text{Softmax}())$, where $d$ is the dimension of the observations and $d_h$ is the number of nodes in the hidden layers. The number of layers and number of hidden nodes may be modified as needed. Batch normalization, i.e. $\text{BatchNorm}(\cdot)$, greatly helps in preventing samples from being static during training. 

\paragraph{Details of Algorithm~\ref{alg:estimating_W_GNN}}
To reduce computational complexity, we use mini-batch gradient descent and a smaller number of samples in the empirical expectations of $\varphi_\theta$ and $\ell_\theta$ during training.

\subsubsection{Sampling the simplex component}\label{app:sampleS}

\paragraph{Details of Algorithm~\ref{alg:sampling_S_GNN}} 
From Definition~\ref{def:sdfS}, the marginals of the simplex component are distributed as $\text{Beta}(1,d-1)$, where $d$ is the dimension. 
To enforce the marginals, we compute the samples of the empirical copula $\mathbf{u}$, then apply the quantile function, i.e. inverse cumulative distribution function (CDF), to obtain $\mathbf{s}$ such that:
\begin{equation}\label{eq:marginalS}
    s_j = 1-(1-u_j)^\frac{1}{d-1} \;\; \text{for} \;\; j=\{1,\cdots,d\}.
\end{equation}

Given $n$ observations $(x_{11},\cdots,x_{1d}),\cdots,(x_{n1},\cdots,x_{nd})$ the samples of the empirical copula $(u_{11},\cdots,\allowbreak u_{1d}),\cdots,(u_{n1},\cdots,u_{nd})$ are coordinate-wise rank-normalized such that:
\begin{equation}\label{eq:def_empiricalcopula_u}
    u_{ij} = \frac{1}{n}\sum_{k=1}^n\mathbbm{1}\{x_{kj}\leq x_{ij}\} \;\; \text{for} \;\; i=\{1,\cdots,n\}, j=\{1,\cdots,d\}.
\end{equation}

\subsection{Archimedean generator inference and sampling}\label{app:inferR_general}

\paragraph{Kendall distribution function} 
The expression of $K(w)$ for Archimax copulas is:
\begin{align*}
    K(w)&=P(C(U_1,\cdots,U_d)\leq w),\\
    &=P(\varphi(\ell(\varphi^{-1}(U_1),\cdots,\varphi^{-1}(U_d)))\leq w), && C(\mathbf{u})
    :=\varphi(\ell(\varphi^{-1}(\mathbf{u})),\\
    &=P(\varphi(\ell(RS_1,\cdots,RS_d))\leq w), && \mathbf{U}\stackrel{d}{=}\varphi(R\mathbf{S}),\\
    &=P(\varphi(R\,\ell(S_1,\cdots,S_d))\leq w), && \ell(c\,\mathbf{s})=c\,\ell(\mathbf{s}),c>0,\\
    &=P(\varphi(RZ)\leq w),&& Z:=\ell(\mathbf{S}),\\
    &=P(\varphi(T)\leq w),&& T:=RZ.
\end{align*}

\paragraph{Reconstruction of radial distribution with non-iid random variables and repetition of elements} To provide an outline for the overall approach, given the current estimate of $\mathcal{R}$, we compute the mapping between $\mathcal{R},\mathcal{Z}$ and $\mathcal{T}$, solve for the probabilities $P_\mathcal{R}$, update the support $\mathcal{R}$, and iterate as needed. The algorithm can be understood as an alternating minimization algorithm, where the map between $\mathcal{R},\mathcal{Z}$ to $\mathcal{W}$ via $\mathcal{T}$, and the support $\mathcal{R}$ are updated in an alternating fashion.

We initialize by computing $\mathcal{W}=\{w_1,\cdots,w_m\}$ and $P_\mathcal{W}=\{p(w_1),\cdots,p(w_m)\}$ of the empirical Kendall distribution function in~\eqref{eq:def_kendall_empiricalk}~and~\eqref{eq:def_kendall_empiricalw}, where $m\leq n_rn_z$. We do not perform the additional (linear) interpolation step of the main paper. We then sort $\mathcal{W}$ in decreasing order. We also initialize $\mathcal{R}=\{r_{n_r}=1,\,r_j=r_{j+1}\alpha_j$ for $j=\{1,...,n_r-1\}\}$, where we select $\mathbf{\alpha}=(0.9,...,0.9)$. The support $\mathcal{Z}$ and probabilities $P_\mathcal{Z}$ are assumed to be given.

Given $\mathcal{Z}$ and the current estimate of $\mathcal{R}$, we compute
\begin{equation}\label{eq:inferR_supportT}
    \mathcal{T}=\{r_jz_l:r_j\in\mathcal{R},z_l\in\mathcal{Z}\}.
\end{equation}

We then sort $\mathcal{T}$ in increasing order and compute the ordering
\begin{equation}\label{eq:inferR_orderingrz}
    (\sigma_r,\sigma_z)(i):\{1,\cdots,m\}\to\{1,\cdots,n_r\}\times\{1,\cdots,n_z\}
\end{equation}
defined as a surjective function such that
\begin{equation}\label{eq:inferR_ti}
    t_i=r_{\sigma_r(i)}z_{\sigma_r(i)}.
\end{equation}

We solve for the probabilities $P_\mathcal{R}=\{p(r_1),\cdots,p(r_{n_r})\}$ by minimizing the residuals
\begin{equation}\label{eq:inferR_pr}
    \sum_{i=1}^m\left(\left(\sum_{(\sigma_r^{-1}(j),\sigma_z^{-1}(l))=i}p(r_j)p(z_l)\right)-p(w_i)\right)^2.
\end{equation}

We solve for the support $\mathcal{R}=\{r_1,\cdots,r_{n_r}\}$, by minimizing the residuals 
\begin{equation}\label{eq:inferR_residuals_app}
\sum_{i=1}^m(\varphi_\theta(t_i)-w_i)^2
\end{equation}
where, following Definition~\ref{def:arch_gen},
\begin{equation}\label{eq:inferR_varphiti_app}
    \varphi_\theta(t_i)=\sum_{j=1}^{n_r}p(r_j)\left(1-\frac{t_i}{r_j}\right)_+^{d-1}.
\end{equation}

\paragraph{Numerical illustration} 
We give the details of the algorithm with a simple numerical example. 

Consider the supports $\mathcal{R}=(r_1, r_2, r_3) = (1,2,3)$, $\mathcal{Z}=(z_1, z_2)=(0.5, 0.75)$ with probabilities $P_\mathcal{R}=(0.4, 0.4, 0.2)$, $P_\mathcal{Z}=(0.25, 0.75)$ and noisy observations $\mathcal{W}=(0.07, 0.19, 0.34, 0.49, 0.67)$, $P_\mathcal{W}=(0.07, 0.33, 0.14,0.31,0.15)$.

In this case, $(t_1, t_2, t_3, t_4, t_5) = (0.5, 0.75, 1.0, 1.5, 2.25)$ corresponds to $(r_1z_1, r_1z_2, r_2z_1, r_2z_2 \cup r_3z_1, r_3z_2)$ with probabilities $(0.1, 0.3, 0.1, 0.3+0.05, 0.15)$.

For $P_\mathcal{R}$, we solve the overdetermined system of linear equations: 
\begin{equation}
\begin{pmatrix}
0.25 &  &  \\
0.75 &  &  \\
 & 0.25 &  \\
 & 0.75 & 0.25 \\
 &  & 0.75 \\
\end{pmatrix}
\begin{pmatrix}
p_1 \\
p_2 \\
p_3
\end{pmatrix} = 
\begin{pmatrix}
0.07 \\
0.33 \\
0.14 \\
0.31 \\
0.15
\end{pmatrix}
\end{equation}
and obtain the solution as $(0.43, 0.37, 0.20)$, where the solution has been normalized to sum to 1. 

For $\mathcal{R}$, we minimize the residuals in~\eqref{eq:inferR_residuals_app}. Since scaling such that $\mathcal{R}=\{cr_1,\cdots,cr_{n_r},c>0\}$ does not change the copula, we solve for $\mathcal{R}$ in terms of ratios $(\alpha_1,\cdots,\alpha_{n_r-1})$, recursively defined such that $r_{n_r}=1$ and $r_j=r_{j+1}\alpha_j$ for $j=\{1,\cdots, n_r-1\}$.  

The full technique is presented in Algorithm~\ref{alg:estimating_R}, with code attached in the supplementary material.

\begin{algorithm}
\caption{Estimate radial component}\label{alg:estimating_R}
\begin{algorithmic}
\STATE \textbf{input} $\mathcal{W} = \{w_i: i=1,...,m\}, \; P_\mathcal{W} = \{p(w_i): i=1,...,m\}$.
\STATE \textbf{input} $\mathcal{Z} = \{z_l: l=1,...,n_z\},\; P_\mathcal{Z} = \{p(z_l):l=1,...,n_z\}$.
\STATE \textbf{initialize} $(\alpha_1,...,\alpha_{n_r-1})$ for instance $(0.9,...,0.9)$. 
\STATE \textbf{sort} $\mathcal{W}$ in decreasing order.\\
\algorithmicdo \hskip 2pt \algorithmicwhile $\;\text{MSE}=\frac{1}{m}\sum_{i=1}^{m} (\varphi_\theta(t_i)-w_i)^2 > \epsilon$ :
\item \hskip 15pt \textbf{compute} $\mathcal{R}=\{r_{n_r}=1,\,r_j=r_{j+1}\alpha_j$ for $j=1,...,n_r-1\}$. 
\item \hskip 15pt \textbf{compute} $\mathcal{T}=\{r_jz_l, \, r_j\in\mathcal{R},z_l\in\mathcal{Z}\}$.
\item \hskip 15pt \textbf{sort} $\mathcal{T}$ in increasing order.
\item \hskip 15pt \textbf{compute} $\{(\sigma_r,\sigma_z)(i):i=1,...,m\}$ from~\eqref{eq:inferR_ti}.
\item \hskip 15pt \textbf{solve} $P_\mathcal{R}$ from~\eqref{eq:inferR_pr}.
\item \hskip 15pt \textbf{compute} $(\varphi_\theta(t_1),...,\varphi_\theta(t_m))$ from~\eqref{eq:inferR_varphiti_app}.
\item \hskip 15pt \textbf{solve} $\text{argmin}_{\alpha} \sum_{i=1}^{m} (w_i-\varphi_\theta(t_i))^2$ such that $\mathbf{\alpha}\in(\alpha_l,\alpha_u)$ for instance (0.01,1).\\
\algorithmicend \hskip 2pt \algorithmicwhile
\RETURN Estimated support $\mathcal{R}$ and probabilities $P_\mathcal{R}$.
\end{algorithmic}
\end{algorithm}

The ratios $\mathbf{\alpha}$ may be solved iteratively for a unique solution. In our case, motivated by the uniform convergence of the empirical process $\sqrt{n}(K_n-K)$ as $n\to\infty$~\citep{barbe1996kendall}, we optimize for a least-squares solution with bounds $\mathbf{\alpha}\in(0.01,1)$ using \verb|scipy.optimize.least_squares|~\citep{lstsqrSTIR1999}. 

The disadvantage of the above general approach compared to the approach presented in the main paper is the direct relationship between the computation cost and $n_r, n_z$, the sizes of supports for $R, Z$.

\paragraph{Architecture of generative network for radial component} 

$G_R$ is a generic multilayer perceptron with layers: $(\text{Linear}(1,d_h)$, $\text{BatchNorm}(d_h)$, $\text{ReLU}()$, $\text{Linear}(d_h,d_h)$, $\text{BatchNorm}(d_h)$, $\text{ReLU}()$, $\text{Linear}(d_h, 1)$, $\text{Exp}())$, where $d_h$ is the number of nodes in the hidden layers. The number of layers and number of hidden nodes may be modified as needed. Unlike in $G_\mathbf{W}$, the use of batch normalization is not essential in $G_R$. 

\paragraph{Details of Algorithm~\ref{alg:estimating_R_GNN}}
To speed up training, we initially resample $\mathcal{Z}$ only once every $k$ mini-batch iterations, decreasing $k$ until $k=1$ as we approach convergence.

\subsection{Inference and sampling for Archimax copulas}

\subsubsection{Inference for Archimax copulas}\label{app:initialization}

\paragraph{Pre-process for extreme-value dependence for initial estimate of stdf} To determine the block size for the block maximas in~\eqref{eq:def_blockmaxima}, we use the test for extreme-value dependence via the max-stable property by~\citet{kojadinovic2011_evctest}, where the max-stable property is defined as:
\begin{equation}\label{eq:evctest}
    C(u_1,\cdots,u_d) = C^r(u_1^{1/r},\cdots,u_1^{1/r}), \text{ for } r=1,2,\cdots, \mathbf{u}\in[0,1]^d.
\end{equation}

The Cramér–von Mises (CvM) distance in~\eqref{eq:def_CvM} between $C(u_1,\cdots,u_d)$ and $C^r(u_1^{1/r},\cdots,u_1^{1/r})$ is computed using Monte Carlo integration with samples $\mathbf{u}$ drawn uniformly at random from $[0,1]^d$.

\paragraph{Details of Algorithm~\ref{alg:estimating_W_GNN} for initial estimate of stdf} Randomizing the order of observations may help to create different block-maximas in each mini-batch iteration.

\paragraph{Alternative initialization scheme}

We also considered initialization with different one-parameter families of Archimedean generators, with choice of generator based on the highest log-likelihood of transformed observation $\xi$, from equations~\eqref{eq:def_inferW_xi} and~\eqref{eq:def_inferW_xi_nll}. The parameter for each family may be computed from an average of inversion of pairwise Kendall tau, as per the following equation~\citep{caperaa2000archimax2d}, for each pair:
\begin{equation}
    \tau_{\varphi,\ell} = \tau_\ell + (1-\tau_\ell)\tau_\varphi,
\end{equation}

where $\tau_{\varphi,\ell}$ is the Kendall's tau of the Archimax bivariate marginal, $\tau_\ell$ is the Kendall's tau of the extreme-value component and $\tau_\varphi$ is the Kendall's tau of the Archimedean component. An average of inversion of pairwise Kendall tau was employed in~\citep{chatelain2020inference}, with emphasis on the Clayton generator.

Initialization with specific families of Archimedean generators might bias initialization, and thus we suggested initializing via the Archimedean generator first set to $\varphi(x)=\exp\{-x\}$ representing extreme-value copulas and pre-processing the initial data to have extreme-value dependence via block-maximas. This was also motivated by the experiment on extrapolating to extremes.

\paragraph{Identifiability}

It follows from a result of~\citet{chatelain2020inference} that the sources of non-identifiability in modeling Archimax copulas are only in: (i) power transformation of $\varphi$ and $\ell$, and (ii) the scale ambiguity of $\varphi$. 
The power transformation of $\varphi$ and $\ell$ can be illustrated through the following example: Consider both pairs of generators and stdf $(\varphi_a(x) = \exp(-x^{1/\theta})$ and $\ell_a(\mathbf{x})=\|\mathbf{x}\|_1)$ and $(\varphi_b(x) = \exp(-x)$ and $\ell_b(\mathbf{x})=\|\mathbf{x}\|_\theta = (x_1^\theta+\cdots+x^{\theta})^{1/\theta})$. 
Both $(\varphi_a, \ell_a)$ and $(\varphi_b, \ell_b)$ lead to the same Archimax copula.  
The scale ambiguity of $\varphi$ comes from the fact that $\varphi(cx), c >0$ leads to the same Archimax copula. 

We note that these sources of non-identifiability are non-issues in our methods. For the power transformation, there is no ambiguity of $\varphi$ and $\ell$ since the class of $1-\varphi(1/\cdot)$ where $\varphi(\cdot)$ is calculated as the Williamson $d$-transform of $R$ with a finite support is regularly varying with index $-1$~\citep{chatelain2020inference, belzile2017nsd}. 
In addition, we include a regularization term such that $\mathbb{E}_R[R]=1$.

\subsubsection{Sampling for Archimax copulas}

\paragraph{Details of Algorithm~\ref{alg:sampling_archimax_GNN}}
Given learned generative networks $G_R, G_\mathbf{W}$, we can generate many samples from the Archimax copula.

\subsection{Background on Archimax copulas}\label{app:background_Archimax}
Archimax copulas generalize \emph{Archimedean} and \emph{extreme-value} copulas. They allow asymmetry and arbitrary tail dependence. They were initially developed as a tool to study the behaviour of methods used to estimate the joint distribution of extreme events~\citep{caperaa1997cfg}.
The main motivation for Archimax copulas is to model extreme data (e.g. very strong and rare earthquakes) from a mix of moderately less extreme data (e.g. strong earthquakes) and extreme data. 
This in turn can be used to generate samples for further studies and simulations. 

Archimax copulas were applied to applications such as nutrient intake~\citep{mcneil2010liouville}, river flow rates~\citep{mesiar2011mle2d} and rainfall~\citep{chatelain2020inference}, where the dependence is asymmetric and sub-asymptotic. In these applications, the authors noted a better fit when using Archimax copulas over Archimedean and extreme-value copulas.

We provide a few connections between Archimax, Archimedean and extreme-value copulas:
\begin{itemize}
    \item When the \emph{Archimedean generator} $\varphi(x) = \exp(-x)$, and the \emph{radial component} $R\sim\text{Erlang}(d)$, Archimax copulas reduce to  extreme-value copulas.
    \item When the \emph{stable tail dependence function (stdf)} $\ell(\mathbf{x})=(x_1+\cdots+x_d)=\|\mathbf{x}\|_1$ and the simplex component $\mathbf{S}\sim\text{Unif}(\Delta_{d-1})$, Archimax copulas reduce to Archimedean copulas.
\end{itemize}

Archimax copulas have intuitive interpretations, such as scale mixture of extremes, dependent frailties and resource sharing. In the case of resource sharing, $R>0$ is a resource to be distributed randomly among $d$ agents in a way specified by $\mathbf{S}$, where both $R$ and $\mathbf{S}$ are themselves results of independent random processes. For example, $R$ may be profits, and $\mathbf{S}$ may be the way profits is to be divided between stakeholders.

\subsection{Background on multivariate copulas}\label{app:background_multivariate}

Copulas are cumulative distribution functions (CDFs) of dependent uniform random variables. They summarize the dependence described by an arbitrary joint CDF after the marginals have been normalized to be uniform. They provide easy marginalization and calculation of tails. In addition, when used in a graphical model, some conditional independence that cannot be easily represented with Markov random fields or Bayesian networks, can be easily represented with cumulative distribution networks~\citep{huang2008uai_cdn}. They are also particularly convenient in some applications, such as ranking, where the likelihood is a CDF~\citep{huang2008nips_rank}.

For an introduction to copulas, the following textbooks and collection of works are great resources~\citep{genest2007everything, nelsen2007introduction, joe2014dependence, jaworski2010_theoryapp, jaworski2012_mathfinance}.

We also summarize the common multivariate copulas in Table~\ref{tab:common_copulas}.
\begin{table}[h]
\caption{Multivariate copulas\vspace{5pt}}\label{tab:common_copulas}
\centering 
\begin{tabular}{ll}
\toprule
\textsc{Gaussian (GC)}                
& $\Phi_\mathbf{R}(\Phi^{-1}(u_1),\cdots,\Phi^{-1}(u_d))$                                                                                   
    \\
    
\textsc{Vine (RV, CV, DV)}                     
& $\prod_{e\in\mathcal{E}(\mathcal{V})} c_{U_{e_1},U_{e_2}|\{U_{e_d}\}}(F_{U_{e_1}|\{U_{e_d}\}}(u_{e_1}),F_{U_{e_2}|\{U_{e_d}\}}(u_{e_2}))$ 
    \\

\textsc{Archimedean (AC)}              
& $\varphi(\varphi^{-1}(u_1)+\cdots+\varphi^{-1}(u_d))$                                                                                     
    \\

\textsc{H. Archimedean (HAC)} 
& $C_0(C_1(\cdots),\cdots,C_J(\cdots)), \, C_i\in \textsc{AC}$                                                                                                     
    \\

\textsc{Extreme-Value (EV)}            
& $\exp\{-\ell(-\log(u_1),\cdots,-\log(u_d))\}$\\\bottomrule                                                                         
\end{tabular}
\end{table}

The Gaussian copula (GC) has a tractable expression for both the CDF and the density. However, it is independent in the tails, a significant reason why Gaussian copula is not suitable for modeling financial risks. Vine copulas, such as R-vines (RV), C-vines (CV) and D-vines (DV), are computationally intensive and hard to interpret due to repeated conditioning with pair copulas. Archimedean copulas are symmetric in all coordinates, which is an assumption that is usually not held in practice. Hierarchical Archimedean copulas aim to break this symmetry but are difficult to construct due to nesting conditions that are hard to satisfy. Extreme-value copulas are max-stable copulas which results in lower tail independence, an assumption that is sometimes not held in practice. 
Many copulas are not flexible in the extremes, which leads to independence, except in the case of Archimdean copulas which only Gumbel copulas satisfy the tail dependence. In high dimensions, none of the existing copulas typically fit data well. 
Model misspecification is often accepted in return for tractability, and some dependence is better than independence~\citep{hofert2013highdim}.

Inferring the parameters of a copula is usually done via maximum likelihood estimation if a density can be computed, or by using minimum distance estimator and goodness-of-fit tests if a density cannot be computed. In both cases, expectations are usually replaced by their empirical versions. For more background on estimating copulas, see~\citep{charpentier2007estimation}. 

Sampling from a copula using the conditional sampling method with Rosenblatt transform is usually not possible in high dimensions, due to repeated differentiation. In our experiments, the conditional sampling method, using automatic differentiation in \verb+PyTorch+ breaks down at dimension $d=4$. In general, only models with stochastic representation may be easy to sample~\citep{maischerer2017sampling}. 

\subsection{Background on copulas in machine learning }\label{app:background_ml}

Copulas is a rising topic in machine learning, as evident from numerous publications, including but not limited to:
\begin{itemize}
    \item Cumulative distribution networks, modeled as a product of copulas~\citep{huang2008uai_cdn, huang2010aistats_graddsp, huang2011jmlr_cdn}. They can represent some conditional independencies not represented by Markov random fields and Bayesian networks, allow loops~\citep{huang2010nips_loopygraph} and mixed graphs~\citep{silva2011aistats_mixed}, with application to ranking~\citep{huang2008nips_rank} and heavy-tailed distributions~\citep{huang2010nips_loopygraph}.
    \item Copula Bayesian networks, modeled as a product of conditional copulas~\citep{elidan2010nips_copulabayesian}, with application to missing data~\citep{elidan2010uai_missing}, classification~\citep{elidan2012aistats_classifier}, time-series~\citep{eban2013aistats_dynamic} and fast structure learning~\citep{tenzer2013uai_speedystructurelearning, tenzer2014uai_mixedstructurelearning}.
    \item Copula processes~\citep{ghahramani2010nips_copulaproc} and application of copulas to time-series~\citep{lopez2013icml_gpvine, hernandez2013nips_condtime, salinas2019nips_prochighdim, wen2019icml_quantile-copula}.
    \item Copula based dependence measures and distances~\citep{ma2011ieee_mi, samo2021aistats_mi, barnab2012icml_copulaMMD, chang2016aistats_dependence, mazaheri2020nips_cvm}.
    \item Copula variational inference, allowing dependencies between latent variables~\citep{tran2015nips_copulavarinf, han2016aistats_copulavarinf, hirt2019nips_copulavarinf}.
    \item Generative modeling~\citep{tagasovska2019nips_vae, chilinski2020uai_neuralLikelihoods, laszkiewicz2021icml_copulaflows, janke2021nips_GAN, gobet2021tail, engelke2022tail}.
    \item Applications of copula in areas including graph neural networks~\citep{ma2020iclr_copulagnn},
    multi-label learning~\citep{liu2019nips_multilabel},
    multi-agent interactions~\citep{wang2021ecmlkdd_multi},
    bundle pricing~\citep{letham2014icml_bundle},
    missing value~\citep{wang2014aistats_missing, landgrebe2020icml_missing, zhao2020nips_matrix},
    sparse representation~\citep{wieser2018iclr_infobottleneck},
    outlier detection~\citep{li2020icdm_outlierdet},
    causal discovery~\citep{cui2018uai_causallearning},
    domain adaptation and transfer learning~\citep{lopezpaz2012nips_domainadap, salinas2020icml_transferlearning}, structure learning~\citep{change2019aistats_vinestructurelearning} and
    variational gradient descent~\citep{gong2019icml_quantilesvgd}.
    \item Recent work on deep network based copulas, including Archimedean~\citep{ling2020nips_ACNet, ng2021uai_genAC}, extreme-value~\citep{hasan2021uai_deepevc}, autoregressive~\citep{ostrovski2018icml_AIQN, kamthe2021_copulaflows} and transformer-attentional copulas~\citep{drouin2022_tactis}.
\end{itemize}

\section{Experiments}\label{app:experiments}

The metric we use to compare the methods is based on the Cram\'er-von Mises (CvM) statistic~\citep{remillard2007_gof} which is defined as:
\begin{equation}\label{eq:def_CvM}
    \text{CvM} = \int (C_{*,n}(\mathbf{u}) - C_{\theta,n}(\mathbf{u}))^2\,\mathrm{d} \mathbf{u},
\end{equation}
where $C_{*,n}$ is the empirical copula of true samples and $C_{\theta,n}$ is the empirical copula of generated samples and the integral is computed using Monte Carlo integration with 10,000 samples of $\mathbf{u}$ drawn uniformly at random from $[0,1]^{d-1}$.

The empirical copula for $n$ given observations $(u_{11},\cdots,u_{1d}),\cdots,(u_{n1},\cdots,u_{nd})$ is defined as:
\begin{equation}\label{eq:def_empiricalcopula}
    C_n(\mathbf{u}) = \frac{1}{n}\sum_{i=1}^n\mathbbm{1}\{u_{i1}\leq u_1,\cdots,u_{id}\leq u_d\}.
\end{equation}

All timings are with a 2.7 GHz Intel Core i7, 16GB 2133MHz LPDDR3.

\subsection{Inference for Archimedean generator}\label{app:results_estimating_varphi_stdfGT}

We summarize the common Archimedean generators in Table~\ref{tab:common_archimedean}.
\begin{table}[h]
\caption{Archimedean generators\vspace{5pt}}\label{tab:common_archimedean}
\centering 
\begin{tabular}{llll}
\toprule
& $\varphi_\theta(x)$ & $\theta_{\tau=0.2}$ &  $\theta_{\tau=0.5}$   \\\midrule

\textsc{Clayton (C)}                
& $(1+x)^{-1/\theta}$ & 0.5 & 2 \\

\textsc{Frank (F)}                     
& $-\log(1-(1-\exp(-\theta))\exp(-x))/\theta$ & 1.86 & 5.74\\

\textsc{Joe (J)}              
& $1-(1-\exp(-t))^{1/\theta}$ & 1.44 & 2.86 \\

\textsc{Gumbel (G)} 
& $\exp(-t^{1/\theta})$ & 1.25 & 2 \\\bottomrule                                                                         
\end{tabular}
\end{table}

The Clayton (C) generator is lower tail dependent, upper tail independent, the Frank (F) generator is symmetric in both lower and upper tails, the Joe (J) and Gumbel (G) generators are lower tail independent and upper tail dependent. Thus the above generators represent different radial envelopes. 

The map $\lambda$ is commonly used to estimate and evaluate estimates of $\varphi$~\citep{genest1993inference2d, genest2011inferencend}. The map $\lambda$ is defined as:
\begin{equation}
\label{eq:def_lambda}
    \lambda(w) = \varphi^{-1}(w)/(\varphi^{-1}(w))'=\{\varphi'\circ\varphi^{-1}(w)\}\varphi^{-1}(w),
\end{equation}
such that an estimate of $\varphi$ can be recovered from $\lambda$ as
\begin{equation}
\varphi^{-1}(w)=\exp\left\{\int_{w_0}^w1/\lambda(t)dt\right\}.
\end{equation} 
It is more convenient to present results in terms of $\lambda$ since it is scale invariant, unlike $\varphi$, where $\varphi(cx)$ for any $c>0$ lead to the same copula. 
In dimension $d=2$, $\lambda$ is directly related to the Kendall distribution function as $\lambda(w)= w-K(w)$. 
As such, the asymptotic variance for $\lambda_n$ may be computed from the asymptotic variance for $K_n$, where $n$ is the number of observations. This relationship is more complicated in dimension $d>2$ but often used as an approximate, useful for drawing confidence bands around $\lambda_n$ to reject models whose $\lambda_\theta$ fall outside the band.
In addition, the asymptotic variance of the independence copula may be easily computed as:
\begin{equation}\label{eq:def_lambda_sigma}
    \sigma^2_{\lambda_n}(x) = \frac{x(x-\log(x)-1)}{n}.
\end{equation}

\clearpage

\paragraph{Selection of support sizes $n_r,n_z$}\label{app:inferR_nrnz}

\begin{figure}[h]
    \centering
    \includegraphics[width=0.325\textwidth]{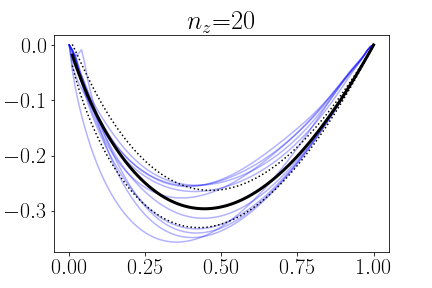}
    \includegraphics[width=0.325\textwidth]{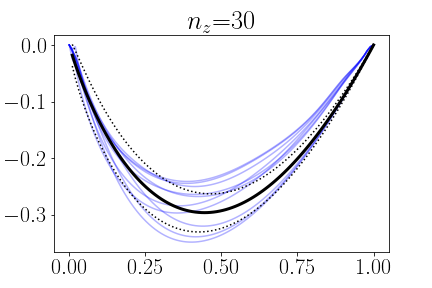}
    \includegraphics[width=0.325\textwidth]{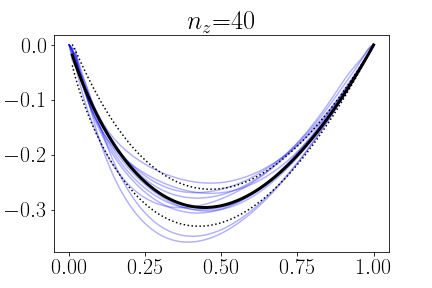}
    \includegraphics[width=0.325\textwidth]{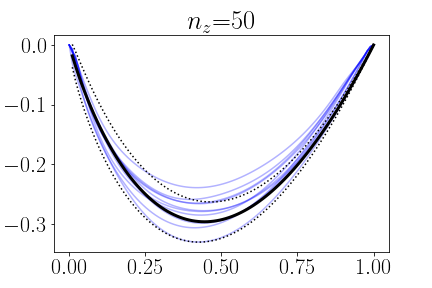}
    \includegraphics[width=0.325\textwidth]{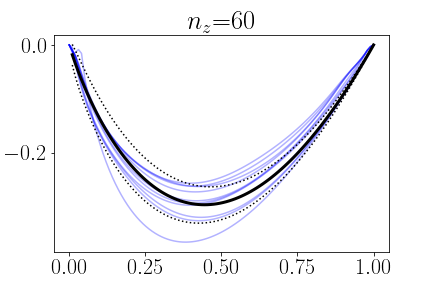}
    \includegraphics[width=0.325\textwidth]{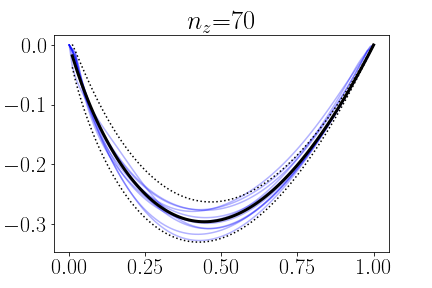}
    \includegraphics[width=0.325\textwidth]{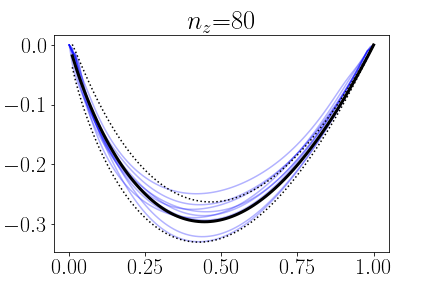}
    \includegraphics[width=0.325\textwidth]{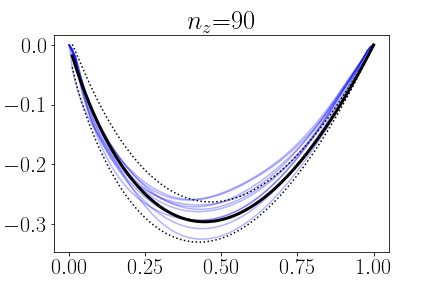}
    \includegraphics[width=0.325\textwidth]{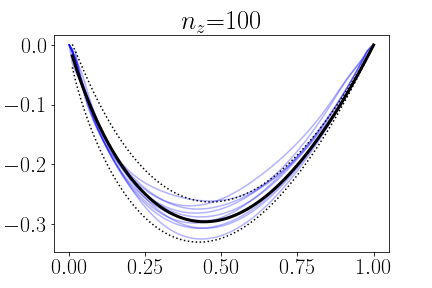}
    \caption{Estimates of $\lambda$ with $n_r=100$ and $n_z\in\{20,\cdots,100\}$. Ground truth in black, approximate confidence bands in dotted black. Estimates from different runs in blue.}
    \label{fig:selecting_nz_lambda}
\end{figure}

\begin{figure}[h]
    \centering
    \includegraphics[width=0.31\textwidth]{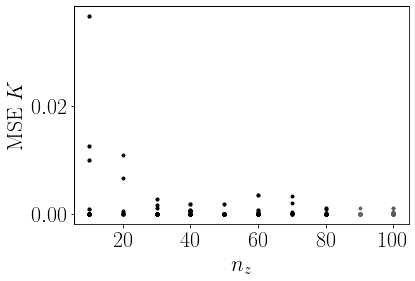}
    \includegraphics[width=0.32\textwidth]{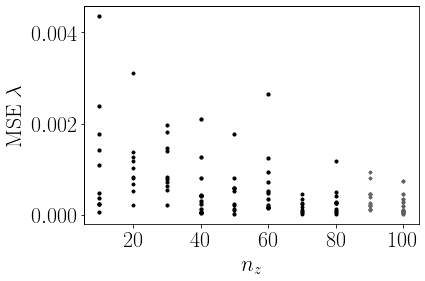}
    \includegraphics[width=0.31\textwidth]{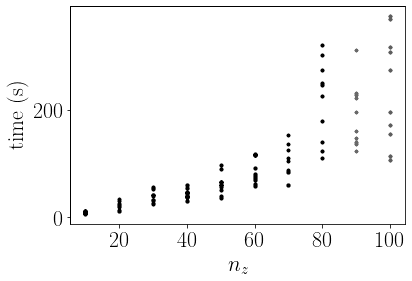}
    \caption{Mean squared error (MSE) in fitting $K$ and $\lambda$ and the time taken.}
    \label{fig:selecting_nz_MSE_time}
\end{figure}

For $n_r=100$ and $n_z\in\{20,30,\cdots,100\}$, we plot the estimates of $\lambda$ in~\figurename~\ref{fig:selecting_nz_lambda}. Results from different runs are in blue. The ground truth is a Clayton generator with $\tau=0.2$ and a negative scaled extremal Dirichlet stdf with $\alpha=(1,1,1,1,2,2,2,3,3,4), \rho=0.69$. The ground truth is plotted in black and approximate confidence bands around the ground truth is in dotted black. The computed the mean squared error (MSE) in fitting $K$ and $\lambda$ and the time taken are given in~\figurename~\ref{fig:selecting_nz_MSE_time} and Table~\ref{tab:selecting_nz_MSE_time}, where the standard deviation is given in parenthesis.

\begin{table}[h]
\caption{Mean squared error (MSE) in fitting $K$ and $\lambda$ and the time taken.\vspace{5pt}}\label{tab:selecting_nz_MSE_time}
\resizebox{\textwidth}{!}{%
\begin{tabular}{lllllllllll}
\toprule
$n_r=100, n_z$             & 10 & 20 & 30 & 40 & 50 & 60 & 70 & 80 & 90 & 100 \\\midrule
\textsc{MSE} $K$ $\times 10^{-3}$      & 5.99(11.08) & 1.84(3.60) & 0.55(0.91) & 0.34(0.56) & 0.21(0.57) & 0.46(1.02) & 0.60(1.06) & 0.23(0.39) & \textbf{0.12}(\textbf{0.31}) & 0.16(0.31)\\
\textsc{MSE} $\lambda$ $\times 10^{-3}$ & 1.24(1.27) & 1.10(0.75) & 1.04(0.56) & 0.57(0.62) & 0.49(0.49) & 0.74(0.72) & \textbf{0.18}(\textbf{0.13}) & 0.32(0.32) & 0.39(0.27) & 0.23(0.21)\\
\textsc{time (sec)}    & 8.59(1.78) & 21.30(7.10) & 40.06(11.11) & 43.46(8.23) & 62.46(18.45) & 86.08(21.91) & 100.62(29.75) & 217.68(71.45) & 190.10(56.29) & 239.04(96.97)    \\\bottomrule
\end{tabular}%
}
\end{table}

From the results, for $n_r=100$, it would be appropriate to use $n_z\in\{70,\cdots,100\}$, with a tradeoff between accuracy and computation time. 

\clearpage

\paragraph{Plots of $\lambda$ for sample size $n=1000$}

While our method infers an arbitrary Archimedean generator and takes the joint dependence across covariates into account, the method in~\citep{chatelain2020inference} infers a Clayton generator from pairwise Kendall taus. Thus the performance gap between our method and the method in~\citep{chatelain2020inference} is expected to increase as the generator differs from the Clayton generator and as the observations become less symmetric.

\begin{figure}[h]
    \centering
    
    \includegraphics[width=0.267\textwidth]{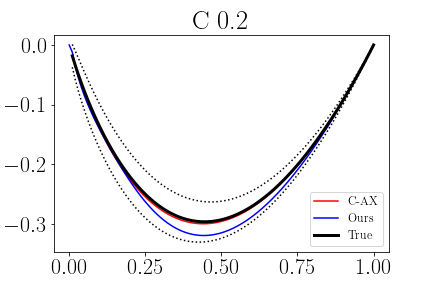}
    \includegraphics[width=0.20\textwidth]{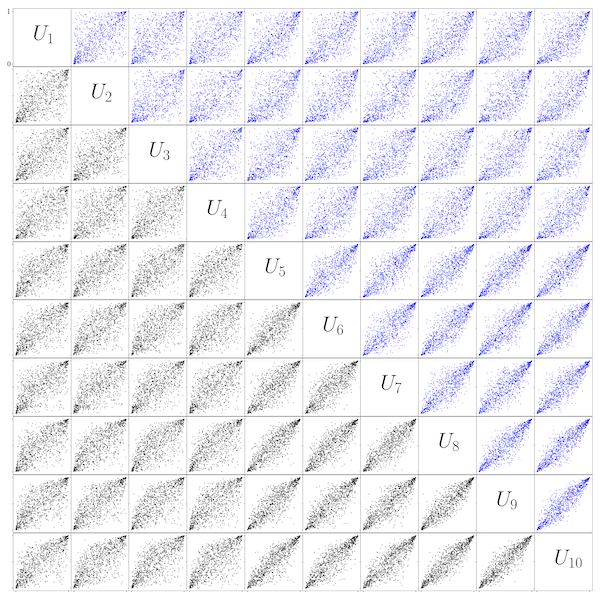}\hspace{5pt}
    \includegraphics[width=0.267\textwidth]{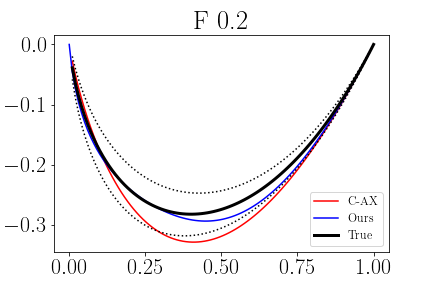}
    \includegraphics[width=0.20\textwidth]{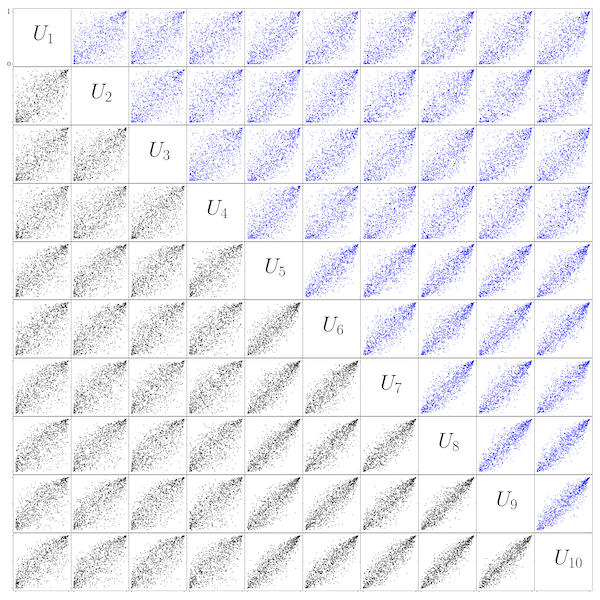}
    
    \includegraphics[width=0.267\textwidth]{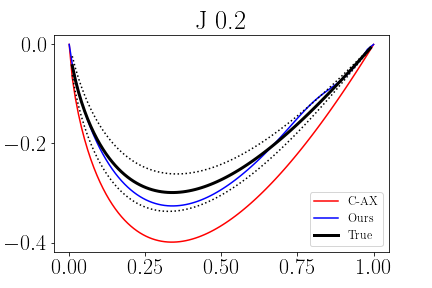}
    \includegraphics[width=0.20\textwidth]{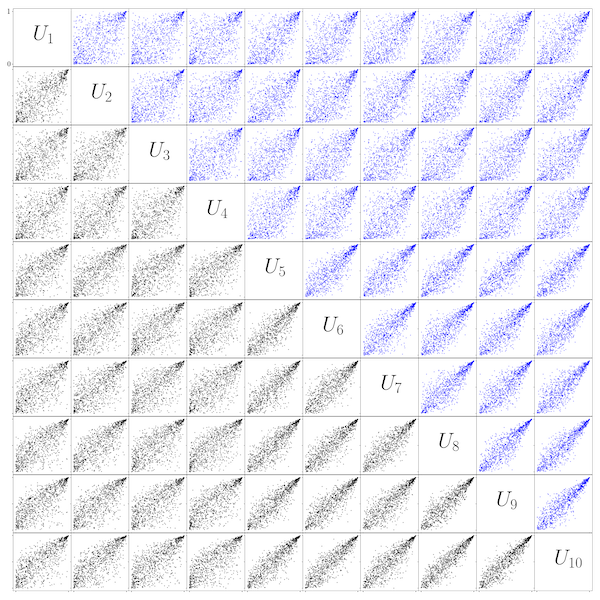}\hspace{5pt}
    \includegraphics[width=0.267\textwidth]{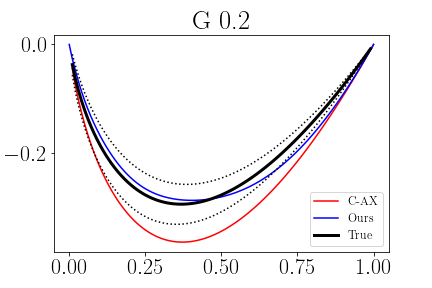}
    \includegraphics[width=0.20\textwidth]{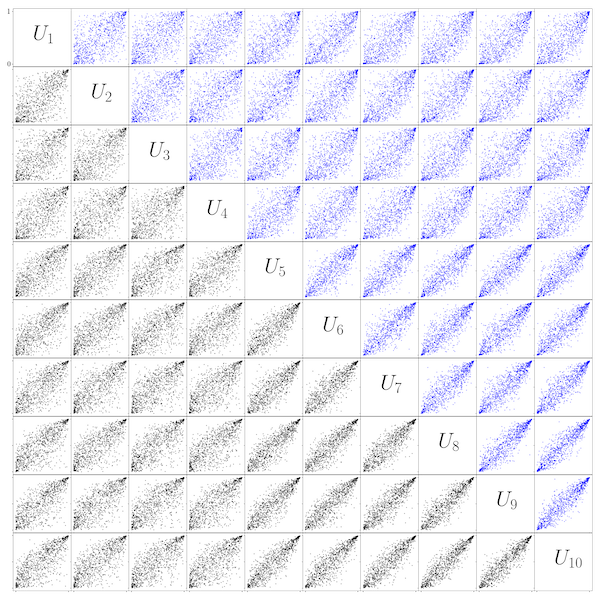}
    
    \includegraphics[width=0.267\textwidth]{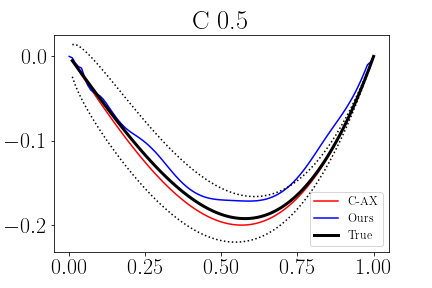}
    \includegraphics[width=0.20\textwidth]{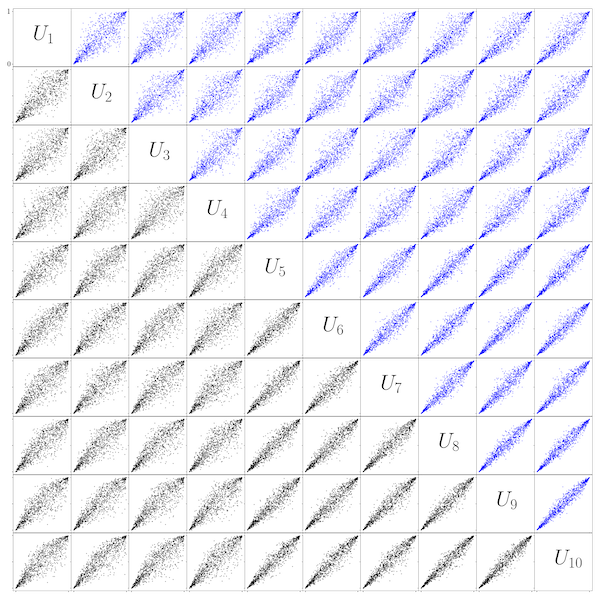}\hspace{5pt}
    \includegraphics[width=0.267\textwidth]{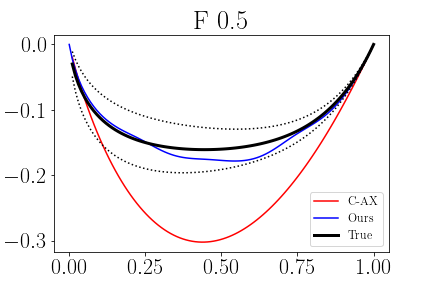}
    \includegraphics[width=0.20\textwidth]{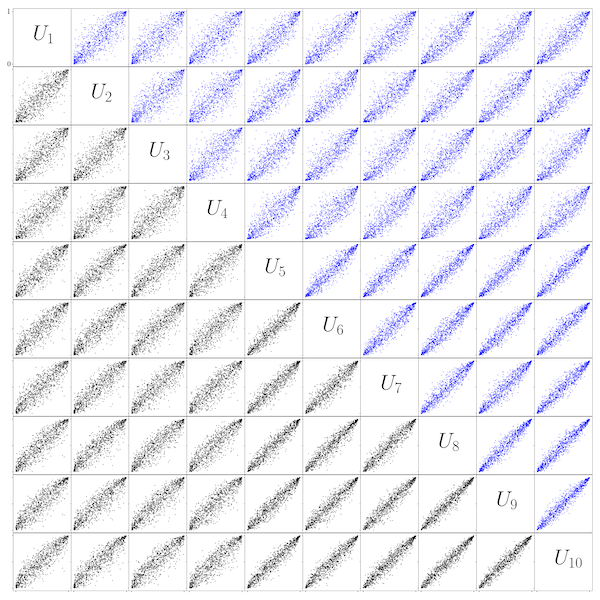}
    
    \includegraphics[width=0.267\textwidth]{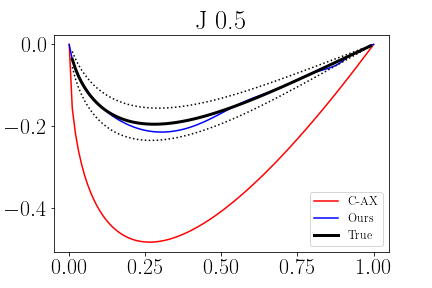}
    \includegraphics[width=0.20\textwidth]{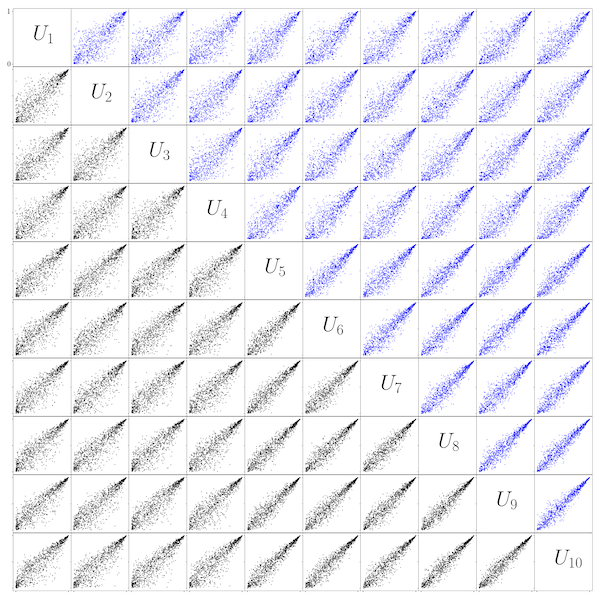}\hspace{5pt}
    \includegraphics[width=0.267\textwidth]{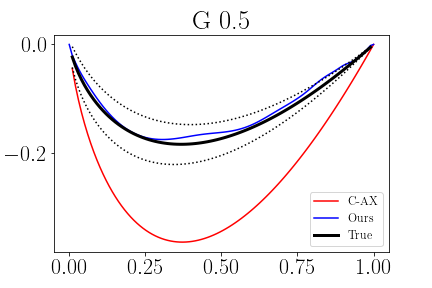}
    \includegraphics[width=0.20\textwidth]{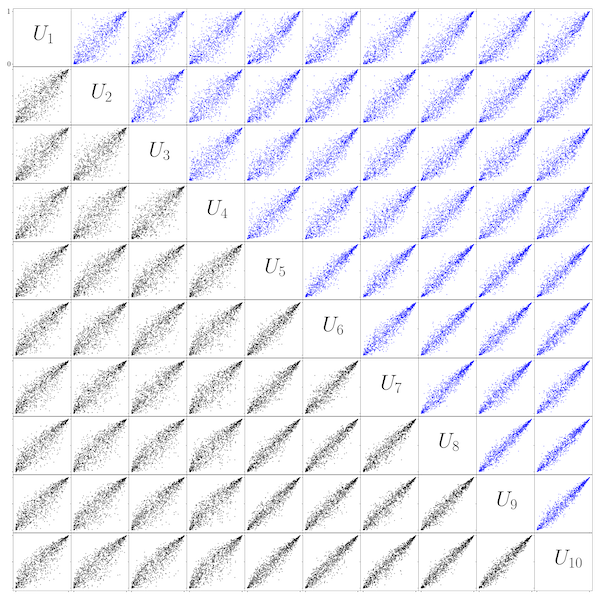}

    \caption{Estimates of $\lambda$ given samples of $\mathbf{S}$ from true $\ell$ for $n=1000$. Ground truth in black, approximate confidence bands in dotted black, the method from~\citep{chatelain2020inference} in red, our method in blue. On the right of each plot of $\lambda$ is a plot of samples from the copula, ground truth below the diagonal in black, our method above the diagonal in blue.}
    \label{fig:estimating_varphi_stdfGT}
\end{figure}

\clearpage

\paragraph{Plots of $\lambda$ for sample size $n=200$}

Increasing the number of observations $n$ improved estimation accuracy, hinting at consistency.

\begin{figure}[h]
    \centering
    
    \includegraphics[width=0.267\textwidth]{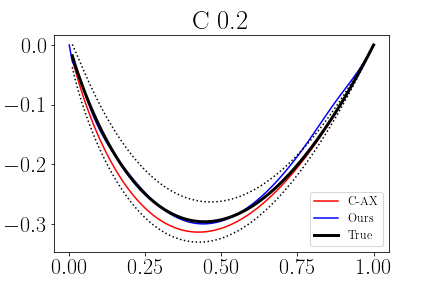}
    \includegraphics[width=0.20\textwidth]{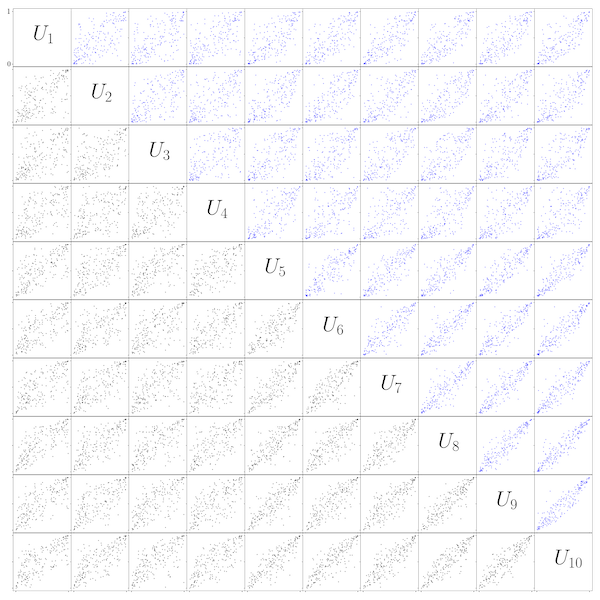}\hspace{5pt}
    \includegraphics[width=0.267\textwidth]{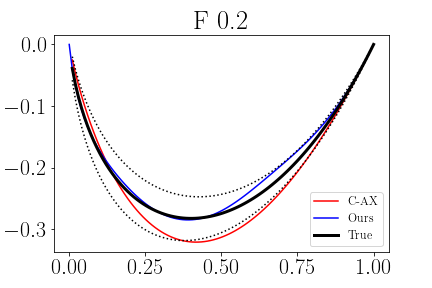}
    \includegraphics[width=0.20\textwidth]{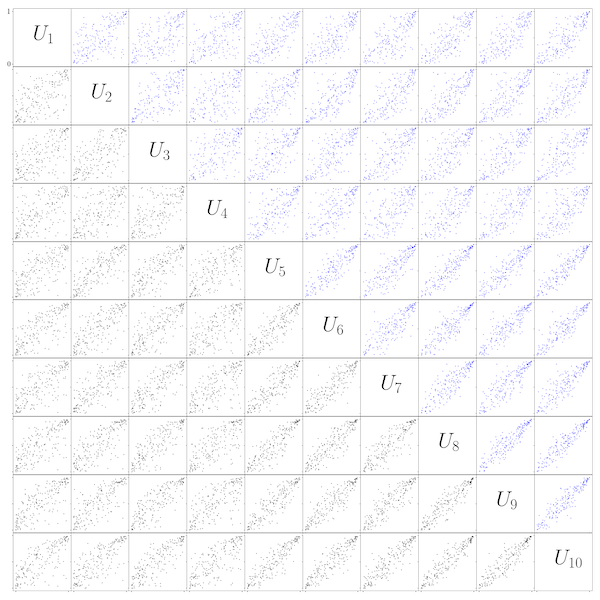}
    
    \includegraphics[width=0.267\textwidth]{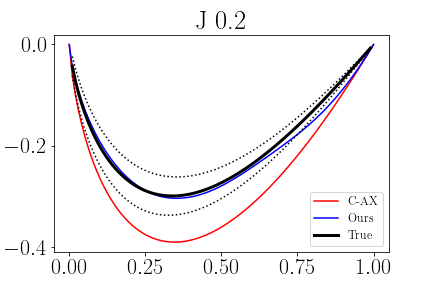}
    \includegraphics[width=0.20\textwidth]{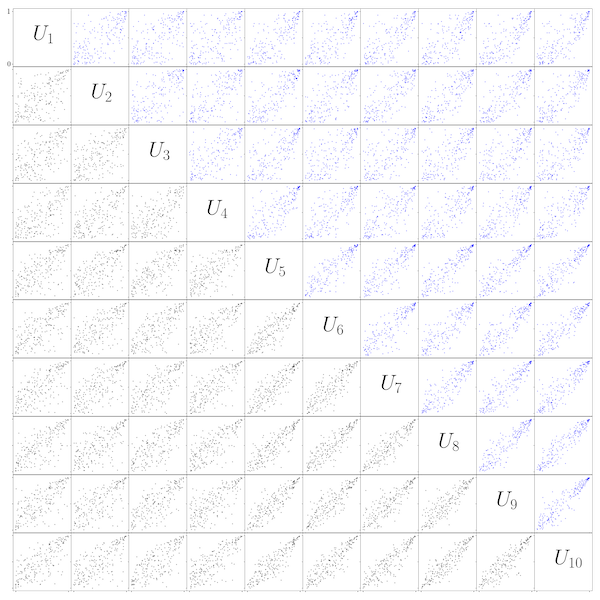}\hspace{5pt}
    \includegraphics[width=0.267\textwidth]{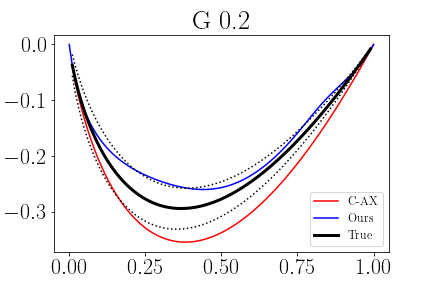}
    \includegraphics[width=0.20\textwidth]{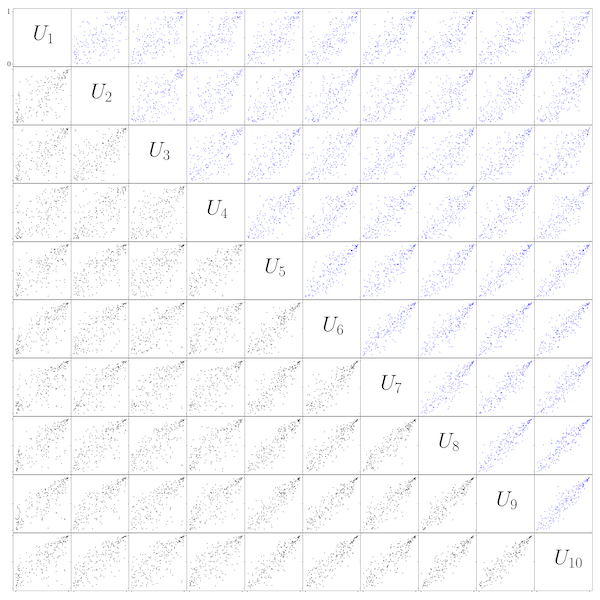}
    
    \includegraphics[width=0.267\textwidth]{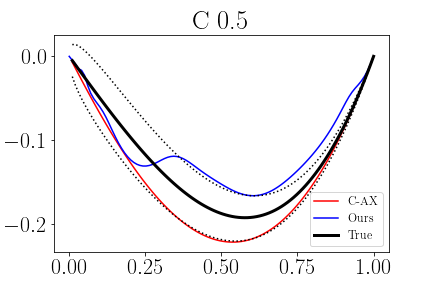}
    \includegraphics[width=0.20\textwidth]{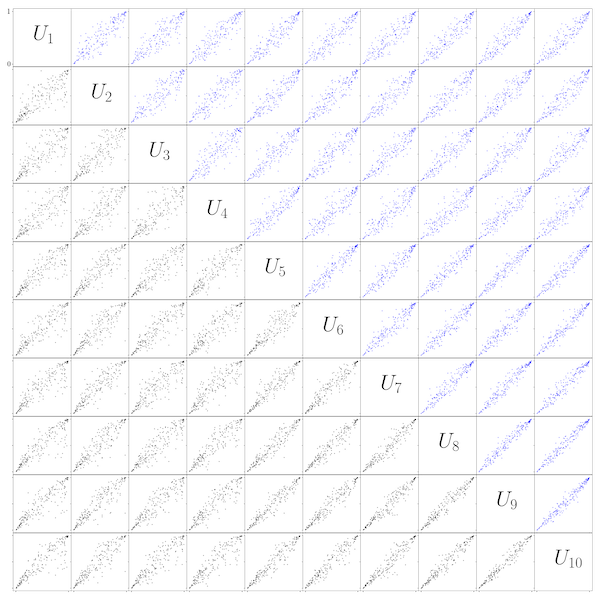}\hspace{5pt}
    \includegraphics[width=0.267\textwidth]{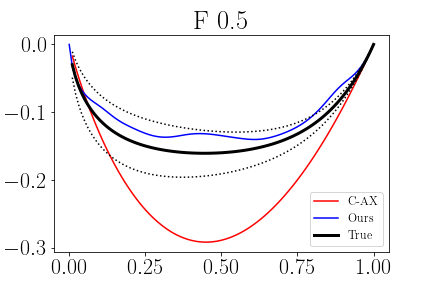}
    \includegraphics[width=0.20\textwidth]{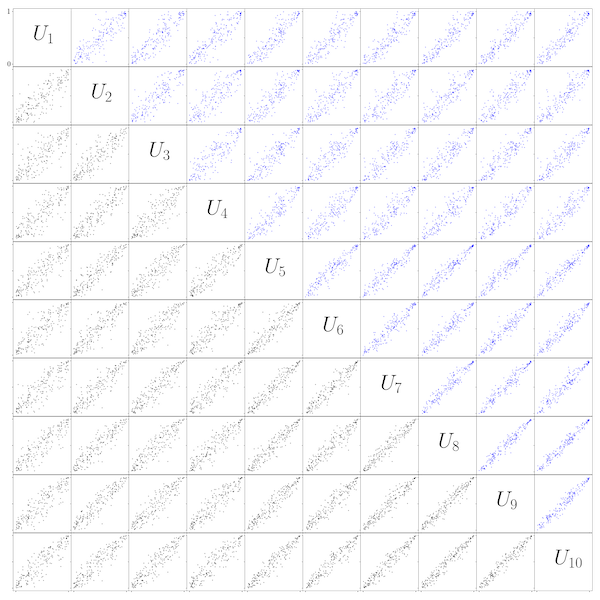}
    
    \includegraphics[width=0.267\textwidth]{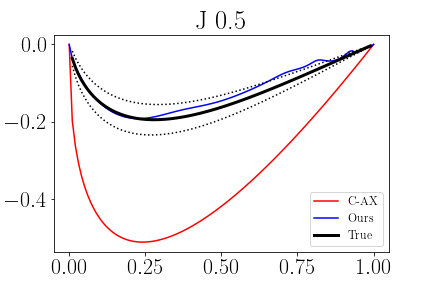}
    \includegraphics[width=0.20\textwidth]{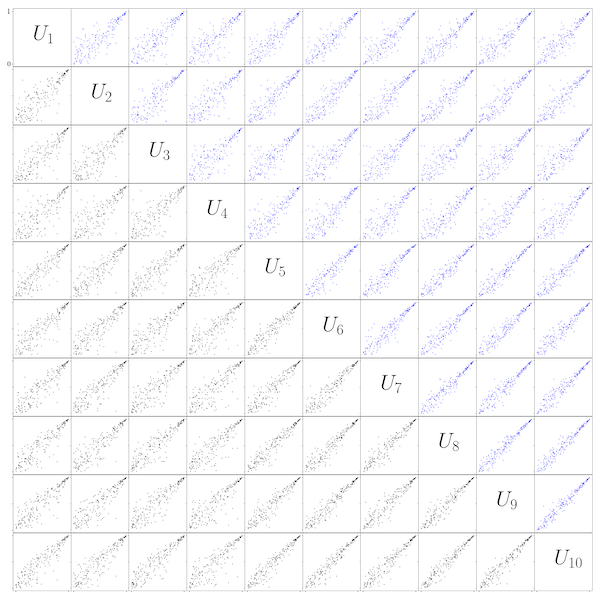}\hspace{5pt}
    \includegraphics[width=0.267\textwidth]{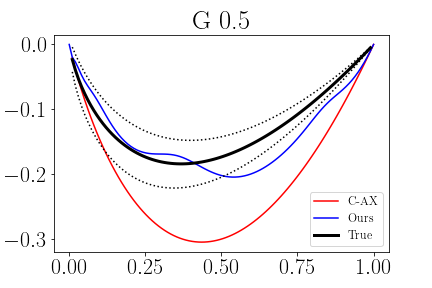}
    \includegraphics[width=0.20\textwidth]{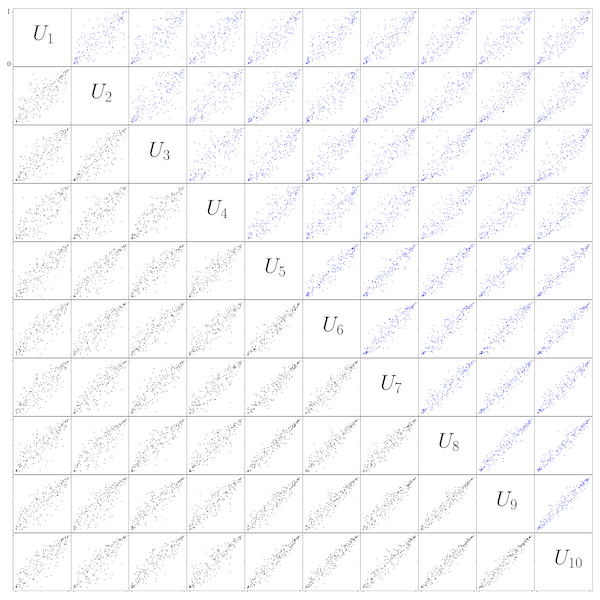}

    \caption{Estimates of $\lambda$ given samples of $\mathbf{S}$ from true $\ell$ for $n=1000$. Ground truth in black, approximate confidence bands in dotted black, the method from~\citep{chatelain2020inference} in red, our method in blue. On the right of each plot of $\lambda$ is a plot of samples from the copula, ground truth below the diagonal in black, our method above the diagonal in blue.}
    \label{fig:estimating_varphi_stdfGT200}
\end{figure}

\clearpage

\subsection{Inference for stable tail dependence function and sampling for simplex component}\label{app:results_estimating_S_varphiGT}

The negative scaled extremal Dirichlet (NSD)~\citep{belzile2017nsd} is a rich class encompassing many parametric models of the stdf and spectral component, including the logistic, asymmetric logistic, negative logistic, and extremal Dirichlet models~\citep{coles1991dirichlet}.

It is specified by:
\begin{equation}\label{eq:def_stdfNSD}
    \ell(\mathbf{x}) = \frac{\Gamma(\alpha_1+\cdots+\alpha_d-\rho)}{\Gamma(\alpha_1+\cdots+\alpha_d)}\mathbb{E}_\mathbf{D}\left[\max_{j=1,\cdots,d}\left(\frac{x_j D_j^{-\rho}\Gamma(\alpha_j)}{\Gamma(\alpha_j-\rho)}\right)\right],
\end{equation}
where $\mathbf{D}=(D_1,...,D_d)$ is distributed as a $\text{Dirichlet}(\alpha_1,...,\alpha_d)$ with $\alpha_1,...,\alpha_d>0$ and  $\rho\in(0,\min(\alpha_1,\cdots,\alpha_d))$.

The integrated relative absolute error (IRAE) is commonly used to evaluate estimates of $\ell$, and is given by~\citep{chatelain2020inference}:
\begin{equation}\label{eq:def_IRAE}
    \text{IRAE}(\ell,\ell_\theta) = \frac{1}{|\Delta_{d-1}|}\int_{\Delta_{d-1}}|\ell(\mathbf{x})-\ell_\theta(\mathbf{x})|/\ell(\mathbf{x}) \, \mathrm{d}\mathbf{x}.
\end{equation}
The IRAE is computed using Monte Carlo integration with 10,000 samples $\mathbf{x}$ drawn uniformly at random from the simplex $\Delta_{d-1}$.

For given true $\varphi$, we report the IRAE of the modified Pickands estimator from~\citep{chatelain2020inference}, the modified CFG estimator from~\citep{chatelain2020inference}, and our method in Table~\ref{tab:IRAE_varphiGT}. The stdf is a NSD with $\alpha=(1,1,1,1,2,2,2,3,3,4),\allowbreak\rho=0.69$.

\begin{table}[h]
\caption{Inference of $\ell$ given true $\varphi$\vspace{5pt}} \label{tab:IRAE_varphiGT}
\centering \footnotesize
\begin{tabular}{lllllllll}
\toprule
 & \textsc{C 0.2}      & \textsc{C 0.5}     & \textsc{F 0.2}      & \textsc{F 0.5}      & \textsc{J 0.2}      & \textsc{J 0.5}    & \textsc{G 0.2}      & \textsc{G 0.5} \\\midrule
\textsc{IRAE $\pm0.01$}~P~\citep{chatelain2020inference}      
  & 0.16       & 1.00      & 0.05       & 0.06       & 0.06       & 0.07     & 0.08       & 0.17   \\
\textsc{IRAE $\pm0.01$}~CFG~\citep{chatelain2020inference}      
  & 0.05       & 0.11      & 0.04       & 0.04       & 0.05       & 1.00     & 0.06       & 0.15   \\
\textsc{IRAE $\pm0.01$ (Ours)}     
  & 0.06 & 0.12 & 0.04 & 0.05 & 0.06 & \textbf{0.07} & 0.08 & 0.15 \\\bottomrule
\end{tabular}
\end{table}

As mentioned in Appendix~\ref{app:inferW}, though our estimator is based on the Pickands estimator, it performs better than the Pickands estimator. This suggests that our direct method of training with the generative network generating the class of valid stdfs would perform better than the alternative method that first estimates $\hat{\ell}$ then train the generative network to match $\hat{\ell}$. As noted in~\citep{chatelain2020inference}, the modified CFG estimator performs better than the modified Pickands estimator for Archimax copulas. A future direction may be to improve the robustness of our estimator with the CFG modification.

\begin{figure}[h]
    \centering
    \includegraphics[width=0.415\textwidth]{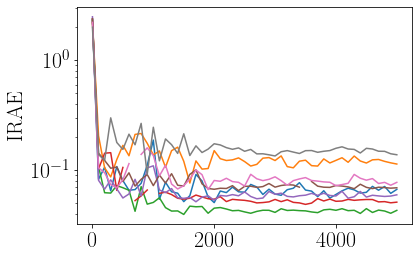}
    \includegraphics[width=0.4\textwidth]{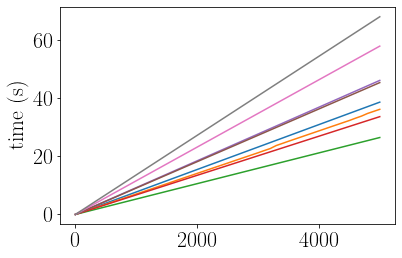}
    \caption{IRAE in fitting $\ell$ and the time taken versus number of mini-batch iterations. Each line is for a different copula setting in Table~\ref{tab:IRAE_varphiGT}.}
    \label{fig:estimating_S_varphiGT_ISE_IRAE_time}
\end{figure}

We also provide the IRAE and time taken versus number of minibatch iterations in~\figurename~\ref{fig:estimating_S_varphiGT_ISE_IRAE_time}. The plots suggest that our algorithm converges and is not computationally intensive.

\clearpage

\subsection{Modeling nutrient intake}\label{app:results_nutrient}

The data and documentation from the study of nutrient intake in women is made available by the U.S. Department of Agriculture (USDA)~\citep{usda1985nutrient}. There are $n=1459$ observations of dimension $d=17$, corresponding to the variables: Energy, Protein, Vitamin A (IU), Vitamin A (RE), Vitamin~E, Vitamin C, Thiamin, Riboflavin, Niacin, Vitamin B6, Folate, Vitamin B12, Calcium, Phosphorus, Magnesium, Iron, Zinc. This dataset was previously studied in~\citep{mcneil2010liouville} where using a Clayton-NSD Archimax copula improved fit over a Clayton Archimedean copula. However, in~\citep{mcneil2010liouville}, the experiment was limited to only $n=737$ and $d=3$, corresponding to Calcium, Iron and Protein.

We compared our method to copula based models and deep network based models from literature using the CvM distance in~\eqref{eq:def_CvM}. The abbreviations for the copula based models in Table~\ref{tab:nutrient} of the main paper correspond to the abbreviations in Table~\ref{tab:common_copulas} summarizing the common multivariate copulas. They are Gaussian (GC), R-vine (RV), C-vine (CV), D-vine (DV), Archimedean (AC *), hierarchical Archimedean (HAC)~\citep{gorecki2017hac} and extreme-value (EV $\dagger$) copulas. The abbreviation (C-AX $\dagger$) corresponds to the state-of-the-art in inferring Archimax copulas with a Clayton generator for $\varphi$ and the modified CFG estimator for $\ell$~\citep{chatelain2020inference}. The abbreviations for the deep network based models in Table~\ref{tab:nutrient} are Wasserstein GAN with gradient penalty (WGAN), masked autoregressive flow (MAF) and variation autoencoders (VAE). Lastly, the abbreviation (Gen-AX *$\dagger$) corresponds to our method. For methods marked with *, we use $\varphi$ described in Algorithm~\ref{alg:estimating_R_GNN} and for methods marked with $\dagger$ we use $\ell$ and the sampling methods described in Algorithms~\ref{alg:sampling_S_GNN}~and~\ref{alg:sampling_archimax_GNN}.

Table~\ref{tab:nutrient} shows our method outperforming the above methods using the CvM distance. 

We additionally provide plots of generated samples versus true samples in~\figurename~\ref{fig:results_appl_nutrient_copula}~(a-l). The generated samples are plotted in blue above the diagonal while the true samples are plotted in black below the diagonal. From the plot of our generated samples in~\figurename~\ref{fig:results_appl_nutrient_copula}~(l), an improvement can be made with hierarchical Archimax copulas~\citep{hofert2018harchimax} to have different Archimedean generators $\varphi$ and thus different radial envelopes for covariates.

\begin{figure}[h]
    \centering
    \includegraphics[width=0.5\textwidth]{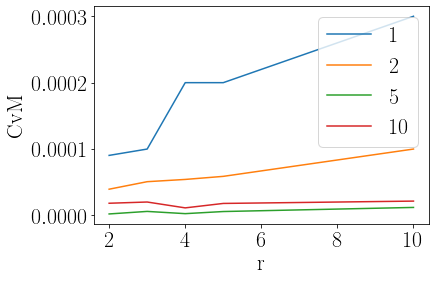}
    \caption{Selection of block size $n/k=5$ using the test for extreme-value dependence~\citep{kojadinovic2011_evctest}.}
    \label{fig:evctest}
\end{figure}

We briefly describe the training process using our proposed method. We first initialize $\ell_\theta$ with Algorithm~\ref{alg:estimating_W_GNN} on block-maximas. Using the test for extreme-value dependence~\citep{kojadinovic2011_evctest}, we chose the block size $n/k=5$. The plots for block sizes $n/k\in\{1, 2, 5, 10\}$ and exponents $r\in\{2,3,\cdots,10\}$ are given in~\figurename~\ref{fig:evctest}. 

We compared our initial estimate $\ell_\theta$ to the state-of-the-art CFG estimator applied on the block maximas $\hat{\ell}$~\citep{caperaa1997cfg}. The plot of IRAE against mini-batch iterations show convergence in 2000 mini-batch iterations, with a duration of 20s, and an IRAE of 0.08. We then learn $\varphi_\theta$ with Algorithm~\ref{alg:estimating_R_GNN} on the full dataset, with samples of $\mathbf{S}$ from Algorithm~\ref{alg:sampling_S_GNN}. The plot of MSE in fitting the empirical Kendall distribution show convergence in 2000 mini-batch iterations, with a duration of 25s, and an MSE of 0.0008. We note that the initialization scheme seems to be performing well since the learnt $\varphi_\theta$ focused on modifying only the lower tail. We then update $\ell_\theta$ with Algorithm~\ref{alg:estimating_W_GNN} on the full dataset, given $\varphi_\theta$. The NLL of transformed observations $\xi$ went from 1.67 in the initialization to -1.12 with the use of $\varphi_\theta$. The IRAE to the state-of-the-art modified CFG estimator on the full dataset was 0.078. 

The estimation of the copula based models was done using the \verb+Copulas+ library~\citep{copulas_python} in Python and the \verb+HACopula+ toolbox~\citep{hac_matlab} in MATLAB.

The architectures of the deep networks are:
\begin{itemize}
    \item WGAN: 3 layers for generator, 3 layers for discriminator, hidden size 128.
    \item MAF: 2 flows, hidden size 128 for each flow.
    \item VAE: 3 layers for encoder, 2 layers for decoder, hidden size 128, latent size 16.
    \item Gen-AX: 3 layers hidden size 30 for $G_\mathbf{W}$, 3 layers hidden size 10 for $G_R$.
\end{itemize}

The implementation was with \verb+PyTorch+, the Adam optimizer was used with learning rate 1e-3.

\begin{figure}[h]
    \centering
    \includegraphics[width=\textwidth]{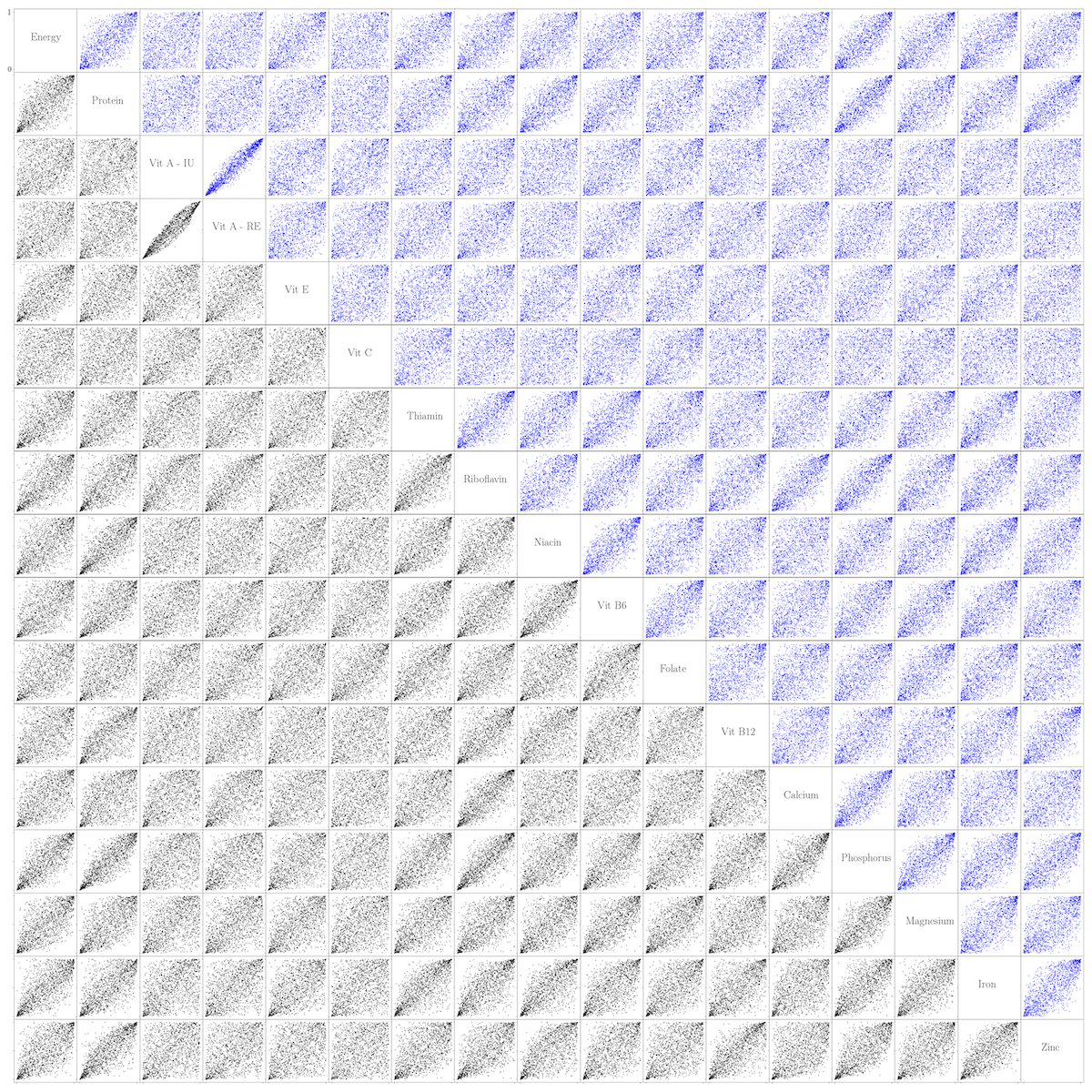} 
    \caption{(a) Gaussian copula (GC).}\label{fig:results_appl_nutrient_copula}
\end{figure}

\begin{figure}[h]\ContinuedFloat
    \centering
    \includegraphics[width=\textwidth]{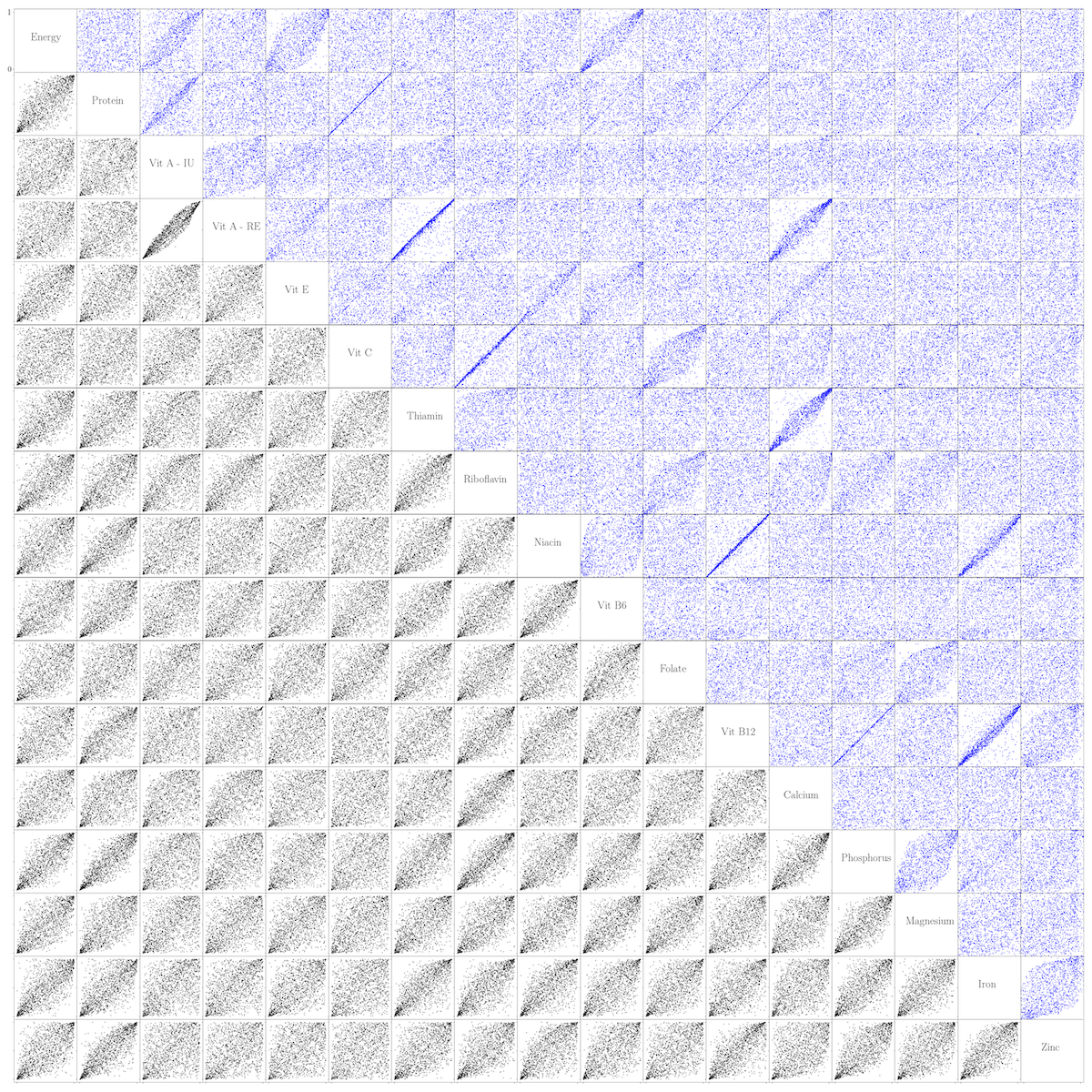} 
    \caption{(b) R-vine (RV).}\label{fig:results_appl_nutrient_copula}
\end{figure}

\begin{figure}[h]\ContinuedFloat
    \centering
    \includegraphics[width=\textwidth]{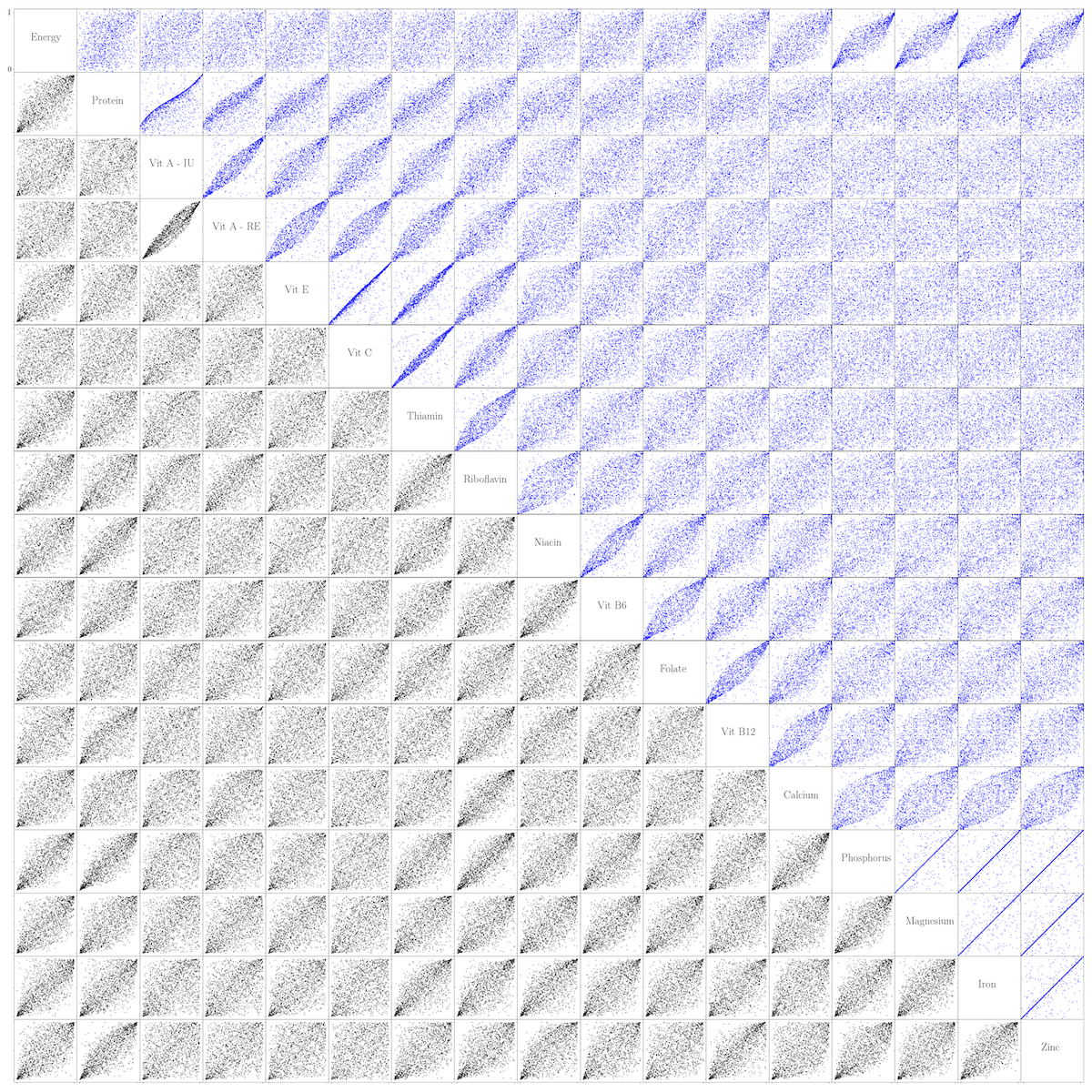} 
    \caption{(c) C-vine (CV).}\label{fig:results_appl_nutrient_copula}
\end{figure}

\begin{figure}[h]\ContinuedFloat
    \centering
    \includegraphics[width=\textwidth]{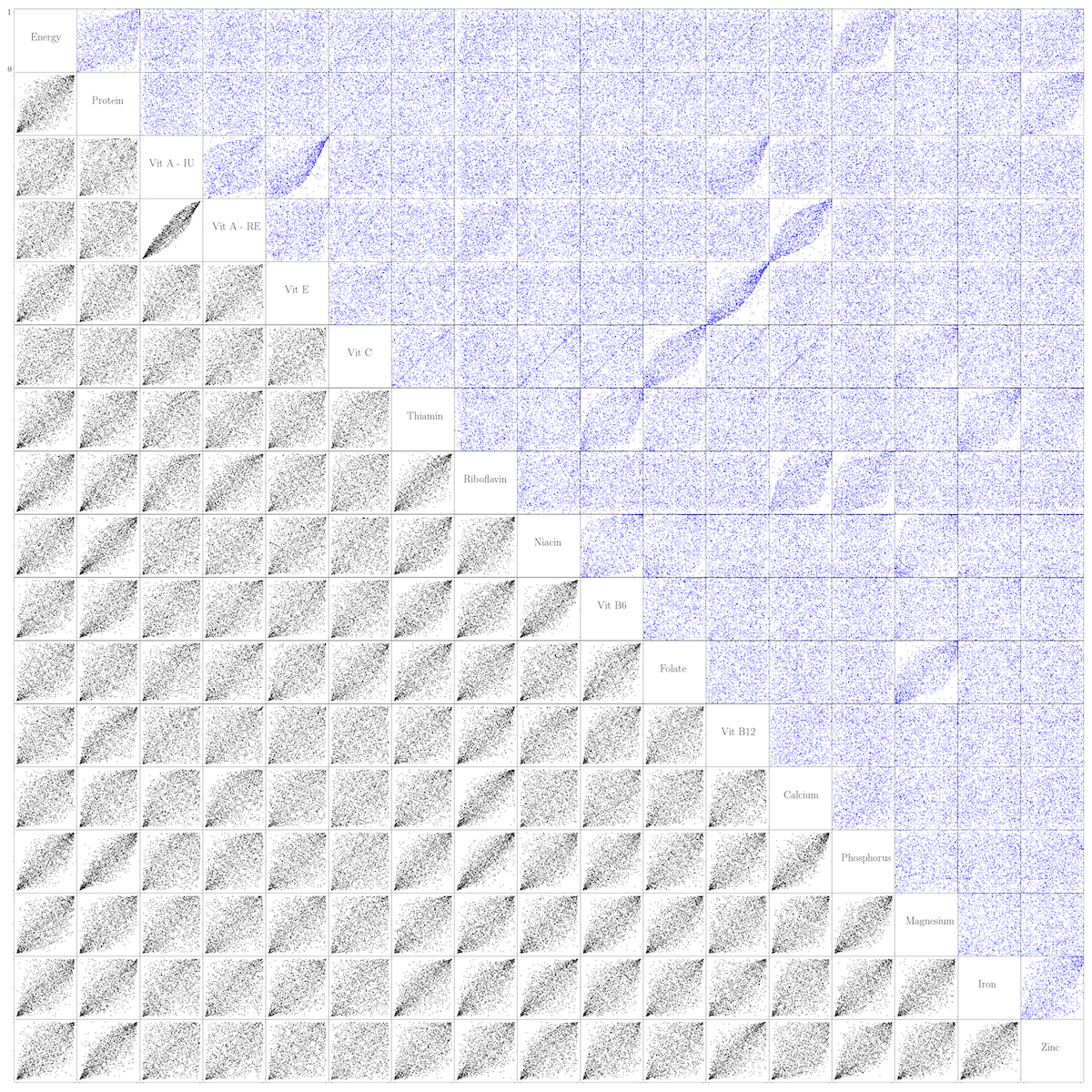} 
    \caption{(d) D-vine (DV).}\label{fig:results_appl_nutrient_copula}
\end{figure}

\begin{figure}[h]\ContinuedFloat
    \centering
    \includegraphics[width=\textwidth]{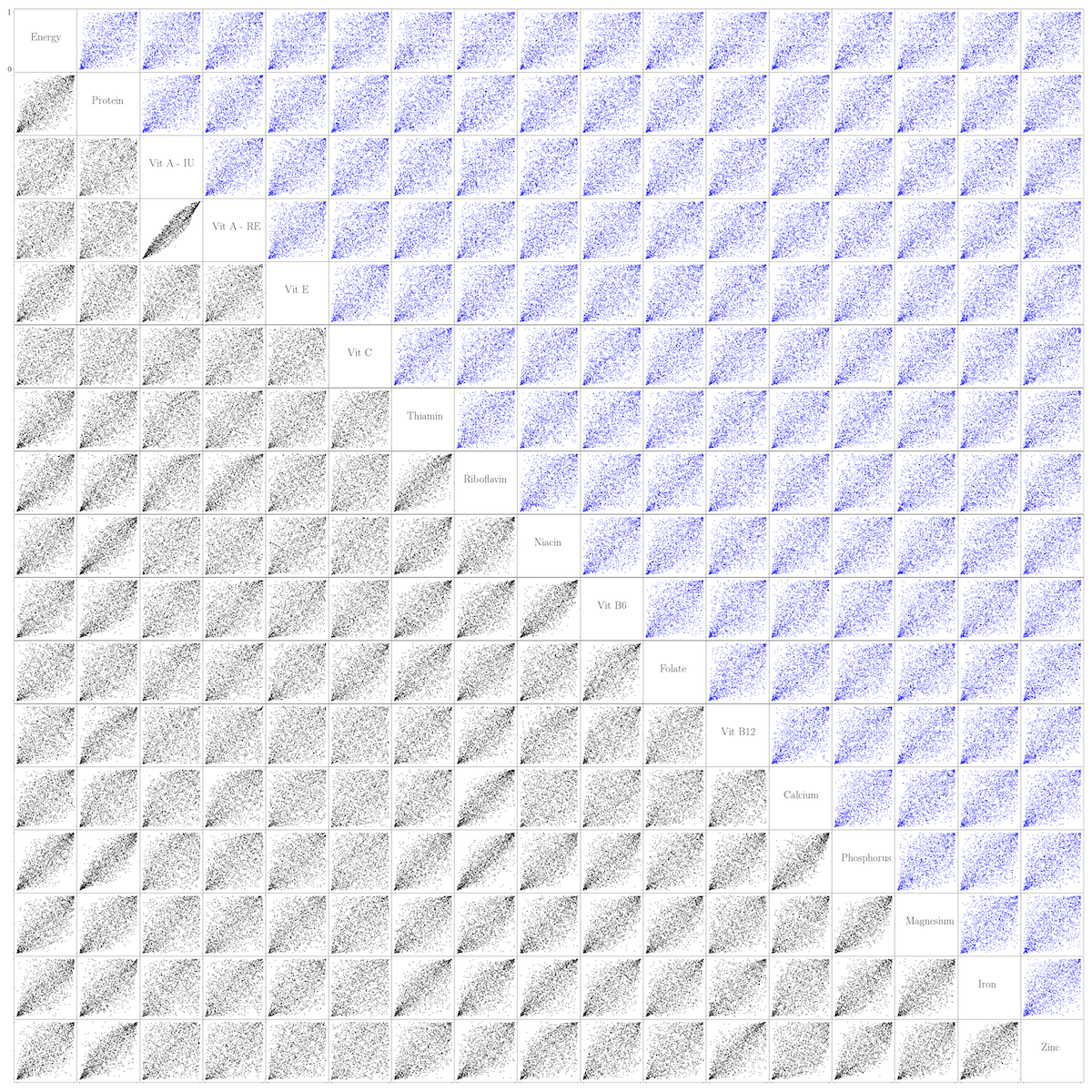} 
    \caption{(e) Archimedean copula (AC *) inferred with Algorithm~\ref{alg:estimating_R_GNN}.}\label{fig:results_appl_nutrient_copula}
\end{figure}

\begin{figure}[h]\ContinuedFloat
    \centering
    \includegraphics[width=\textwidth]{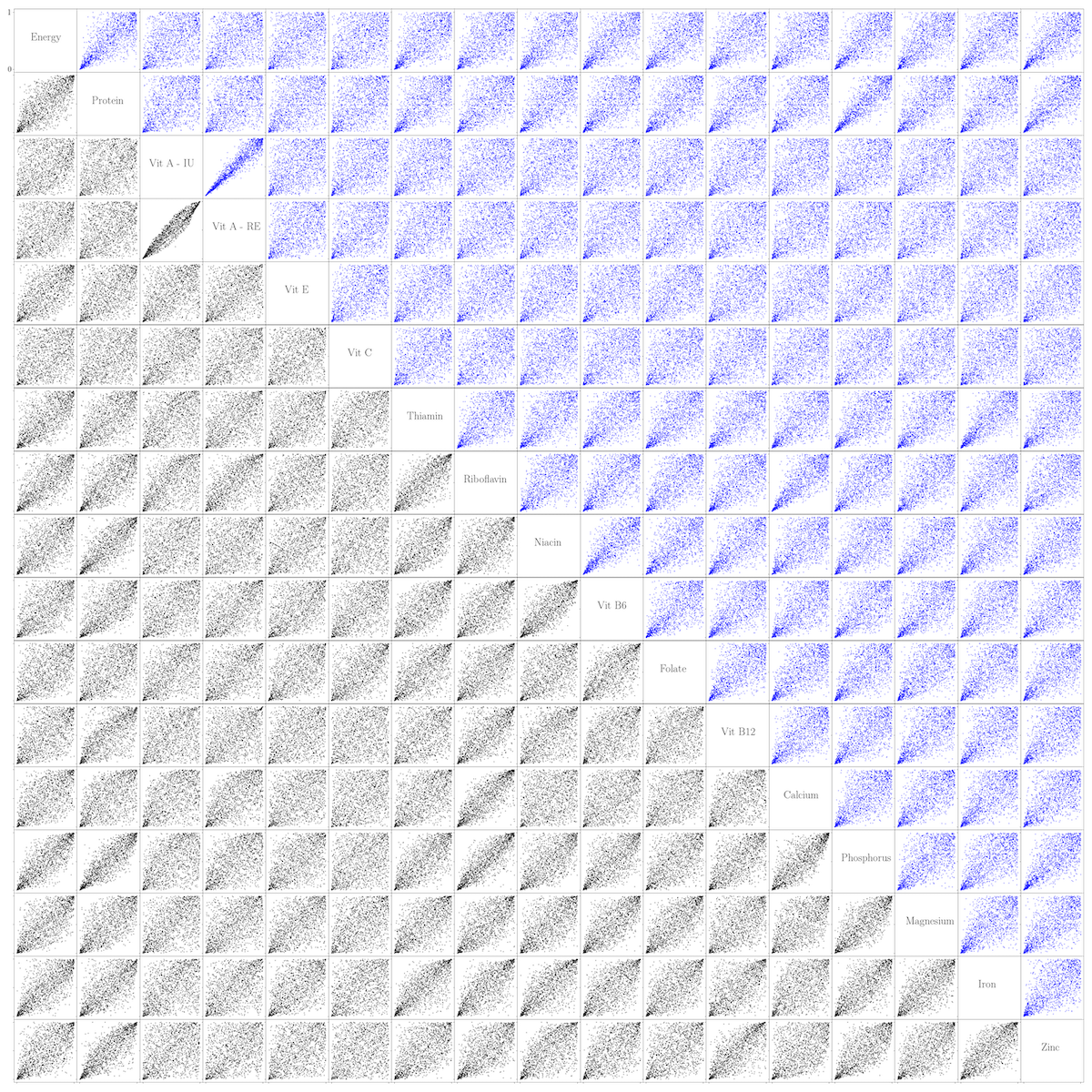} 
    \caption{(f) Hierarchical Archimedean copula (HAC)~\citep{gorecki2017hac}.}\label{fig:results_appl_nutrient_copula}
\end{figure}

\begin{figure}[h]\ContinuedFloat
    \centering
    \includegraphics[width=\textwidth]{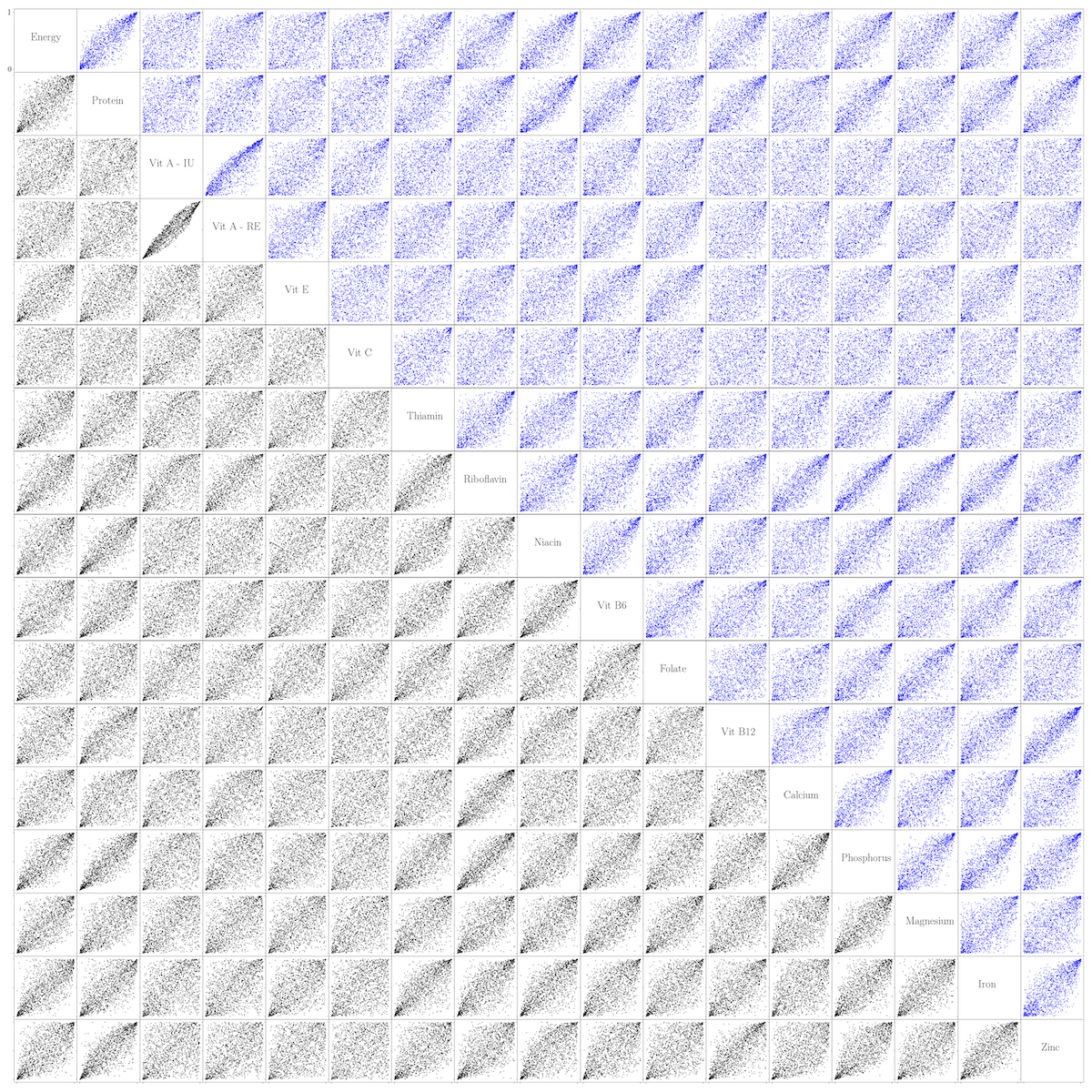} 
    \caption{(g) Extreme-value copula (EV $\dagger$) sampled with Algorithms~\ref{alg:sampling_S_GNN}~and~\ref{alg:sampling_archimax_GNN}. }\label{fig:results_appl_nutrient_copula}
\end{figure}

\begin{figure}[h]\ContinuedFloat
    \centering
    \includegraphics[width=\textwidth]{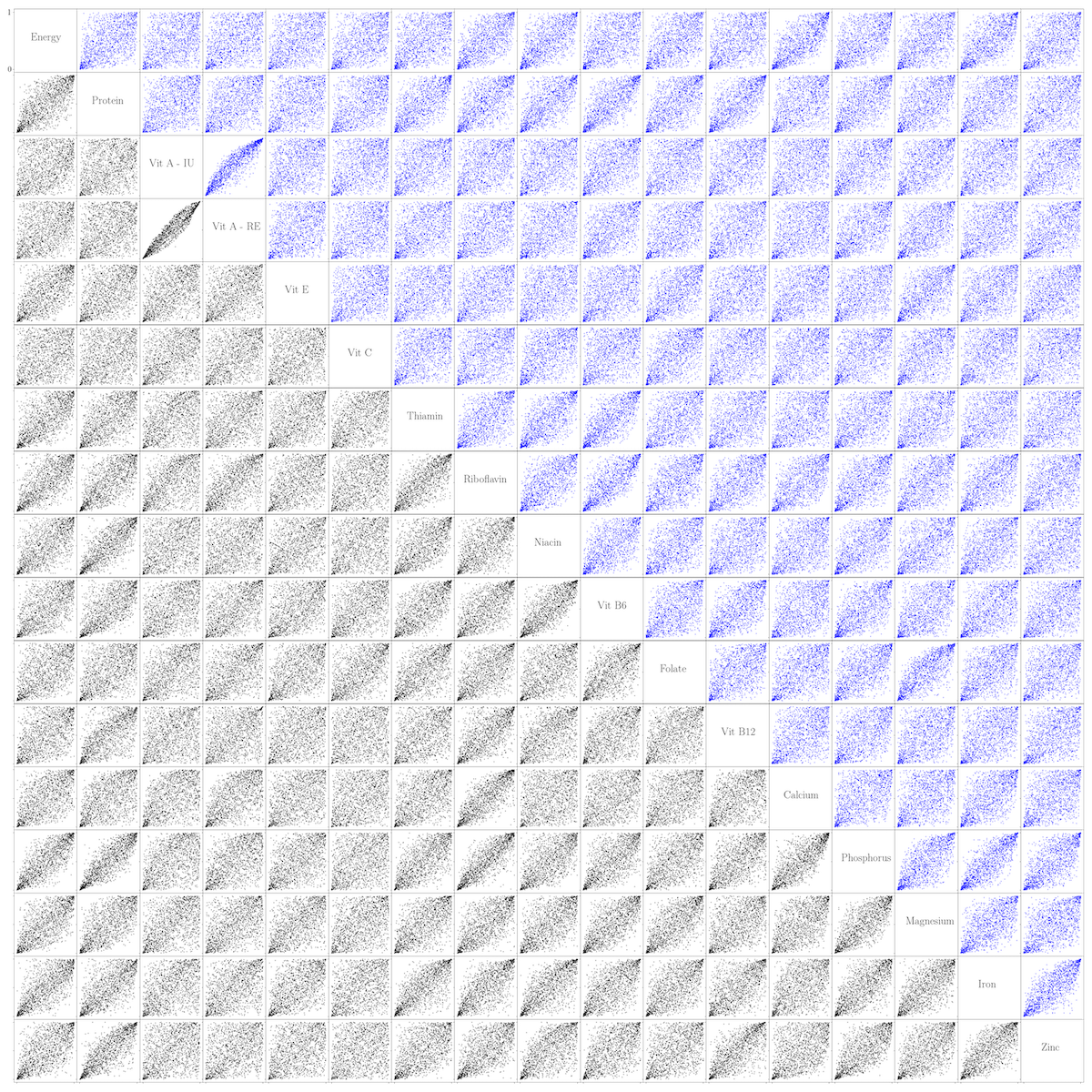} 
    \caption{(h) Archimax copula (C-AX $\dagger$) inferred with the state of the art~\citep{chatelain2020inference}, sampled with Algorithms~\ref{alg:sampling_S_GNN}~and~\ref{alg:sampling_archimax_GNN}.}\label{fig:results_appl_nutrient_copula}
\end{figure}

\begin{figure}[h]\ContinuedFloat
    \centering
    \includegraphics[width=\textwidth]{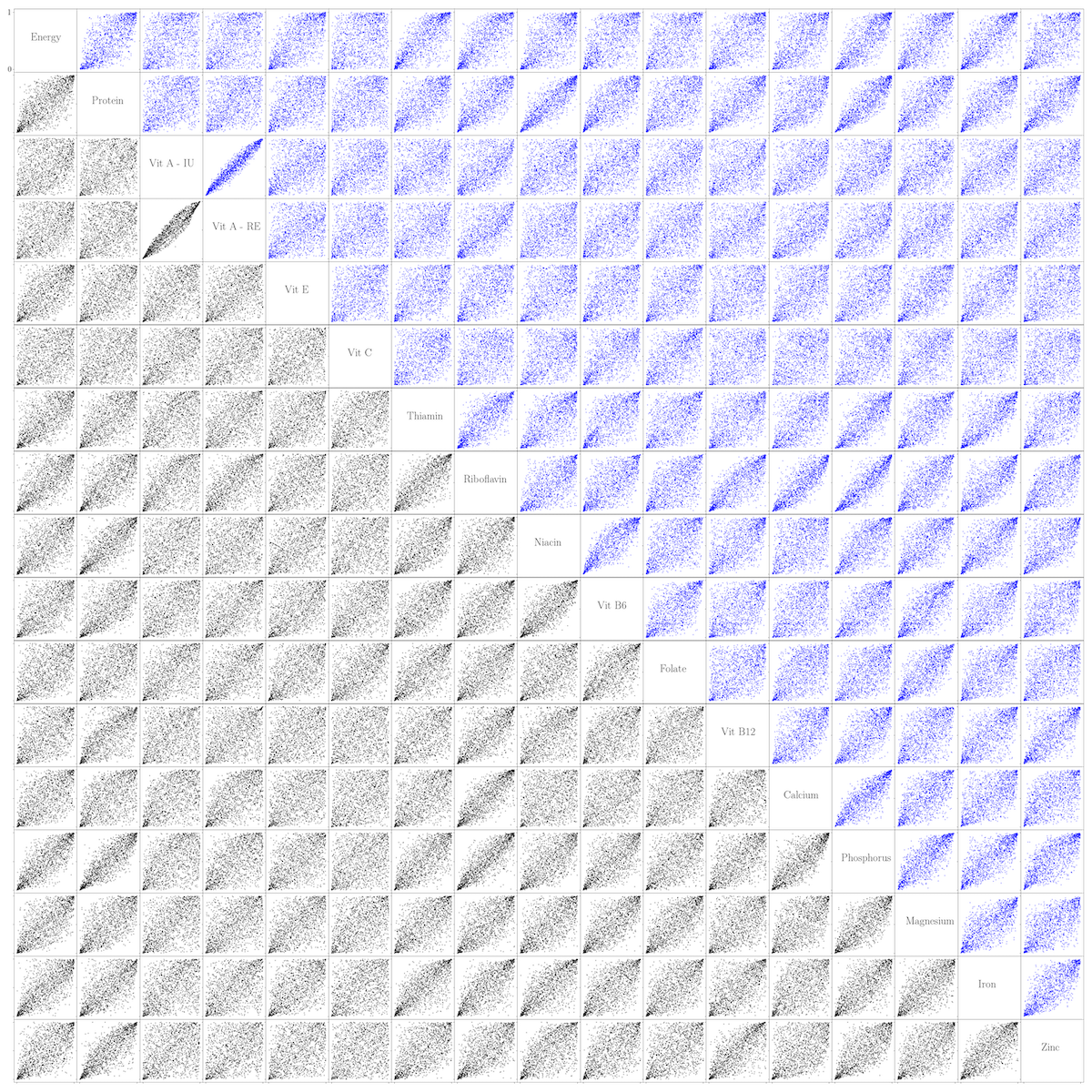} 
    \caption{(i) Wasserstein GAN with gradient penalty (WGAN).}\label{fig:results_appl_nutrient_copula}
\end{figure}

\begin{figure}[h]\ContinuedFloat
    \centering
    \includegraphics[width=\textwidth]{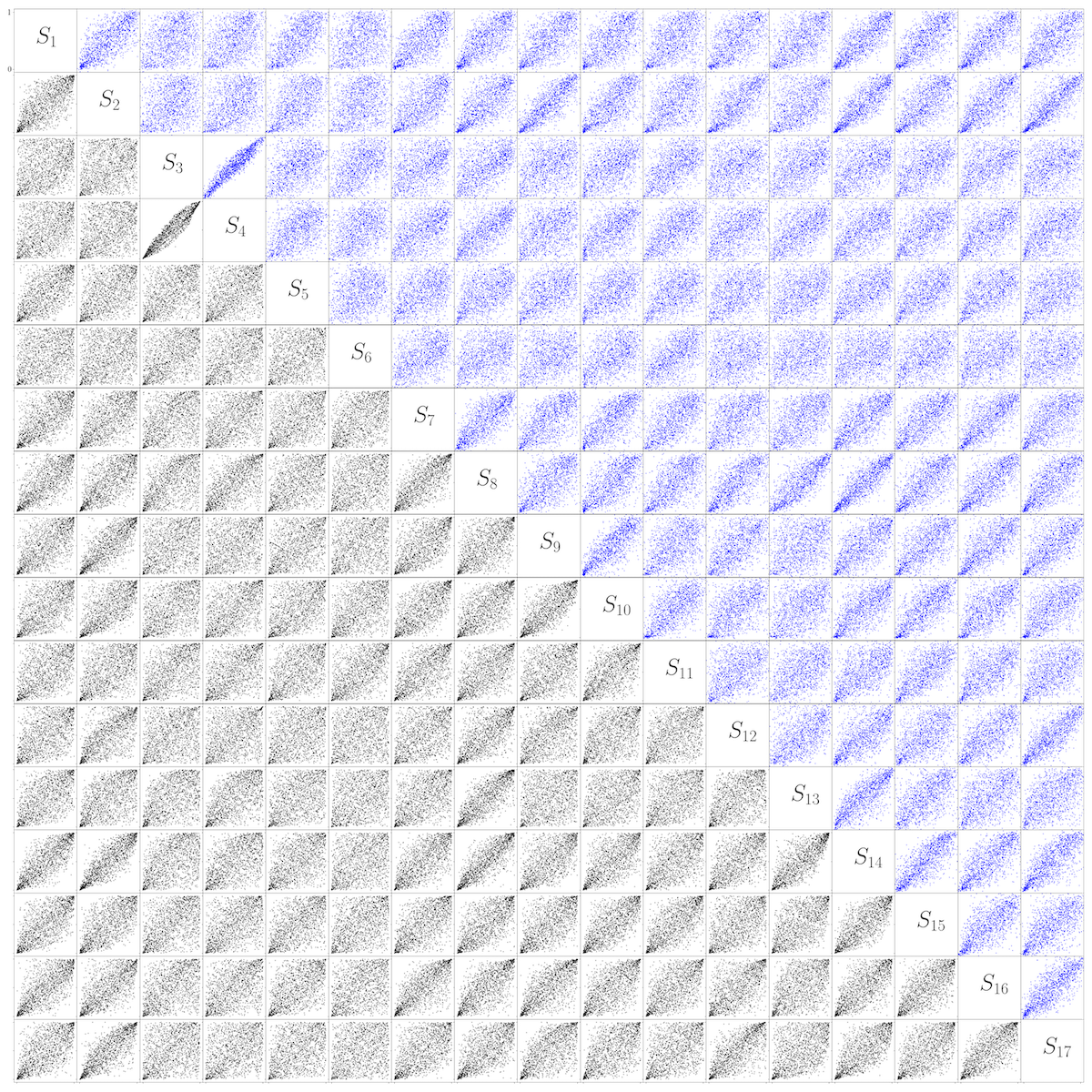} 
    \caption{(j) Masked autoregressive flow (MAF). }\label{fig:results_appl_nutrient_copula}
\end{figure}

\begin{figure}[h]\ContinuedFloat
    \centering
    \includegraphics[width=\textwidth]{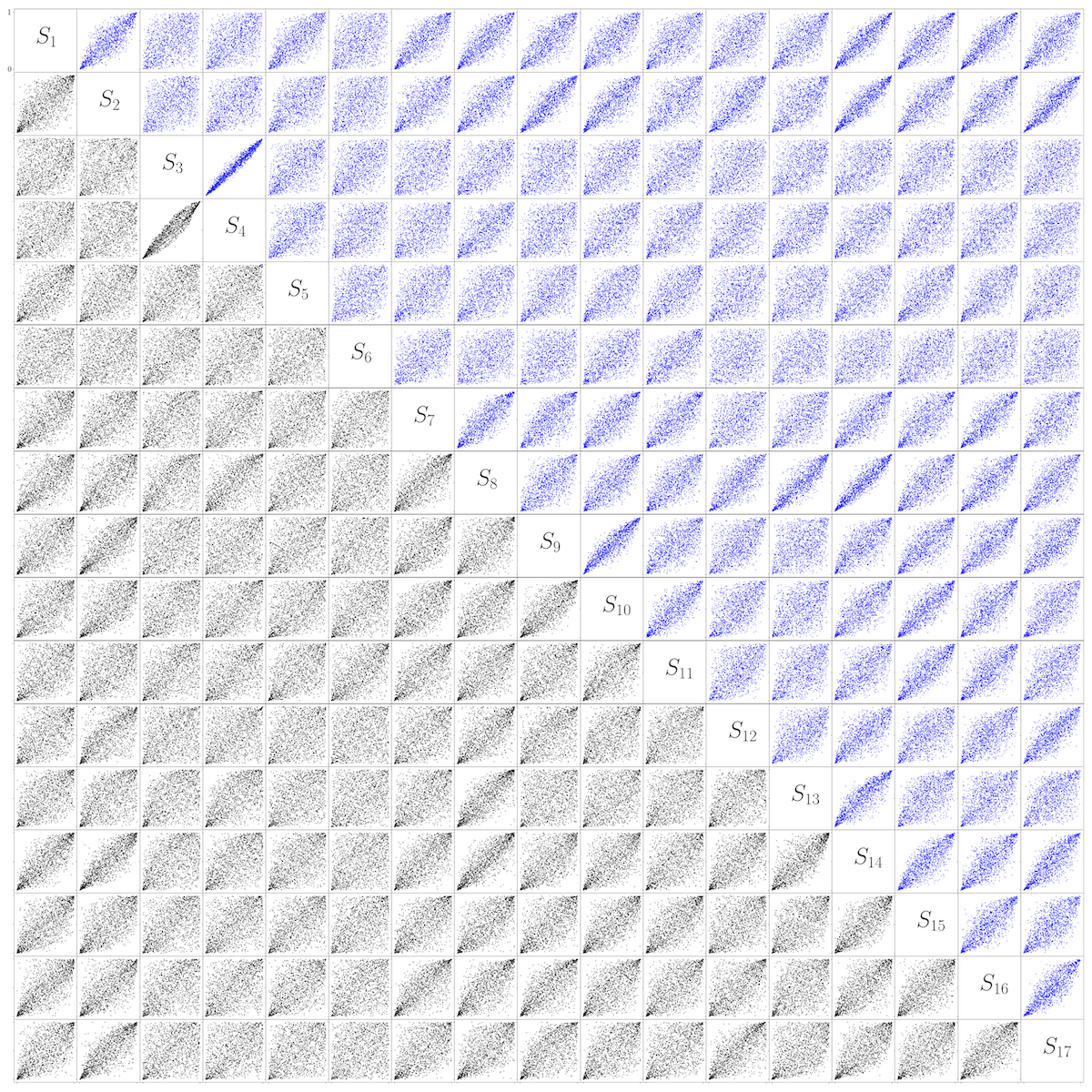} 
    \caption{(k) Variational autoencoder (VAE).}\label{fig:results_appl_nutrient_copula}
\end{figure}

\begin{figure}[h]\ContinuedFloat
    \centering
    \includegraphics[width=\textwidth]{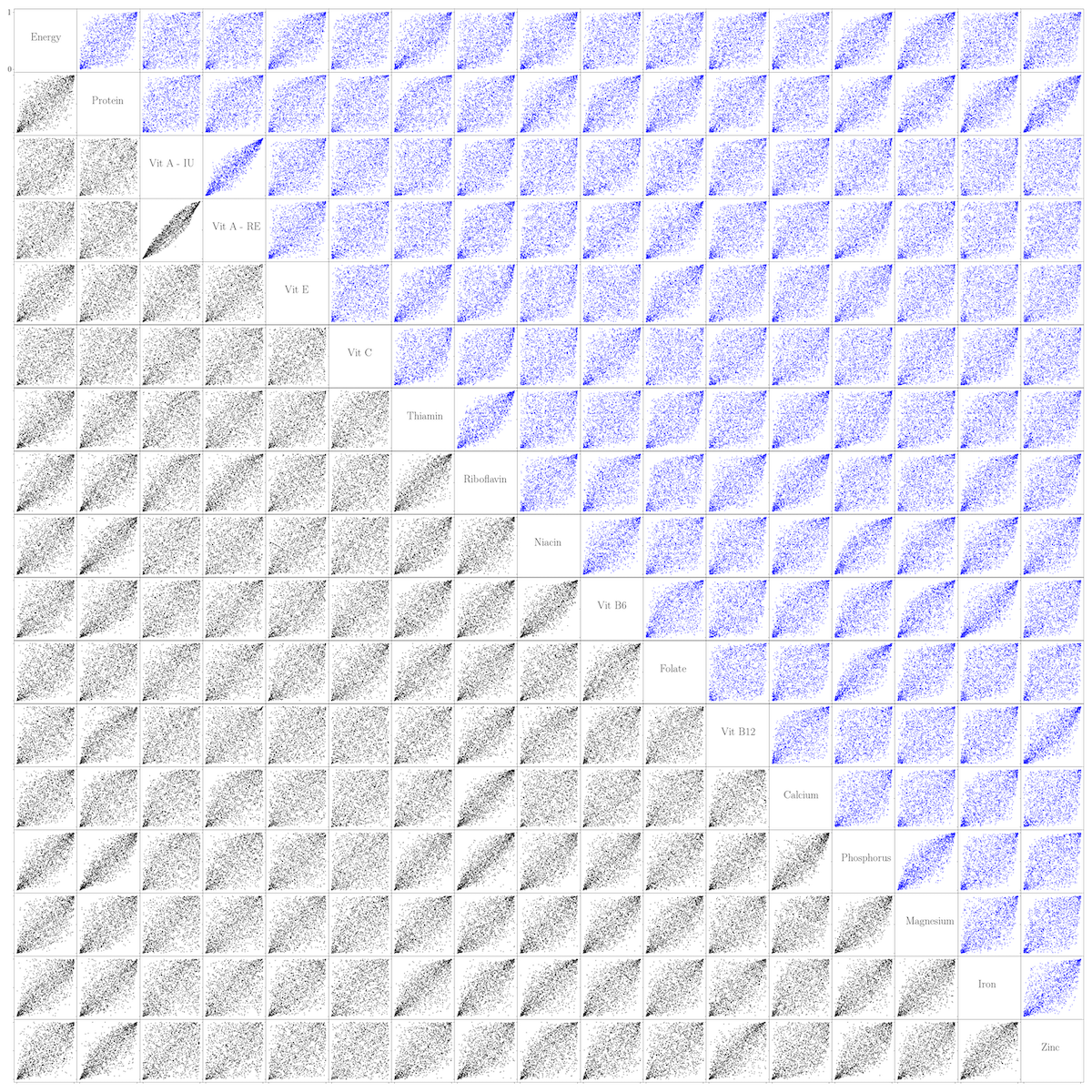} 
    \caption{(l) Archimax copula (Gen-AX *$\dagger$) inferred and sampled with our methods.}\label{fig:results_appl_nutrient_copula}
\end{figure}

\clearpage

\subsection{Extrapolating to extreme rainfall}\label{app:results_CNSD}

This data simulates the monthly rainfall for 3 locations (Belle-Ile, Groix, and Lorient) in French Brittany, where the dependence is asymmetric and non-extreme. We follow the Archimax model in~\citep{chatelain2020inference} and extend the experiment with different Archimedean generators of the same Kendall tau $\tau$. The methods used for comparison are Wasserstein GAN with gradient penalty (WGAN), masked autoregressive flow (MAF), variational autoencoder (VAE) and the state-of-the-art Clayton-Archimax copula (C-AX). All methods were first trained on all observations $n=240$ with dimension $d=3$. Many samples were then generated from the trained model to estimate $\hat{\ell}$ from block maximas using the state-of-the-art CFG estimator for extreme-value copulas~\citep{caperaa1997cfg}. As mentioned in the main paper, for extrapolating to extremes, we did not compare to other copula based models as many classical copulas are independent or Gumbel in the extremes. 

\begin{figure}[h]
    \centering
    \resizebox{\textwidth}{!}{%
    \begin{tabular}{ccccc}
    \textsc{GAN}  & \textsc{MAF}  & \textsc{VAE}  & \textsc{Ours} & \textsc{Ground Truth}\\
    
    \includegraphics[width=0.195\textwidth]{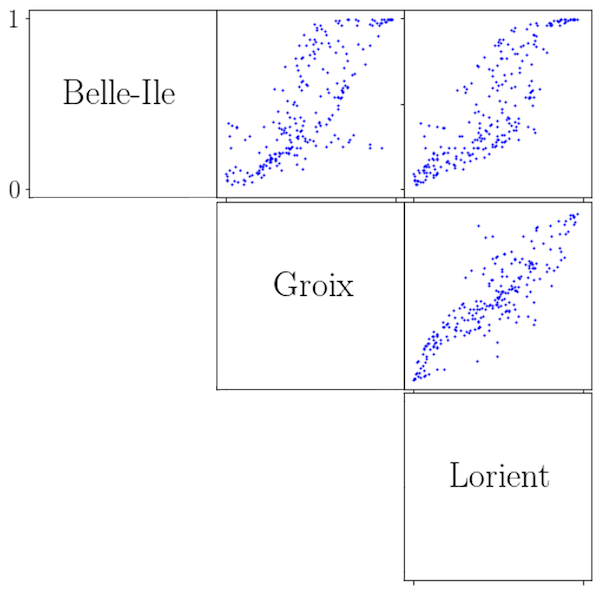}
    & \includegraphics[width=0.195\textwidth]{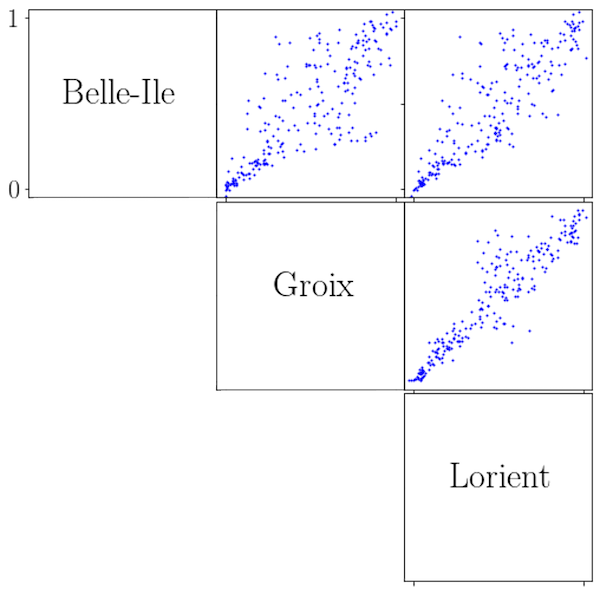}
    & \includegraphics[width=0.195\textwidth]{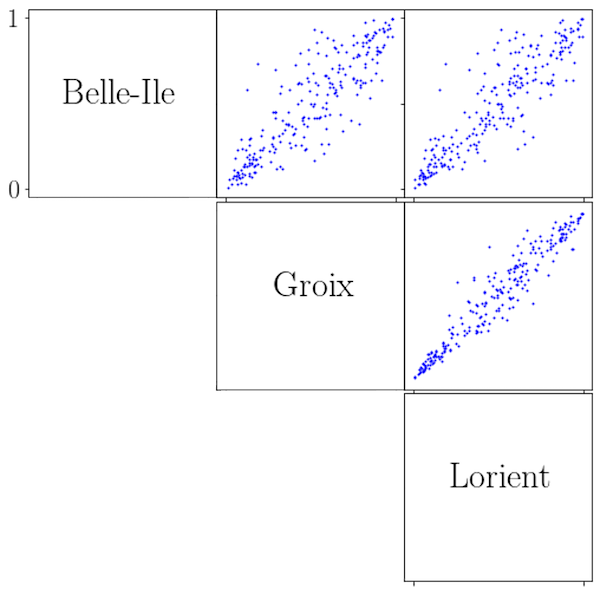}
    & \includegraphics[width=0.195\textwidth]{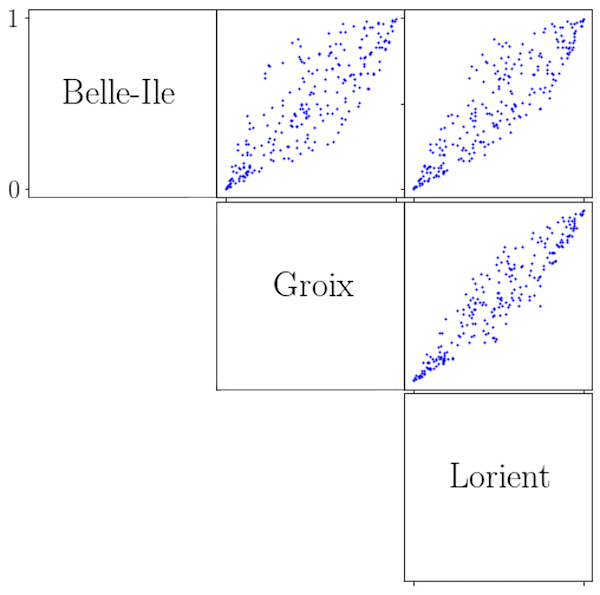}
    & \includegraphics[width=0.195\textwidth]{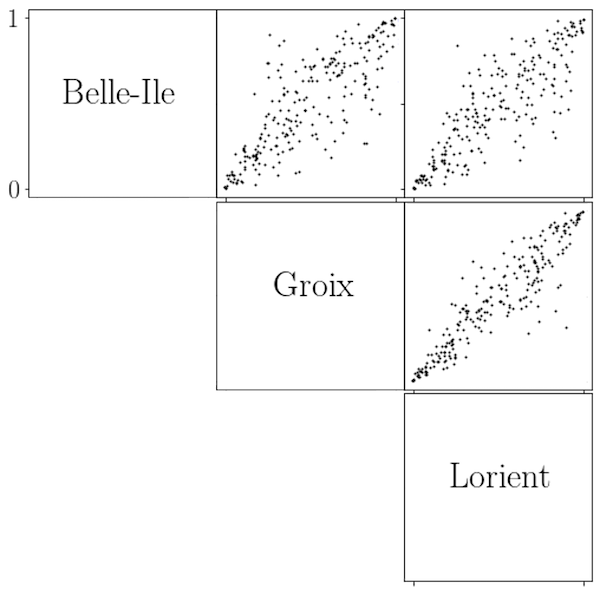}
    \\
    
    \includegraphics[width=0.195\textwidth]{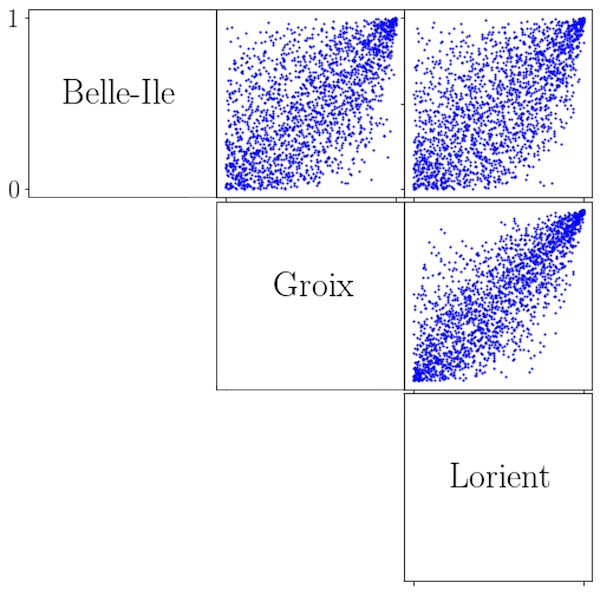}
    & \includegraphics[width=0.195\textwidth]{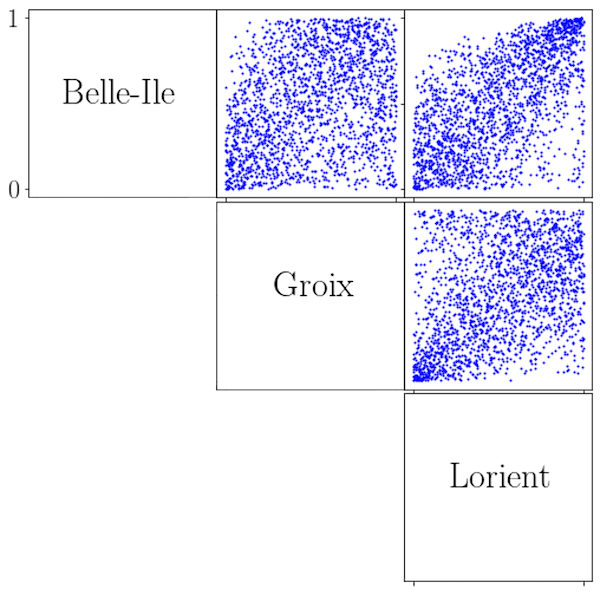}
    & \includegraphics[width=0.195\textwidth]{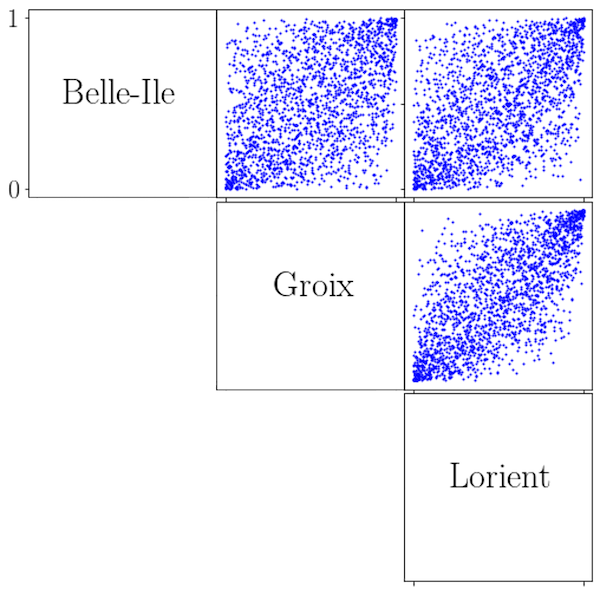}
    & \includegraphics[width=0.195\textwidth]{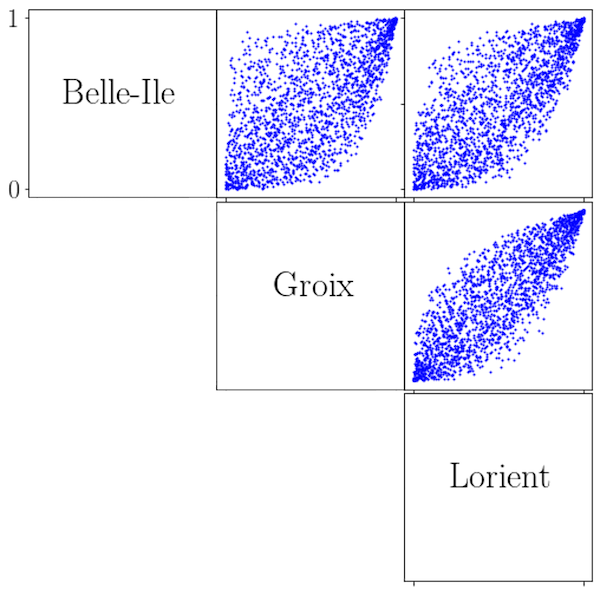}
    & \includegraphics[width=0.195\textwidth]{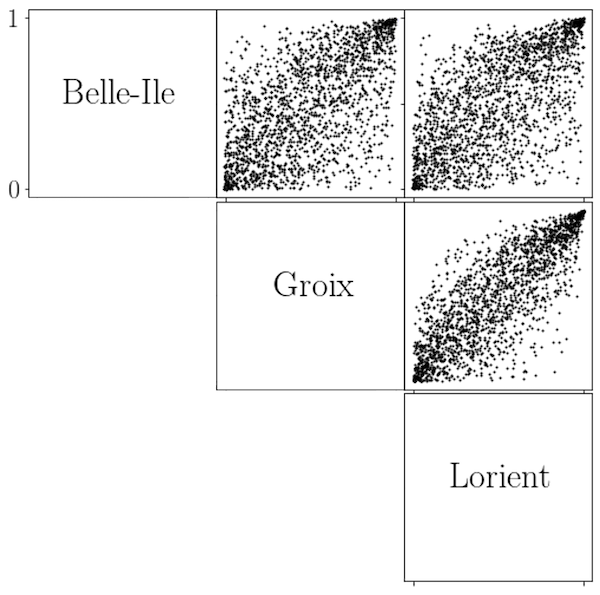}
    \\
    
    \includegraphics[width=0.195\textwidth]{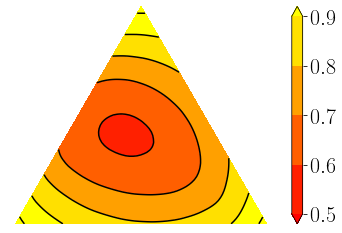}
    & \includegraphics[width=0.195\textwidth]{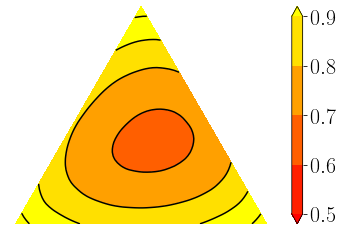} 
    & \includegraphics[width=0.195\textwidth]{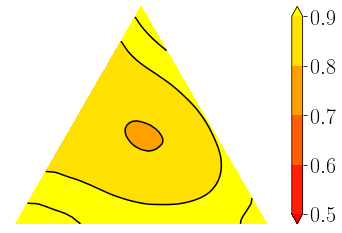} 
    & \includegraphics[width=0.195\textwidth]{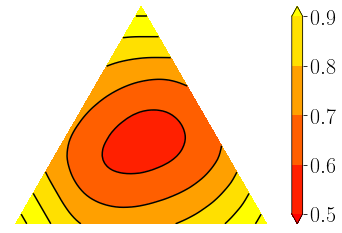} 
    & \includegraphics[width=0.195\textwidth]{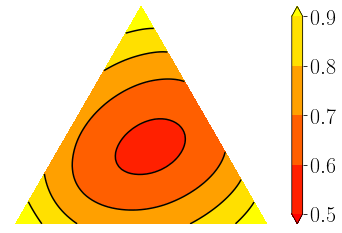} \\
    
    \end{tabular}
    }
    \caption{Extrapolating to dependence in the extremes. Plots of generated samples (top), generated samples in the extremes (middle) and $\ell(\mathbf{w}),\mathbf{w}\in\Delta_{3-1}$ (bottom).}
    \label{fig:results_appl_CNSD_extrapolate_app}
\end{figure}

Table~\ref{tab:extreme} of the main paper show our method performs consistently across the different Archimedean generators. In addition, our method always performs the best and in the case of ties, always among the best. Plots of the generated samples and generated samples in the extremes are given in~\figurename~\ref{fig:results_appl_CNSD_extrapolate_app}, with the Clayton-NSD Archimax copula experiment setting.

\clearpage

\subsubsection{Modeling monthly rainfall}~\label{app:results_CNSDmontly}

Although we did not compare to other copula based models for extrapolating to extremes, we compared to other copula based models for modeling monthly rainfall. In particular, the skew-t copula is a good addition to one's arsenal for flexible asymmetrical copulas~\citep{demarta2005tcopula, kollo2010skewtcopula, yoshiba2018skewtmle, smith2012skewtmcmc}.

For the Clayton-NSD Archimax copula experiment setting, the CvM distance for Archimax, Gaussian, extreme-value, t and skew-t copulas are: \textbf{0.0003} (lower is better), 0.0005, 0.0006, 0.0008, 0.0027.

For this scenario, the Archimax and Gaussian copula performed better than the extreme-value, t and skew-t copulas. This may be because the Clayton-NSD copula does not exhibit extreme-value dependence, i.e. it fails the test of extreme-value dependence~\citep{kojadinovic2011_evctest}. The skew-t copula might have performed not as well as the t copula due to over-parameterization. In addition, maximum likelihood estimation for the skew-t copula was extremely time consuming even for three dimensions and thus intractable for higher dimensions.

\subsection{Out-of-distribution detection}\label{app:results_ood}

\begin{figure}[h]
    \centering
    \resizebox{\textwidth}{!}{%
    \begin{tabular}{ccc}
    \textsc{MAF}  & \textsc{VAE}  & \textsc{Ours}\\
    \includegraphics[width=0.3\textwidth]{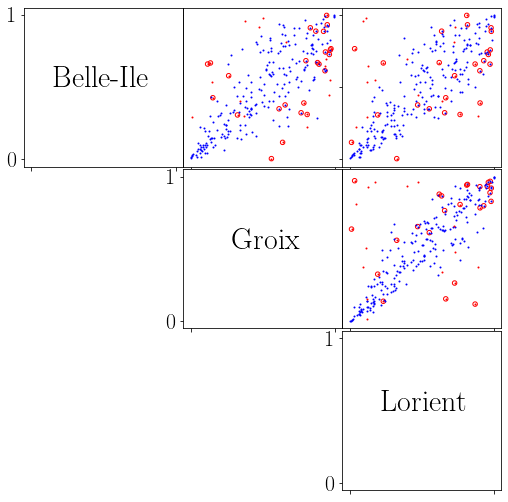} &
    \includegraphics[width=0.3\textwidth]{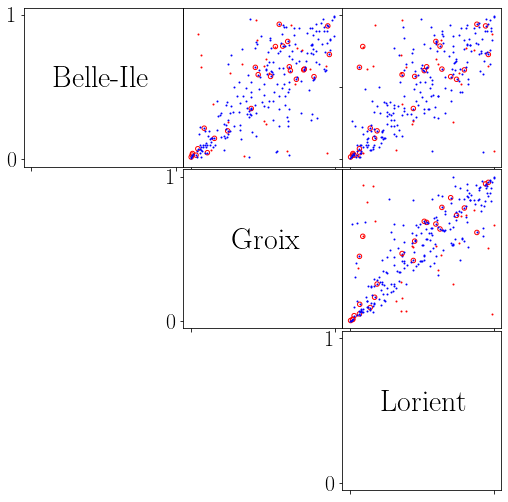} &
    \includegraphics[width=0.3\textwidth]{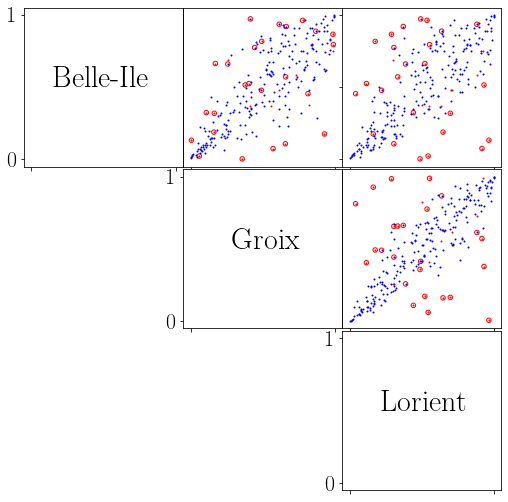}\\
    \end{tabular}}
    \caption{Out-of-distribution detection based on likelihoods. Inliers are represented with blue dots, outliers with red dots, and detected points are circled in red.}
    \label{fig:outlier}
\end{figure}

The inliers were generated from the Clayton-NSD copula representing the monthly rainfall of French Britanny and the outliers were generated uniformly at random on the unit cube. The number of inlier observations was $n_{in}=225$, the number of outlier observations was $n_{out}=25$, corresponding to 10\% data contamination. A visual comparison of the results is given in~\figurename~\ref{fig:outlier}. 

The masked autoregressive flow (MAF) provided an explicit likelihood. The likelihood of the variational autoencoder (VAE) was approximated from the reconstruction error. Including the KL divergence to the latent prior made results worse. The likelihood of the Archimax copula was approximated from the inclusion-exclusion scheme, checked to converge using various interval sizes.

\subsection{High dimensional modeling}\label{app:results_100d}

We infer and sample a 100-dimensional Clayton-NSD Archimax copula, with parameters $\theta=2,\;\alpha = (\alpha_0,\cdots,\alpha_0),\; \alpha_0=(1,1,1,1,2,2,2,3,3,4),\allowbreak \rho=0.69$ using our method. The inference results are given in Table~\ref{tab:nutrient} of the main paper. We plot the samples in~\figurename~\ref{fig:results_appl_100d}, 20 coordinates at a time, with coordinates $(0, 1, 2, 3, 4, 5, 6, 7, 8, 9, 20, 21, 22, 23,\allowbreak 24, 25, 26, 27, 28, 29)$ in~\figurename~\ref{fig:results_appl_100d}~(a) and coordinates $(0, 10, 20, 30, 40, 50, 60, 70, 80, 90, 9, 19, 29, 39, 49,\allowbreak 59, 69, 79, 89, 99)$ in~\figurename~\ref{fig:results_appl_100d}~(b). The generated samples are in blue above the diagonal, the true samples are in black below the diagonal.

\begin{figure}[h]
    \centering
    \includegraphics[width=\textwidth]{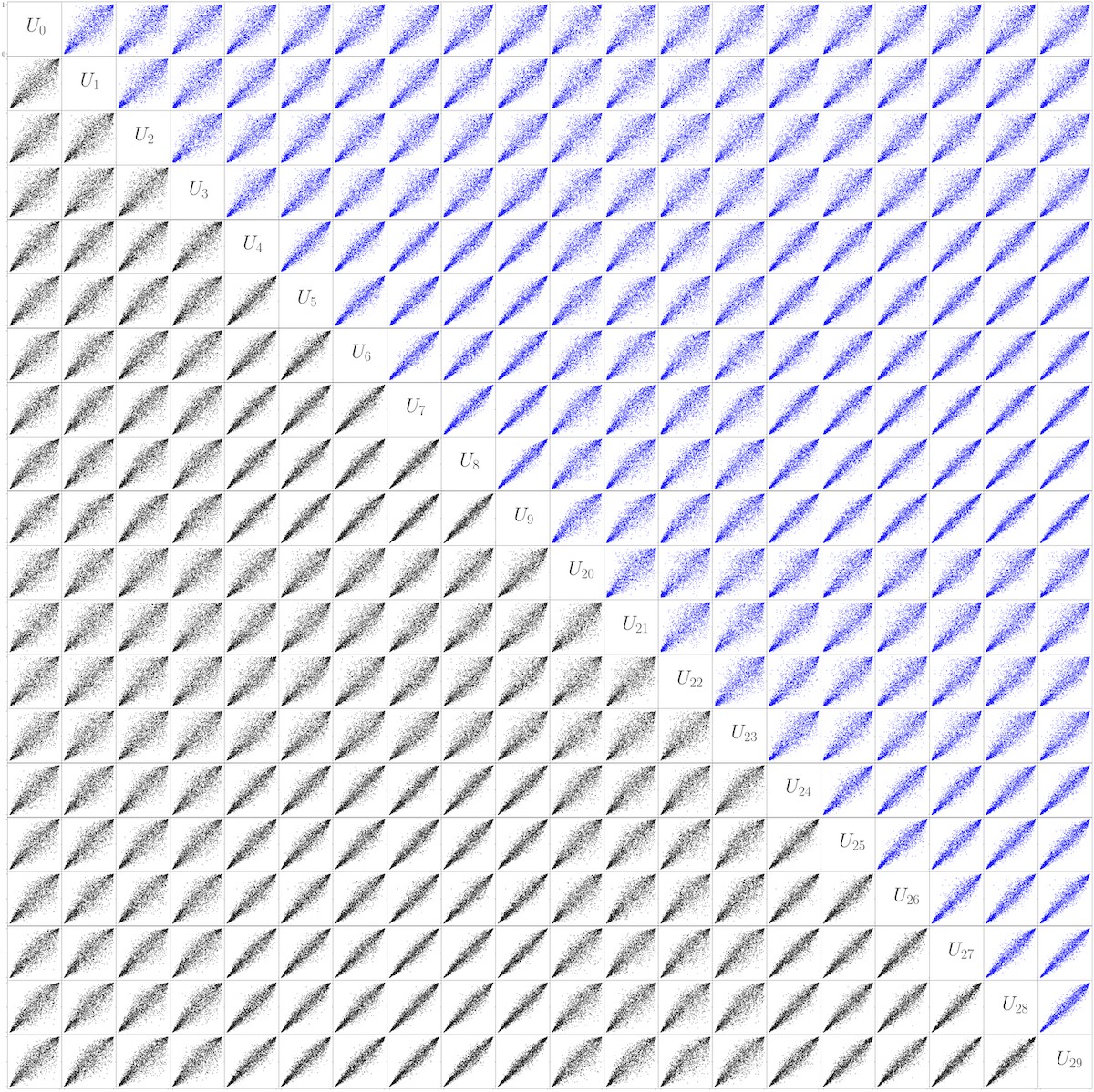}
    \caption{(a) Samples from 100-dimensional Clayton-NSD copula, showing coordinates $(0,1,2,3,4,5,6,7,8,9,20,21,22,23,24,25,26,27,28,29)$.}
    \label{fig:results_appl_100d}
\end{figure}

\begin{figure}[h]\ContinuedFloat
    \centering
    \includegraphics[width=\textwidth]{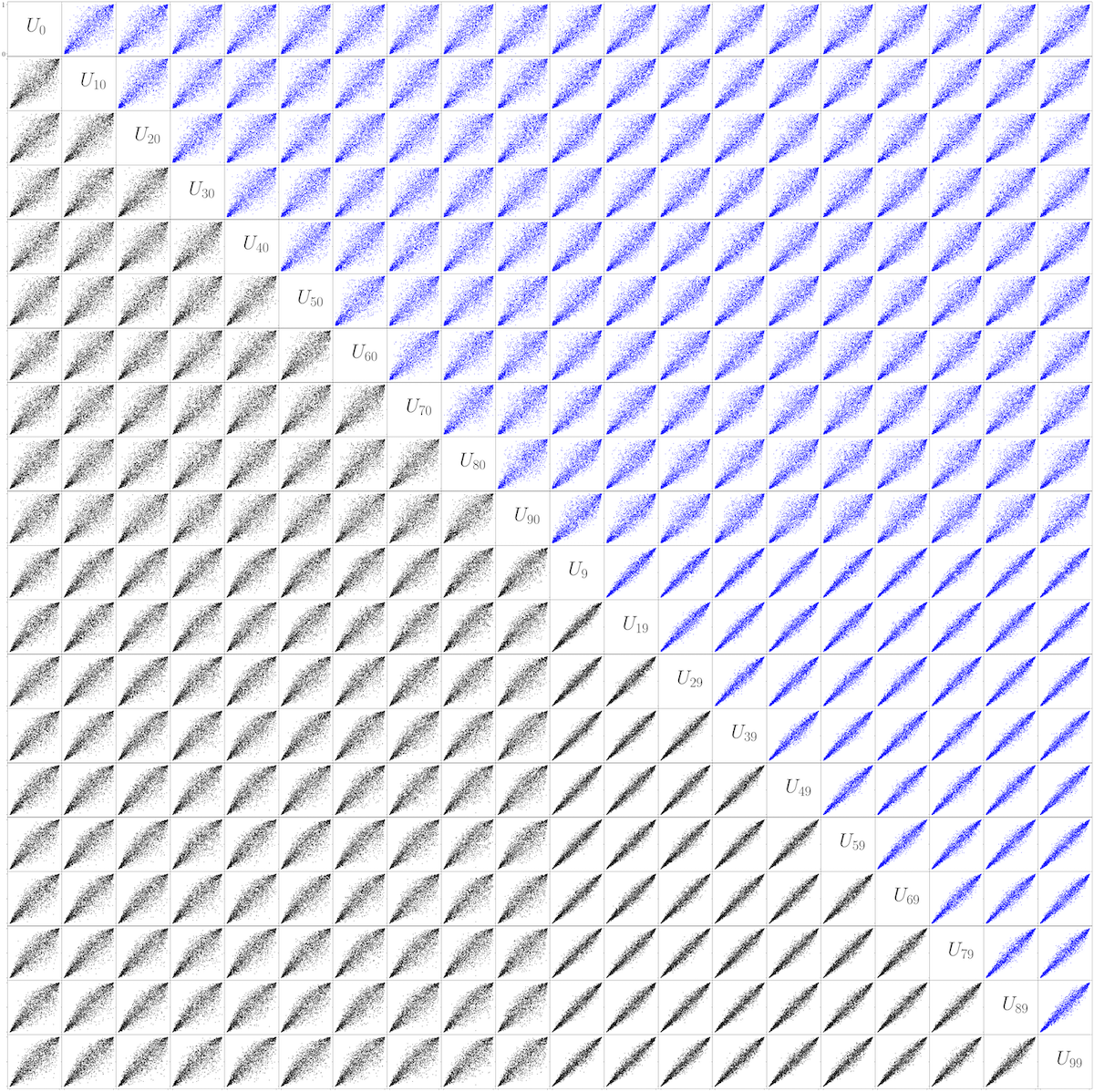}
    \caption{(b) Samples from 100-dimensional Clayton-NSD copula, showing coordinates $(0, 10, 20, 30, 40, 50, 60, 70, 80, 90, 9, 19, 29, 39, 49, 59, 69, 79, 89, 99)$.}
    \label{fig:results_appl_100d}
\end{figure}

\clearpage

\bibliography{main}

\clearpage

\section*{Checklist}

\begin{enumerate}

\item For all authors...
\begin{enumerate}
  \item Do the main claims made in the abstract and introduction accurately reflect the paper's contributions and scope?
    \answerYes{}
  \item Did you describe the limitations of your work?
    \answerYes{}
  \item Did you discuss any potential negative societal impacts of your work?
    \answerYes{}
  \item Have you read the ethics review guidelines and ensured that your paper conforms to them?
    \answerYes{}
\end{enumerate}

\item If you are including theoretical results...
\begin{enumerate}
  \item Did you state the full set of assumptions of all theoretical results?
    \answerYes{}
        \item Did you include complete proofs of all theoretical results?
    \answerYes{See Appendix A.}
\end{enumerate}

\item If you ran experiments...
\begin{enumerate}
  \item Did you include the code, data, and instructions needed to reproduce the main experimental results (either in the supplemental material or as a URL)?
    \answerYes{See Appendix B, and the code in the supplemental material is to be made available online.}
  \item Did you specify all the training details (e.g., data splits, hyperparameters, how they were chosen)?
    \answerYes{See Appendix B.}
        \item Did you report error bars (e.g., with respect to the random seed after running experiments multiple times)?
    \answerYes{See Appendix B.}
        \item Did you include the total amount of compute and the type of resources used (e.g., type of GPUs, internal cluster, or cloud provider)?
    \answerYes{See Appendix B.}
\end{enumerate}

\item If you are using existing assets (e.g., code, data, models) or curating/releasing new assets...
\begin{enumerate}
  \item If your work uses existing assets, did you cite the creators?
    \answerYes{Nutrient intake data~\citep{usda1985nutrient}, Copulas \url{https://github.com/sdv-dev/Copulas}, HAC toolbox \url{https://github.com/gorecki/HACopula}, PyTorch, SciPy, NumPy, Python.}
  \item Did you mention the license of the assets?
    \answerNo{Most information presented on the USDA Web site is considered public domain information. Public domain information may be freely distributed or copied, but use of appropriate byline/photo/image credits is requested. Attribution may be cited as follows: "U.S. Department of Agriculture."}
  \item Did you include any new assets either in the supplemental material or as a URL?
    \answerYes{The code in the supplemental material is to be made available online.}
  \item Did you discuss whether and how consent was obtained from people whose data you're using/curating?
    \answerNo{Data was collected only after consent was given.}
  \item Did you discuss whether the data you are using/curating contains personally identifiable information or offensive content?
    \answerNo{Personally identifiable information has been removed.}
\end{enumerate}

\item If you used crowdsourcing or conducted research with human subjects...
\begin{enumerate}
  \item Did you include the full text of instructions given to participants and screenshots, if applicable?
    \answerNo{The instructions given to participants is available at~\citep{usda1985nutrient}.}
  \item Did you describe any potential participant risks, with links to Institutional Review Board (IRB) approvals, if applicable?
    \answerNo{The description of potential participant risks is available at~\citep{usda1985nutrient}.}
  \item Did you include the estimated hourly wage paid to participants and the total amount spent on participant compensation?
    \answerNo{The description of participant compensation is available at~\citep{usda1985nutrient}.}
\end{enumerate}

\end{enumerate}


\end{document}